\documentclass{aa}  

\usepackage{graphicx}

\usepackage{xcolor}

\usepackage{placeins}           % useful with \FloatBarrier, to keep 

\usepackage{siunitx}
\newcommand{\micron}{$\mu$m}
\newcommand{\msun}{M$_{\odot}$}
\newcommand{\lsun}{L$_{\odot}$}
\usepackage{tikz,lipsum,lmodern}
\usepackage[most]{tcolorbox}
\newcommand{\tdust}{$\rm T_{d}$}
\newcommand{\mdust}{$\rm M_{d}$}

\newcommand{\lbol}{$\rm L_{bol}$}

\newcommand{\lfir}{$\rm L_{FIR}$}
\newcommand{\mh}{$\rm M_{H2}$}
\newcommand{\lcii}{$\rm L_{[CII]}$}
\newcommand{\lco}{$\rm L^{\prime}_{CO}$}
\newcommand{\CII}{\mbox{[C\,{\sc ii}]}}

\newcommand{\uaco}{$\rm M_{\odot}\ pc^{-2}\ (K\ km\ s^{-1})^{-1}$}
\newcommand{\aco}{$\alpha_{CO}$}

\usepackage{txfonts}
\begin{document}

   \title{Molecular gas and dust properties in $z>7$ quasar hosts}

   \subtitle{}

   \author{Francesco Salvestrini
          \inst{1, 2}
          \and
          Chiara Feruglio
          \inst{1, 2}
          \and
          Roberta Tripodi
          \inst{3}
          \and
          Fabio Fontanot
          \inst{1, 2}
          \and
          Manuela Bischetti
          \inst{4, 1, 2}
          \and
          Gabriella De Lucia
          \inst{1, 2}
          \and
          Fabrizio Fiore
          \inst{1, 2}
          \and
          Michaela Hirschmann
          \inst{5, 1}
          \and
          Umberto Maio
          \inst{1, 2}
          \and
          Enrico Piconcelli
          \inst{6}
          \and
          Ivano Saccheo
          \inst{7, 6}
          \and
          Alessia Tortosa
          \inst{6}
          \and
          Rosa Valiante
          \inst{6, 8}
          \and
          Lizhi Xie
          \inst{9, 1}
          \and
          Luca Zappacosta
          \inst{6}
          }

   \institute{
         INAF - Osservatorio Astronomico di Trieste, Via G. Tiepolo 11, I-34143 Trieste, Italy
         \and
         IFPU - Institute for Fundamental Physics of the Universe, via Beirut 2, I-34151 Trieste, Italy
         \and
         University of Ljubljana, Department of Mathematics and Physics, Jadranska ulica 19, SI-1000 Ljubljana, Slovenia
         \and
         Dipartimento di Fisica, Universit\'a di Trieste, Sezione di Astronomia, Via G.B. Tiepolo 11, I-34131 Trieste, Italy
         \and
         Institute for Physics, Laboratory for Galaxy Evolution and Spectral Modelling, Ecole Polytechnique Federale de Lausanne, Observatoire de Sauverny, Chemin Pegasi 51, CH-1290 Versoix, Switzerland 
         \and
         INAF-Osservatorio Astronomico di Roma, Via Frascati 33, I-00040 Monte Porzio Catone, Italy
         \and
         Dipartimento di Matematica e Fisica, Universit\'a Roma Tre, Via  della Vasca Navale 84, I-00146 Roma, Italy
         \and
         INFN-Sezione Roma1, Dipartimento di Fisica, Universit\'a di Roma La Sapienza, Piazzale Aldo Moro 2, I-00185 Roma, Italy
         \and
         Tianjin Normal University, Binshuixidao 393, Xiqing, 300387, Tianjin, People's Republic of China
         }

   \date{}

\abstract{Observational campaigns hunting the elusive reservoirs of cold gas in the host galaxies of quasars at the epoch of reionization (EoR) are crucial for studying the formation and evolution of the first massive systems at early epochs.
   We present new Northern Extended Millimeter Array (NOEMA) observations tracing CO(6--5) and CO(7--6) emission lines as well as the underlying continuum in five of the eight quasars at redshift $z>7$ known to date, thus completing the survey of the cold molecular gas reservoir in the host galaxies of the first quasars.
   Combining NOEMA observations with archival Atacama Large Millimeter/submillimeter Array (ALMA) data, we modeled the far-infrared spectral energy distribution with a modified blackbody function to measure dust properties and star formation rates.
   We used CO and [CII] lines to derive molecular gas masses, which we compared with results from semi-analytic models and observations of galaxies at different epochs.
   No statistically significant detection of CO emission lines was reported for the five quasars in this sample, resulting in a relatively low amount of cold molecular gas in the host when compared with galaxies at later epochs.
   Nonetheless, gas-to-dust ratios are consistent with the local value, suggesting that the scaling relation between dust and cold gas holds up to $z>7$.
   Quasars at the EoR show star formation efficiencies that are among the highest observed so far and comparable with those observed in luminous quasars at Cosmic Noon and those predicted for the brightest ($L_{bol}>3\times10^{46}$ erg s$^-1$) quasar objects drawn from the semi-analytic model GAEA.
   Quasar host galaxies at the EoR are undergoing an intense phase of star formation, which suggests a strong coupling between the luminous phase of the quasar and the rapid growth of the host.}

   \keywords{Quasars: emission lines - Galaxies: ISM - Galaxies: evolution - Galaxies: high-redshift}

   \maketitle
%
%-------------------------------------------------------------------

\section{Introduction}
\label{sec:intro}
The significant role of quasars in reionizing the Universe at $z>6$ has been known for many years \citep{GunnPeterson65}.
However, our understanding has advanced considerably over the past couple of decades, paralleling the growth in the number of known quasars, which now includes approximately 300 identified at the epoch of reionization \citep[EoR;][]{Fan23}.
This has been achieved thanks to the combination of near-infrared photometric surveys and spectroscopic follow-up observations, which have proven to be reliable tools for discovering the quasar population at high redshifts ($z>5$; e.g., \citealt{Banados16, Shen19}).

Luminous quasars at the EoR are also crucial probes for testing the coevolution between supermassive black holes (SMBHs) and their host galaxies.
In this regard, modern interferometric facilities have been essential for confirming the redshift of the objects and constraining the star formation activity and properties of the cold phase of the interstellar medium (ISM) of their host galaxies (e.g., \citealt{Riechers06, Riechers07, Riechers09, WangR07, WangR10}).
Far-infrared (FIR)/submillimeter observations have revealed the presence of highly star-forming host galaxies, with star formation rates (SFRs) of up to 1000-3000 \msun~yr$^{\rm -1}$, and copious amount of dust (M$_{\rm dust}>10^8$ \msun; \citealt{Maiolino05, WangR13, Feruglio18, Pensabene22, Venemans17a, Venemans20}).
These massive quasar hosts are crucial for benchmarking gas and dust content and star formation efficiencies (SFEs) during the mid-point and terminal stages of the EoR, but observations are still limited to a few tens of quasar hosts at these epochs.
Furthermore, the scaling relation between gas and dust is not clear yet, and the methods that are currently adopted to derive molecular gas masses can be improved with more data.
Given its brightness, the 158\micron~emission line of the singly ionized carbon atom (\CII~hereafter) is the most used tracer of the ISM at $z>4$.
\begin{table*}[th]
\vspace{0.2cm}
                \caption{NOEMA observations.}
                \centering
                \begin{tabular}{llllcccc}
                        \hline\hline
Name & Ra,Dec & $z_{\rm \CII~}$ & Freq. & Beam & R.m.s. & S$_\nu$ & Size\\
   &   h:m:s, d:m:s (J2000.0)    &                 & [GHz] & [arcsec$^2$] & [$\mu$Jy/beam] & [$\mu$Jy] & [arcsec$^2$] \\
\hline
J0313-1806 & 03:13:43.84, $-$18:06:36.40 & 7.6423 & USB 92.568  & $13.1\times 3.7$ & 48.5 &  \phantom{0}$<100.5^{(a)}$ &\\
           &                            &        & LSB 78.756  & $16.4\times 4.3$ & 46.3 & \phantom{0}-- & \\
           &                            &        & B6 219.911     & \phantom{0}$0.62\times 0.44$ & 19 & \phantom{0}$0.50\pm0.05$ & $0.71\times 0.55$ \\
J1243+0100 & 12:43:53.93, +01:00:38.50   & 7.0749 & USB 100.310 & \phantom{0}$5.8\times2.3$ & 16.1 & $119\pm16$ & \\
           &                            &        & LSB 86.756  & \phantom{0}$6.3\times2.8$ & 13.0 &\phantom{0} $49.8\pm12.5$ & \\
           &                            &        & B6 235.364     & \phantom{0}$0.62\times 0.50$ & 19 & \phantom{0}$2.07\pm0.09$ & $0.78\times 0.72$ \\
J0038-1527 & 00:38:36.10, $-$15:27:23.60 & 7.034  & USB 100.821 & \phantom{0}$6.2\times3.8$ &22.3 & \phantom{0}$<44.4^{(a)}$ & \\          
           &                            &        & LSB 87.256  & \phantom{0}$7.3\times 4.4$ & 19.8 & \phantom{0}-- & \\
           &                            &        & B6 236.562     & \phantom{0}$0.62\times 0.55$ & 37 & \phantom{0}$1.8\pm0.7$ & $1.9\times 1.2$ \\
J2356+0017 & 23:56:46.33, +00:17:47.30   & 7.01  & USB 101.373 & \phantom{0}$4.6\times3.5$ & 15.8 & \phantom{0}$54\pm15$ & \\
           &                            &        & LSB 87.756  & \phantom{0}$5.4\times4.1$  & 13.4 & \phantom{0}$<40.2$ & \\
J0252-0503 & 02:52:16.64, $-$05:03:31.80 & 7.0006 & USB 101.491 & \phantom{0}$5.6\times4.2$ & 19.0 & $123.4\pm17.8$ & \\
           &                            &        & LSB 87.756  & \phantom{0}$6.6\times4.9$ & 15.6 & \phantom{0}$76.9\pm14.9$ & \\
           &                            &        & B6 237.852     & \phantom{0}$0.56\times0.40$ & 35 & \phantom{0}$1.04\pm0.10$ & $0.72\times 0.55$  \\
\hline
                \end{tabular}
                \label{tab:observations}
                \flushleft 
                \footnotesize {{\bf Notes.} NOEMA fluxes are derived from a fit of the visibilities with a point source model. ALMA band 6 (B6) fluxes are the integrated emission of best-fit 2D Gaussian function model of the cleaned continuum map. Upper limits are at the 3$\sigma$ significance level. R.m.s. is estimated using a channel width of 20 MHz. $^{(a)}$ upper limits are computed for the merged USB+LSB continuum data.}
\end{table*}
While the \CII~emission is undoubtedly useful for dynamical studies (e.g., \citealt{WangR13, Decarli17, Venemans19, Venemans20, Neeleman21, WangF24}), it may not effectively trace the dense and cold components of the molecular gas, nor the star-forming regions (e.g., \citealt{Pineda13, HerreraCamus18, Neeleman19, Bischetti24, Izumi24}).
On the other hand, the rotational excited transitions of carbon monoxide (CO) are much more reliable probes of the dense cold $H_2$ and are pivotal when investigating the star formation process \citep{Kaasinen24}. 
Despite being fainter than \CII, CO lines have been successfully detected in massive quasar hosts at $z\sim6$ (e.g., \citealt{WangR13, Decarli22}); however, they remain poorly sampled at the highest redshifts ($z>7$) of quasars, with only one confirmed individual detection \citep{Feruglio23} and a possible detection suggested by the stacking of multiple lines \citep{Venemans17a, Novak19} at $z\sim7.1-7.5$.

In this work we present new observations targeting CO(6-5) and (7-6) emission lines and the underlying continuum in five quasars at $z > 7$, obtained with the Northern Extended Millimeter Array (NOEMA), thus providing a complete census of the molecular gas in the population of $z>7$ quasars known to date \citep{Fan23}.
We also investigated the star formation activity and dust properties for the five quasars by modeling the FIR spectral energy distribution (SED) using NOEMA and archival observations of the Atacama Large Millimeter/submillimeter Array  (ALMA).
Based on these measurements, we studied the growth of host galaxies and the implications for their coevolution with SMBHs in the reionization era.

This paper is organized as follows.
Section~\ref{sec:obs} describes the observations and data reduction.
Section~\ref{sec:results} describes the procedures adopted to model the FIR SED and, in turn, measure the dust and cold molecular gas content, including a new calibration using the [CII] luminosity.
In Sect.~\ref{sec:discussion} we discuss the results and their implication for the coevolution of the host galaxy of quasars at the EoR, and compare them with objects at lower redshifts and the results from semi-analytic models (SAMs).
We summarize the paper in Sect.~\ref{sec:concl}.

Throughout the paper, we adopt a standard flat ${\rm \Lambda}$ cold dark matter cosmology with the matter density parameter ${\rm \Omega_{M}} = 0.30$, the dark energy density parameter ${\rm \Omega_{\Lambda}}= 0.70$, and the Hubble constant H$_{\rm 0} = 70$ km s$^{\rm -1}$.

\section{Sample and observations}
\label{sec:obs}

We focused on the quasars with $z\geq7$, which are eight in total according to \cite{Fan23}, aiming to obtain a complete census of the molecular gas content in these objects.
The five quasars for which we acquired new NOEMA observations under project S23CX are listed in Table~\ref{tab:observations}, while the entire list of $z>7$ quasars with their names and identifications are reported in Table~\ref{tab:properties}.

Regarding project S23CS, receivers were tuned to detect the CO(6--5) and CO(7--6) and the underlying continuum in the lower and upper side band (LSB and USB), respectively. 
We calibrated the visibilities using the CLIC pipeline of the GILDAS software\footnote{\url{www.iram.fr/IRAMFR/GILDAS}}.
Imaging was performed with MAPPING in GILDAS, using a natural weighting scheme for visibilities, with a detection threshold equal to the noise of the pre-imaging visibilities.
The observation targeting quasar J0313 was executed in bad weather conditions, so it did not reach the requested sensitivity to detect either continuum emission or CO lines in both LSB and USB.
Given the low declination of J0038 (coordinates are listed in Table~\ref{tab:observations}), the observation of this target was limited by shadowing, which reduced the total observing time on-source, hence the final sensitivity.
In these two cases, we merged LSB and USB to increase the signal-to-noise ratio, and we used the merged cube to extract an upper limit on the continuum emission.  
Continuum-subtracted cubes were binned to spectral resolutions of 20 MHz, corresponding to $\sim60$ km s$^{-1}$, to optimize the sensitivity.

To extend the sampling of the FIR emission of the quasar host galaxies, we collected archival ALMA observations for four of the five quasars\footnote{Project IDs of the dataset analyzed are listed in the Acknowledgements. J2356 observation with project ID 2023.1.00443.S is still in a proprietary period at the moment of this analysis}, which cover sky frequencies $\sim220-240$ GHz, where the \CII~emission line falls.
ALMA-calibrated visibilities of observations covering \CII~and the underlying continuum were retrieved from the science archive (see the acknowledgements for the list of ALMA project IDs). 
The imaging was performed through the Common Astronomy Software Applications (CASA; \citealt{McMullin07}), version 6.5.5-21.
We ran \texttt{tclean} procedures using natural weighting, a 3$\sigma$ cleaning threshold, and a channel width of 20 MHz.
Continuum-subtracted cubes and continuum maps were created using the \emph{imfit} procedure in CASA, fitting a constant function to all line-free channels.

The synthesized beams, root-mean-square (rms) noise levels, and representative frequencies of the continuum maps are reported in Table~\ref{tab:observations}.
Continuum maps of ALMA and NOEMA observations are shown in Fig.~\ref{fig:cont_maps}.

\begin{figure}[t]
\includegraphics[width=0.5\textwidth]{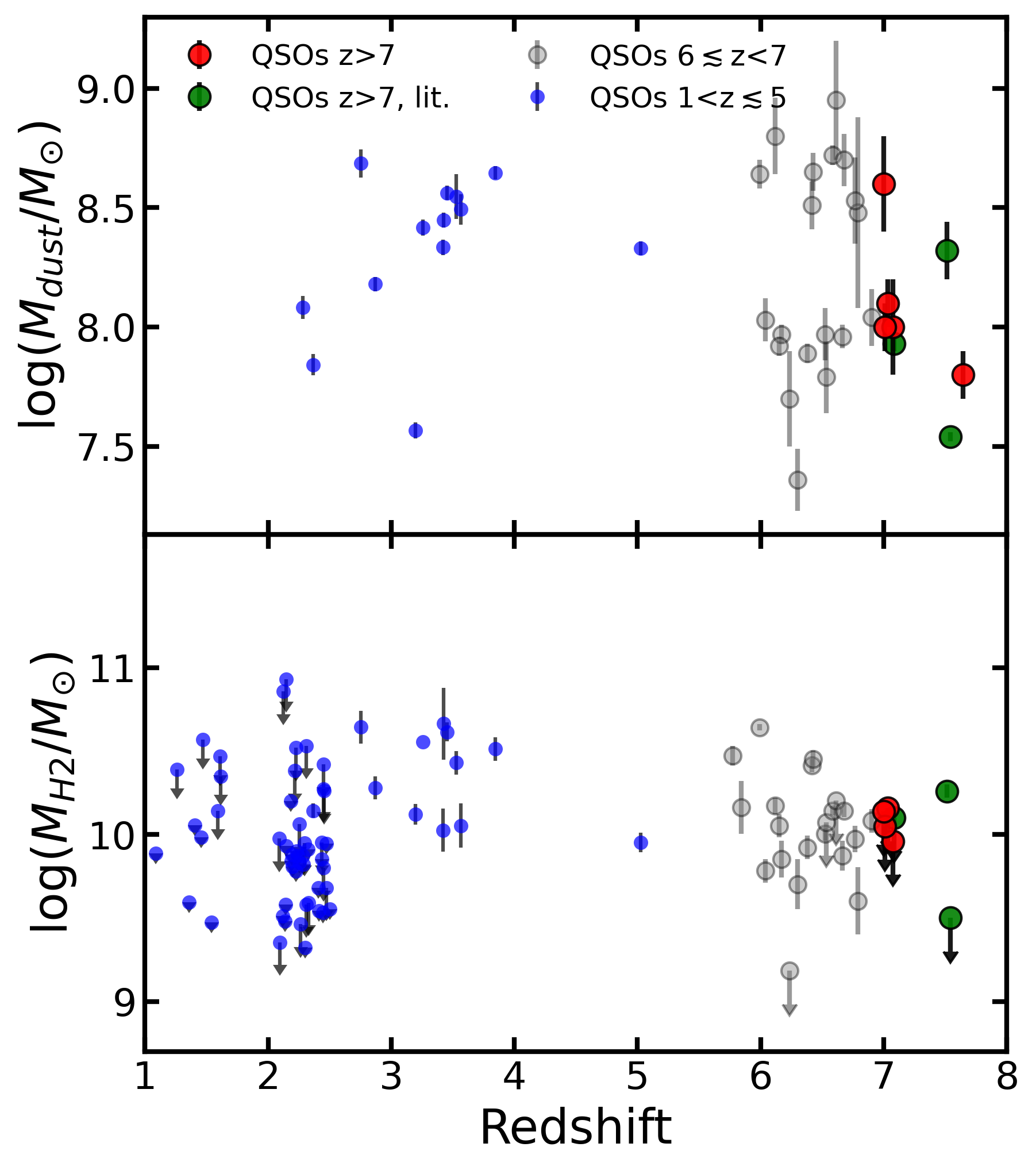}
\caption{Dust mass and molecular gas mass vs. redshift. 
\emph{Upper panel:} \mdust\  vs. redshift. Red circles are the $z>7$ quasars presented in this work. Green circles are for J1120 from \cite{Venemans17a}, J1007 from \cite{Feruglio23}, and J1342 from \cite{Novak19}. Gray circles are the $5<z<7$ quasars from \cite{Venemans17b}, \cite{Izumi21}, \cite{Decarli22}, and \cite{Tripodi24b}, while blue dots are for quasars at intermediate redshifts ($1<z<5$) from \cite{Bischetti21} and Salvestrini et al. (in prep.).
\emph{Lower panel:} \mh~vs. redshift. Data are color-coded as in the upper panel. We also include the quasars at intermediate redshifts ($1<z<5$) from \cite{Bertola24}.}
          \label{fig:mass_redshift}
\end{figure}

\begin{figure*}[t]
\includegraphics[width=\textwidth]{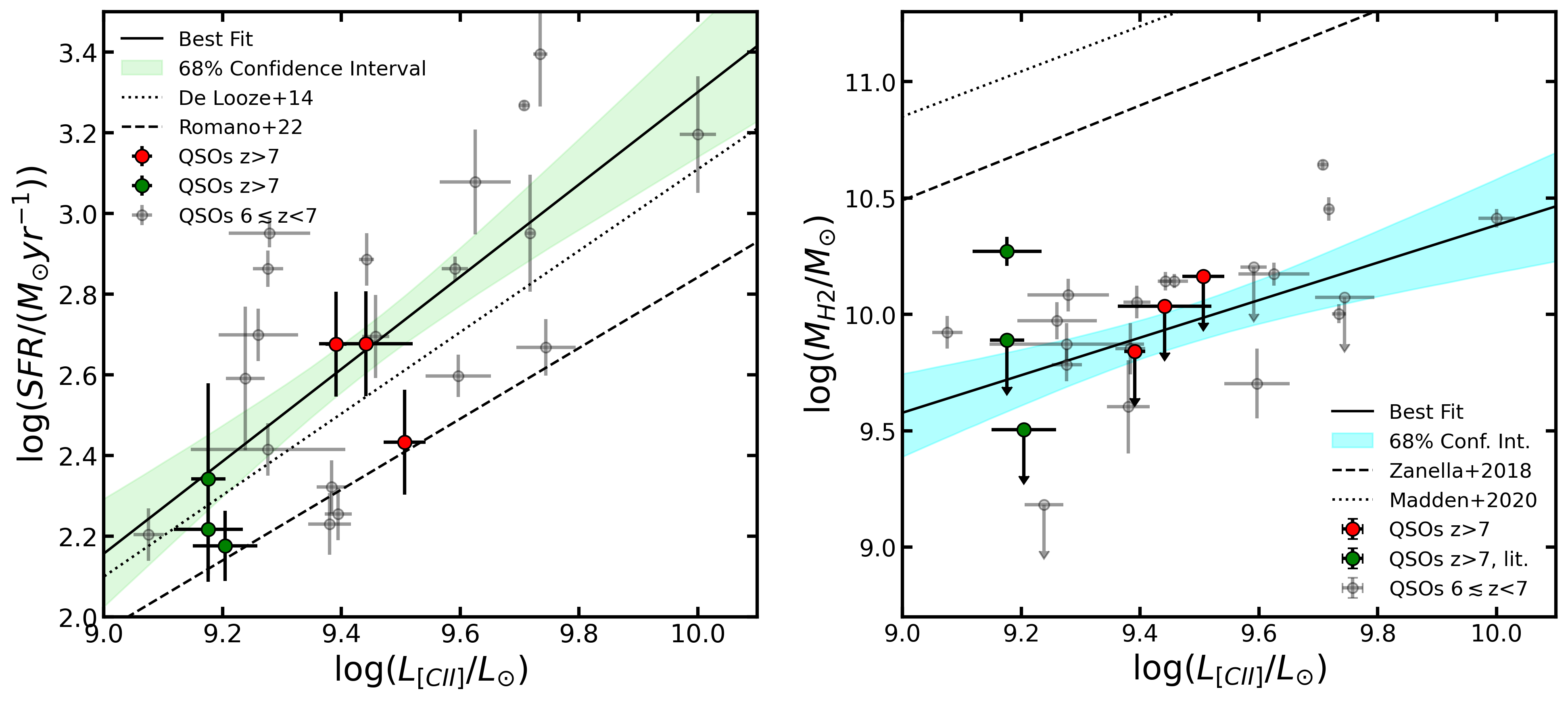}
\caption{SFR and molecular gas mass vs. [CII] luminosity. 
\emph{Left panel:} $z>7$ quasars with $L_{[CII]}$ measurements: J1234, J0038, and J0252 (red circles) and J1120, J1342, and J1007 (green circle).  
Gray dots are the  $z\gtrsim6$ quasars from the literature (see Table~\ref{tab:comp_sample_z6}).
The best-fit relation and the relative 68\% confidence interval are represented as the solid black line and green-shaded region, respectively.
The best-fit parameters of the relation are: slope $1.15^{+0.28}_{-0.27}$, normalization $-8.16^{+0.06}_{-0.06}$, and intrinsic dispersion $0.07^{+0.03}_{-0.02}$~dex.
The dotted line is the relation from \cite{DeLooze14} for low-metallicity dwarf galaxies in the local Universe.
The dashed line is the best-fit relation from \cite{Romano22} for galaxies from the ALPINE survey ($4<z<6$), including both detections and stacked non-detections.
\emph{Right panel:} $z>7$ quasars with $L_{[CII]}$ measurements and \mh~upper limits: J1234, J0038, and J0252 (red circles) and J1342 (green circle); J1007 has both $L_{[CII]}$ and \mh~measurements.
\mh~measurements and upper limits for $z\gtrsim6$ quasars from the literature have been homogenized to those presented in this work by assuming a common $\alpha_{CO}$ of $0.8$\uaco~and a CO SLED.
The  best-fit relation and the relative 68\% confidence interval are represented as the solid black line and gray-shaded region, respectively.
The best-fit intrinsic dispersion of the relation is $\delta_{intr}=0.10^{+0.06}_{-0.03}$ dex.
The dashed line is the scaling relation derived by \cite{Zanella18} for main-sequence (MS) and luminous infrared galaxies, up to z$\sim6$.
We also report the scaling relation provided by \cite{Madden20} for local dwarf galaxies (dotted line).}
\label{fig:cii_zanella}
\end{figure*}

\section{Results}
\label{sec:results}
\subsection{Dust properties and star formation rates}
\label{sec:dustsed}
The continuum maps of the five targets are presented in Fig.\ref{fig:cont_maps}.
We measured continuum flux densities of NOEMA maps with a fit of the visibilities with a point source model.
The associated error accounts for both the statistical uncertainty of the fit and the calibration error ($\lesssim10\%$\footnote{IRAM NOEMA Data Reduction CookBook, \url{https://iram-institute.org/science-portal/noema/documentation/}}).
For J0252, J1243, and J2356, we also measured the continuum flux by integrating the 2D Gaussian function that fits the cleaned continuum emission map.
The two approaches provide flux estimates that are consistent within the uncertainties.
In the case of the NOEMA observations, the observing setup did not allow us to spatially resolve the target extension, all best-fit functions are consistent with a point-like source having a size comparable with the beam (see Table~\ref{tab:observations}).
Flux upper limits are estimated at a 3$\sigma$ significance level using the rms noise and assuming a size equal to the beam size.

As visible in Fig.\ref{fig:cont_maps}, ALMA observations of J0038, J0313, J0252, and J1243 spatially resolve the extent of the continuum emission.
This allowed us to measure the size of the continuum emitting region in quasar host galaxies by modeling the continuum maps with a 2D Gaussian function.
In Table~\ref{tab:observations}, we report the integrated flux and host-galaxy size, which is assumed to be equal to the full width half maximum (FWHM) of the best-fit 2D Gaussian function.

Using {\sc EOS-Dustfit}\footnote{{\sc EOS-Dustfit} is a publicly available tool for fitting the cold dust SED of galaxies (\url{https://github.com/roberta96/EOS-Dustfit}), which has been used in this work and in \citet{Tripodi24b}.
Details about the modeling can be found on the GitHub page and in \citet{Tripodi24b}.}, we fitted the cold dust SED of the target galaxies.
In {\sc EOS-Dustfit}, following the prescription by \citet[][see also \citealt{Carniani19, Tripodi24b}]{DraineLi07}, the cold dust continuum emission is modeled with a modified blackbody (MBB) function, assuming an optically thin regime.
The MBB model can have up to three free parameters, namely dust temperature ($T_d$), dust mass ($M_d$), and the emissivity index ($\beta$).
However, for the five sources of our sample, the limited amount of data prevents us from modeling the FIR SED with three free parameters.
In particular, the peak of the SED is not sampled by current data, making it challenging to constrain the dust temperature. Consequently, we adopted a fixed temperature, $T_d=55$ K, for all five quasars based on the mean dust temperature in a sample of 10 $z>6$ quasars \citep{Tripodi24b}.
This assumption is further supported by recent literature results on $z>6$ quasars (e.g., \citealt{Decarli22},  \citealt{Sommovigo22a},  \citealt{Shao22},  \citealt{Witstok23}, and  \citealt{Tripodi23b}) and lower-z quasars with measured $T_{d}$ (e.g., \citealt{Bischetti21}), as it would be expected in the case of compact, and dense star-forming regions (dust sizes of few kiloparsecs; e.g., \citealt{Venemans17a},  \citealt{Decarli18},  \citealt{Decarli22}, and \citealt{Tripodi24b}).
Their intense star formation still stands even after accounting for the active galactic nucleus (AGN) contribution to the dust heating, which can be substantial, especially in the nuclear region  (\citealt{Duras17},  \citealt{DiMascia21},  \citealt{Walter22}, and  \citealt{Tsukui23}).

Combining NOEMA and ALMA detections in the observed-frame $\sim$80-230 GHz range, we estimated the dust emissivity parameter, $\beta,$ for two objects in our sample (J0252 and J1243).
For the others, we fixed $\beta=1.6$, the local value (\citealt{Beelen06} and  \citealt{Witstok23}) and similar to that measured in J0252 and J1243.
Since NOEMA observations do not allow us to resolve the source size, we assumed that the size does not vary significantly with frequency in the observed range $\sim80$ GHz to $\sim 230$ GHz.
Quasar J2356 only has an unresolved observation at 3~mm (Fig.~\ref{fig:cont_maps}).
Then, we assumed a size of $0.75\times0.64$ arcsec$^{\rm 2}$, based on the median value of the FWHM dimensions measured for the four quasars of our sample with ALMA observations.

EOS-Dustfit explores the parameter space for each SED using a Markov chain Monte Carlo algorithm implemented in the \textit{emcee} package \citep{emcee1}. 
A uniform distribution for priors is assumed for fitting parameters in the range: $5<\log(M_{d}/M_{\odot})<9$ for all the quasars, and $0.5<\beta<3.0$ for both J0252 and J1243. 
We ran 40 chains, with 3000 trials and a burn-in phase of 150 steps for each dataset; we added in quadrature a 10\% calibration uncertainty to the continuum flux errors.
The upper limits of the continuum emission were included in the fit as a 1$\sigma$ detection with a 2$\sigma$ error. There is not yet a standard approach for dealing with upper limits in the fitting procedure, one could either treat them as detections with large errors (as we do; see also, e.g., \citealt{Witstok22} and \citealt{Ronconi24}) or change the fitting code properly to ensure that above the upper limit level, the likelihood is zero. We adopt the first method given that it is commonly employed in many fitting routines (e.g., GalaPy and MERCURIUS) and gives us reasonable results.
We adopted the 50th percentile of the posterior distribution as the best-fit value, while the errors are calculated considering the 16th and 84th percentiles (corner plots of the posterior distribution of free parameters are shown in Fig.~\ref{fig:corner_plots}).
In the case of the modeling of FIR SED assuming a fixed $\beta$, the uncertainties on \mdust~are relatively small (see below) when compared to the fit where $\beta$ is left free to vary.
This is because of the marginalization over the distribution of $\beta$. 
To determine with accuracy to which extent the uncertainty on \mdust~and \lfir~may be underestimated due to marginalization on $\beta$, we performed additional fits for J0252 and J1243, this time setting $\beta=1.6$. 
We then obtained uncertainties on \mdust~and \lfir~lower by $\sim0.05$~dex.
A similar conclusion can be reached for the assumption of $T_d$, leaving it free to vary over a limited range of temperature ($T_d=40-70$~K). 
In Sect.~\ref{sec:results} we take care of this systematic by including an additional contribution to the error on \mdust~and \lfir. 
The results from the SED fitting are reported in Tab.~\ref{tab:properties}, while the best-fit models are shown in Fig.~\ref{fig:fir_sed}.
We measured the FIR luminosity ($L_{FIR}$) by integrating the 40-1000~\micron~emission of the best-fit model produced by {\sc EOS-Dustfit} in the five quasars.
SFR is computed with the relation by \cite{Kennicutt98}, $SFR/(M_{\odot}\,yr^{-1}) = 10^{-10}\,L_{FIR}/L_{\odot}$, assuming a Chabrier initial mass function (IMF; \citealt{Chabrier03}).
The same prescription is commonly assumed in high-z quasars (e.g., \citealt{Duras17},  \citealt{Bischetti21},  and \citealt{Bertola24}), and is roughly consistent with the SFR derived assuming a \cite{Kroupa01} IMF (within $7\%$).
A Chabrier IMF was also adopted for J1007 by \cite{Feruglio23} and J1342 by \cite{Novak19}, while \cite{Venemans17a} assumed a Kroupa IMF for J1120.\\
In two cases we modeled the dust emissivity index, for J1243 the best-fit value derived is fully consistent with the typical value observed in high-redshift quasars ( \citealt{Beelen06},  \citealt{Venemans17a},  \citealt{Venemans20}, and  \citealt{Tripodi24b}). 
For J0252, we derived $\beta=0.93^{+0.20}_{-0.21}$, which is significantly lower (4$\sigma$) than $\beta$ values measured in high-z quasars.
A potential reason for the flat FIR SED in J0252 could be the presence of contaminant sources within the beam of the NOEMA observation (see Fig.~\ref{fig:cont_maps}).
Indeed, the ALMA continuum map of J0252 (see Fig.~\ref{fig:cont_maps}) shows a marginally resolved source surrounded by elongated emission features, each detected with $>2\sigma$ significance level.
These emission features may either hide the presence of close companion galaxies or interloper sources.
However, the relatively low $\beta$ value obtained from the fit does not significantly affect the \mdust~estimate.
Assuming $\beta=1.6$ for J0252, we obtained a \mdust~fully consistent with the one presented in Table~\ref{tab:properties}.\\
We tested the consistency of our assumptions (namely, $T_d$, $\beta$) on the results of the SED fitting.
Using a higher (lower) dust temperature such as $T_d=70$~K (40~K) would have led to an average decrease (increase) in \mdust~by a factor of $\sim1.5$ ($\sim0.18$~dex), while the value of 
\lfir~would have been higher (lower) by a factor of $\sim2$ ($\sim0.3$~dex).
Regarding $\beta$, assuming a steeper emissivity index ($\beta=2$) for the targets with $\beta$ fixed (namely, J0038, J0313, and J2356) would have resulted in a larger \mdust~(lower \lfir) by a factor of $\sim2$.
The results of the modeling assuming different $\beta$ and $T_d$ values are shown in Fig.~\ref{fig:fir_sed_multi_par}.
Conversely, choosing a flatter $\beta$ ($\beta=1.2$) results in a mean increase (decrease) in the value of \mdust~(\lfir) by a factor of $\leq1.2$.
Eventually, we tested the impact of assuming a different source size: increasing the source area up to a factor of 10 provided results consistent within the uncertainties with those listed in Table~\ref{tab:properties}; if the source area was overestimated by a factor of 10 or more, this would have led to \mdust~and \lfir~being lower by a factor of $\geq2$.

The five quasars in our sample show \mdust~in the range $0.6-4\times10^{8}$ \msun~(see Fig.~\ref{fig:mass_redshift} and Table~\ref{tab:properties}), consistent with the dust content derived in the population of luminous quasars at the EoR (e.g., \citealt{Izumi21, Witstok23, Tripodi24b}) and at different cosmic epochs (e.g., \citealt{Duras17},  \citealt{Bischetti21}, and  \citealt{Bertola24}), and star-forming galaxies at lower redshifts (e.g., \citealt{Mancini15},  \citealt{Lesniewska19},  \citealt{Pozzi21},  \citealt{Hygate23},  \citealt{Algera24}).
As shown in Fig.\ref{fig:mass_redshift}, comparing quasars at $z>7$ and sources at later epochs, there is no clear evolution of \mdust~as a function of the redshift.
The quasars in our sample have already built large dust masses in a relatively short time (the Hubble time at $z\sim7$ is $\sim800$ Myr), suggesting that in high-z quasar hosts the physical processes that drive the formation of dust grains are very efficient and overtake those processes destroying dust particles (e.g., \citealt{Popping17}).
We continue the discussion on the dust formation processes in Sect.~\ref{sec:gdr}.
\begin{table*}[ht]
\vspace{0.2cm}
                \caption{Quasar properties.}
                \centering
            \sisetup{separate-uncertainty,
             table-column-width=6em}
                \begin{tabular}{llccccccccc}
                        \hline
Name & ID & logM$_{BH}$ & logL$_{\rm bol}$ & $\beta$ & logM$_{\rm dust}$ & log(L$_{FIR}$) & SFR & logM$_{\rm H2}$ & Ref.\\
 &  & M$_{\odot}$ & L$_{\odot}$ &  & M$_{\odot}$ & L$_{\odot}$ &  M$_{\odot}$ yr$^{\rm -1}$ & M$_{\odot}$ & \\
\hline
J0313-1806 & J0313 & 9.2 & 13.6 & 1.6 & $7.8^{+0.1}_{-0.1}$ & $12.20^{+0.05}_{-0.05}$  & $160^{+20}_{-20}$ & & 1 \\
J1243+0100 & J1243 & 8.2 & 12.6 & $1.97^{+0.19}_{-0.21}$ & $8.0^{+0.2}_{-0.1}$ & $12.68^{+0.10}_{-0.09}$  & $475^{+170}_{-160}$  & $<9.96$ & 1\\
J0038-1527 & J0038 & 9.3 & 13.7 & 1.6 & $8.1^{+0.1}_{-0.1}$ & $12.43^{+0.07}_{-0.07}$  & $270^{+75}_{-70}$ & $<10.16$ & 1\\
J2356+0017 &  J2356 &  & & 1.6 & $8.0^{+0.1}_{-0.1}$ & $12.56^{+0.11}_{-0.11}$  & $370^{+120}_{-110}$ & $<10.05$ & 1\\
J0252-0503 & J0252 & 9.3 & 13.5 & $0.93^{+0.20}_{-0.21}$ & $8.6^{+0.2}_{-0.1}$ & $12.68^{+0.03}_{-0.03}$ & $476^{+36}_{-37}$ & $<10.14$ & 1\\
\hline
J1120+0641 & J1120 & 9.4 & 13.8 &  & $7.93\pm0.03$ & $12.2^{+0.03}_{0.03}$ & $220^{+15}_{-15}$ & $<10.1$ & 2 \\
J1007+2115 & J1007 & 9.2 & 13.7 &  & $8.32\pm12$ & 12.2 & 165 & $10.26\pm0.04$  & 3,4\\
J1342+0928 & J1342 & 8.9 & 13.6 &  & $7.54\pm0.02$ & 12 & 150 & $<9.5$ & 5 \\
\hline
                \end{tabular}
                \label{tab:properties}
                \flushleft 
                \footnotesize {{\bf Notes.} Quasars at $z>7$ known to date \citep{Fan23}. From left to right, columns include source name, ID adopted throughout the paper, logarithm of the black hole mass, logarithm of bolometric luminosity, dust emissivity index, logarithm of the dust mass, and 40-1000 $\mu$ infrared luminosity, SFR and logarithm of the molecular gas mass. We refer to Table~1 in \cite{Fan23} for the references for $M_{BH}$ and $L_{bol}$ for all $z>7$ quasars.  References for $M_{H2}$, [CII], and IR measurements are (1) This work; (2) \cite{Venemans17a}; (3) \cite{Feruglio23}; (4) \cite{Yang20}; (5) \cite{Novak19}}
\end{table*}

\subsection{Molecular gas mass}
\label{sec:mh}
For four out of five quasars observed with NOEMA we derived 3$\sigma$ upper limits on the cold $H_2$ mass using the rms at the expected frequency of CO(6--5) and (7--6) emission lines.
We assumed a line width of FWHM = 300~km/s, as expected for $z>6$ quasars (e.g., \citealt{Decarli22} and  \citealt{Feruglio23}), and a galaxy size equal to the beam of the observations.
We adopted the CO spectral line energy distribution (SLED) correction from \cite{Kaasinen24}, that is, $r_{6,1}=0.92$ for CO(6-5) and $r_{7,1}=0.65$.
Assuming a luminosity to mass conversion factor $\alpha_{co}=0.8$ \msun~(K km s$^{\rm -1}$ pc$^{\rm 2}$)$^{\rm -1}$ (see \citealt{Bolatto13} and  \citealt{CarilliWalter13}), we derive an upper limit $M(H_2)$ in the range $\sim 0.9-1.5\times 10^{10}$ \msun~(see Table~\ref{tab:properties}).
The same $\alpha_{co}$ and CO SLED correction factor are adopted for the upper limits and estimate of the $H_2$ masses of J1007, J1120, J1342, for which we collected CO luminosity from the literature and included in Table~\ref{tab:properties}.
Of the known quasars at $z \gtrsim 7$, five have only upper limits on their $H_2$ masses as traced by CO emission lines, with J1007 being the sole source with a statistically significant detection ($>6\sigma$) of molecular gas mass \citep{Feruglio23}.
The molecular gas mass and the upper limits for the $z>7$ quasars (see Table~\ref{tab:properties} and Fig.~\ref{fig:mass_redshift}) are consistent with the mean value measured in the population of $5<z<7$ quasars ($\log(M_{H2}/M_{\odot})\sim10\pm0.3$; e.g., \citealt{Venemans17b, Decarli22, Kaasinen24}).

To test the consistency of our \mh~upper limits, we compared the CO-derived measurement with that obtained from the [CII]158\micron~([CII], hereafter) luminosity.
\lcii~is a viable way to derive molecular gas masses for $z\gtrsim4$ non-AGN galaxies given its brightness and scaling relation between \lcii~and \mh~calibrated on lower redshift samples.
[CII] is expected to trace multiple gas phases (e.g., \citealt{Maio22},  \citealt{Maio23}, and  \citealt{Casavecchia24b}), but the corresponding emission is mostly due to the star formation activity in the host.
As shown in the left panel of Fig.~\ref{fig:cii_zanella}, \lcii~is tightly correlated (Pearson correlation coefficient $\sim0.71$) with the SFR in quasar host galaxies at $z>6$.
The data points shown in Fig.~\ref{fig:cii_zanella} include all quasar at $z>7$ and a collection of objects at $z\gtrsim6$ from the literature with both CO and [CII] measurements (see Table~\ref{tab:comp_sample_z6} for the full list of objects and references).
For this reason, several proposed relations use the \lcii~to trace the molecular gas content based on the Kennicutt-Schmidt relation \citep{Kennicutt98}.
As an example, the relation by \cite{Zanella18} calibrated on MS and starburst galaxies at redshift up to $z\sim6$, with the bulk of the sample at $z\sim2$. 
Alternatively, the relation by \cite{Madden20} allows us to measure the total \mh, including the potential contribution due to CO-dark clouds, and is calibrated on local dwarf galaxies.
As shown in the right panel of Fig.~\ref{fig:cii_zanella}, both relations overestimate significantly the \mh~measurement and upper limit derived from \lco~in quasars at $z>5$.
Indeed, several works in the literature studying the ISM properties of $z>6$ quasars (e.g., \citealt{Neeleman21},  \citealt{Decarli22},  \citealt{Kaasinen24}, and  \citealt{Tripodi24b}) observed a large discrepancy between \mh~derived with the prescription by \cite{Zanella18} and \cite{Madden20} and CO-derived ones, with [CII]-based \mh~that overestimates CO-based \mh~from a factor of a few to one order of magnitude (e.g., \citealt{Kaasinen24}).
The discrepancies in the determination of \mh~from \lcii~are likely due to the different physical properties of the ISM between the sample collected by \cite{Zanella18} and \cite{Madden20} and the host galaxies of high-z quasars.
Indeed, when considering galaxies at high redshifts ($z\gtrsim4$) with an intense star formation activity \citep{Rizzo20, Rizzo21}, the measured \mh-to-\lcii~ratio is considerably lower ($\sim4-8$\msun$/$\lsun) than the value $\sim30$ \msun$/$\lsun~measured by \cite{Zanella18}.
This is also supported by the results of post-processed FIR emission lines from the SIMBA cosmological hydrodynamical simulations \citep{Vizgan22}, suggesting that the \mh-to-\lcii~ratio should be lowered by up to a factor of $\sim2$ for synthetic star-forming galaxies at $z\simeq6$.

Given that, we combined $L_{[CII]}$ and \mh~measurements and upper limits for the $z>7$ quasars presented in Table~\ref{tab:properties} with measurements for 21 $z\gtrsim6$ quasars from the literature (see Table~\ref{tab:comp_sample_z6}) to provide a new calibration for the \mh-\lcii~relation.
We fitted data points of the resulting quasar sample of 25 objects with a linear regression based on a Bayesian approach, using the Python package {\it linmix} \citep{linmix}.
This package allows us to consider errors on both variables, $L_{[CII]}$ and \mh, and upper limits on \mh.
The best-fit relation is 
\begin{equation}
    \log(M_{H2}/M_{\odot}) = (0.75^{+0.31}_{-0.31} \log(L_{[CII]}) + (2.87^{+0.07}_{-0.07})
    \label{eq:mhcii}
\end{equation}
and is shown in Fig.~\ref{fig:cii_zanella}.
The same relation corresponds roughly to a \mh-to-\lcii~ratio of $\sim3.9^{+1.7}_{-2.3}$ in the \lcii~regime $10^{9-10}$ \lsun, almost one order of magnitude lower than that predicted by \cite{Zanella18}.
We note that even if we considered a Milky Way like CO-H$_{\rm 2}$ conversion factor ($\alpha_{CO}=4.3$ \uaco) to derive \mh~for high-z quasars, the corresponding \mh~would be lower by a factor of 1.5-2 than those derived assuming the calibration of \cite{Zanella18}.
This relation provides \mh~estimates for luminous quasars ($L_{bol}>10^{46}$ erg/s) at $z>5$ that are way more accurate than those provided by scaling relation calibrated from star-forming or dwarf galaxies at lower redshifts.

While applying this relation to high-redshift quasars, it is crucial to consider the following factors: (i) large-scale ionized winds driven by the accreting SMBH could significantly boost the observed \lcii~(e.g., \citealt{Bischetti24}); (ii) tidally stripped gas due to merging companions close to the quasar (e.g., \citealt{Tripodi24a, Decarli24}) could enhance \lcii~measurements, especially in observations with limited angular resolution.
Both (i) and (ii) cases could significantly overestimate the molecular gas masses in the quasar's host galaxy.

\begin{figure}[th]
\includegraphics[width=0.5\textwidth]{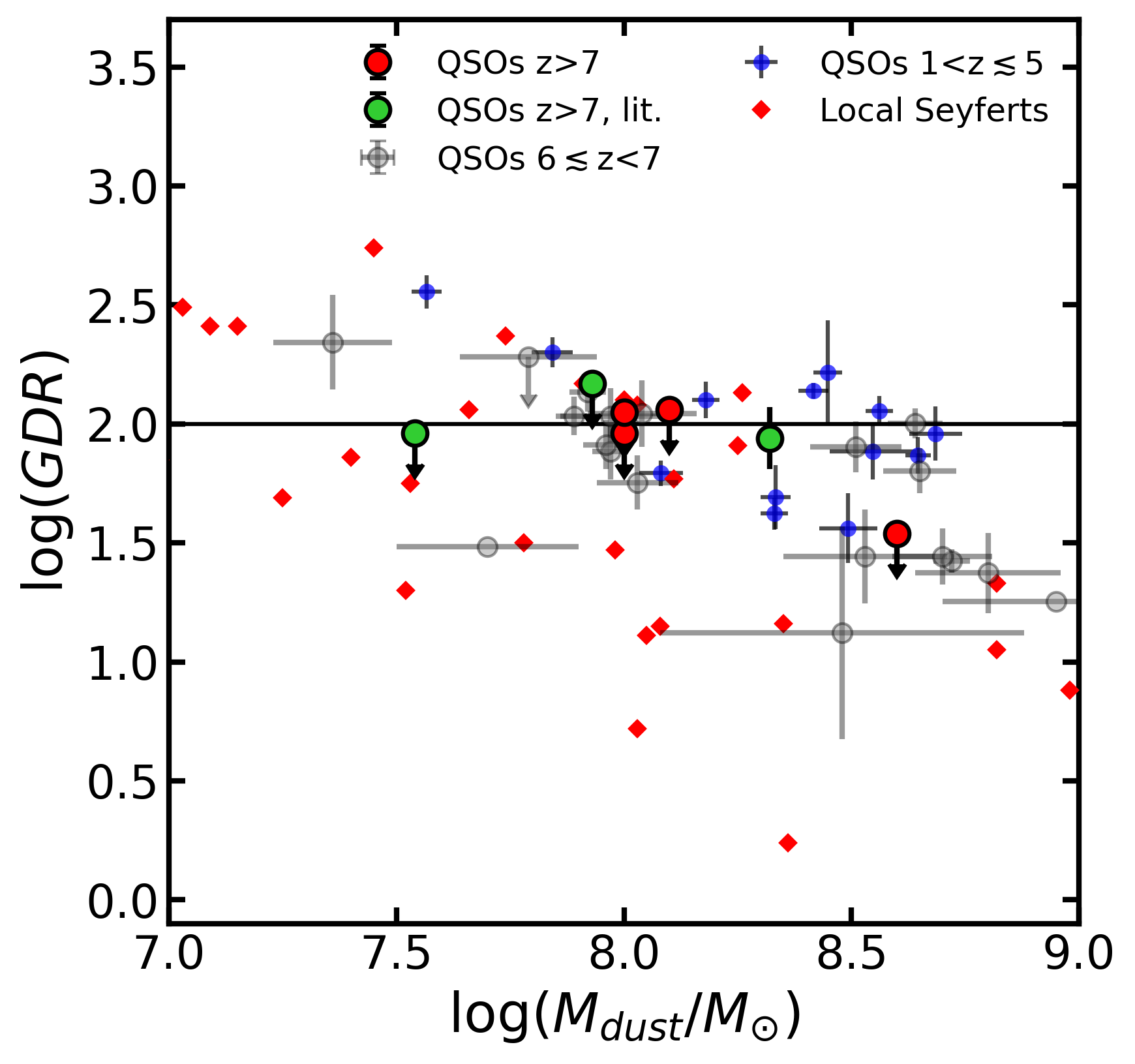}
\caption{GDR vs. dust mass. The horizontal black line represents GDR=100.
Red circles are the $z>7$ quasars presented in this work with \mh~upper limits. Green circles are for J1007 from \cite{Feruglio23}, J1120 from \cite{Venemans17a}, and J1342 from \cite{Novak19}. Gray circles are the $6\lesssim z<7$ quasars from Table~\ref{tab:comp_sample_z6}, while blue dots are for quasars at lower redshifts from \citet{Bischetti21}, \citet{Decarli22}, and Salvestrini et al. (in prep.). Local Seyfert galaxies are shown as red diamonds.}
\label{fig:gdr}
\end{figure}

\section{Discussion}
\label{sec:discussion}

\subsection{Dust enrichment at high redshifts}
\label{sec:gdr}
The origin of large dust reservoirs in the early Universe is widely debated (see \citealt{SchneiderMaiolino24} for a comprehensive review on this topic), since there is no consensus on the dominant formation mechanisms.
However, among the different channels for forming dust in the first few hundred million years of the Universe, asymptotic giant branch (AGB) stars and supernova (SN) ejecta are expected to contribute significantly even at $z>7$ (e.g., \citealt{Sommovigo22a}).
Even considering the maximally efficient yield from SN ejecta, this process can produce dust masses up to a few times $10^7$ \msun~at the redshift of our targets \citep{Mancini15, Valiante11}.
Nonetheless, we must also consider that a significant fraction of the dust created by SN ejecta (from 20\% up to the total amount, depending on model assumptions; e.g., \citealt{Micelotta18} and  \citealt{Kirchschlager19}) could be destroyed by the reverse shocks that may follow the initial explosion \citep{SchneiderMaiolino24}.
Regarding AGB stars, their contribution to dust production depends mostly on the gas metallicity and the IMF (e.g., \citealt{Ventura18, DellAgli19b}), nevertheless it can reach up to 40\% of the total dust at $z\gtrsim6$ (e.g., \citealt{SchneiderMaiolino24}).

Moreover, two more processes are expected to significantly contribute to dust production in a pre-enriched ISM: aggregation and evolution of grains within the ISM, and quasar-driven winds.
The aggregation of dust particles within the ISM usually requires timescales on the order of 1 Gyr and is observed to contribute up to 20-50\% of the total dust production in the local Universe \citep{Saintonge18, Galliano21}. 
However, the extremely dense and turbulent ISM in the host galaxy of quasars ( \citealt{Gallerani10},  \citealt{Valiante11},  \citealt{Valiante14},  \citealt{Mancini15},  \citealt{Decarli23}) could favor the formation of massive dust particles in a fraction of that time, making it the dominant channel in massive systems (e.g., \citealt{Mancini15}).
Indeed, local studies suggest that the aggregation of dust grains in the ISM primarily depends on gas density and temperature (e.g., \citealt{Draine03},  \citealt{Draine09}, and  \citealt{Galliano21}).
Given the luminosity of the quasar at the EoR included in this study, we cannot exclude that a non-negligible amount of dust grains that are produced and continue to grow within the outflowing clouds that are ejected due to the SMBH feedback \citep{Elvis02}.
\begin{figure*}[ht!]
\centering
\includegraphics[width=\textwidth]{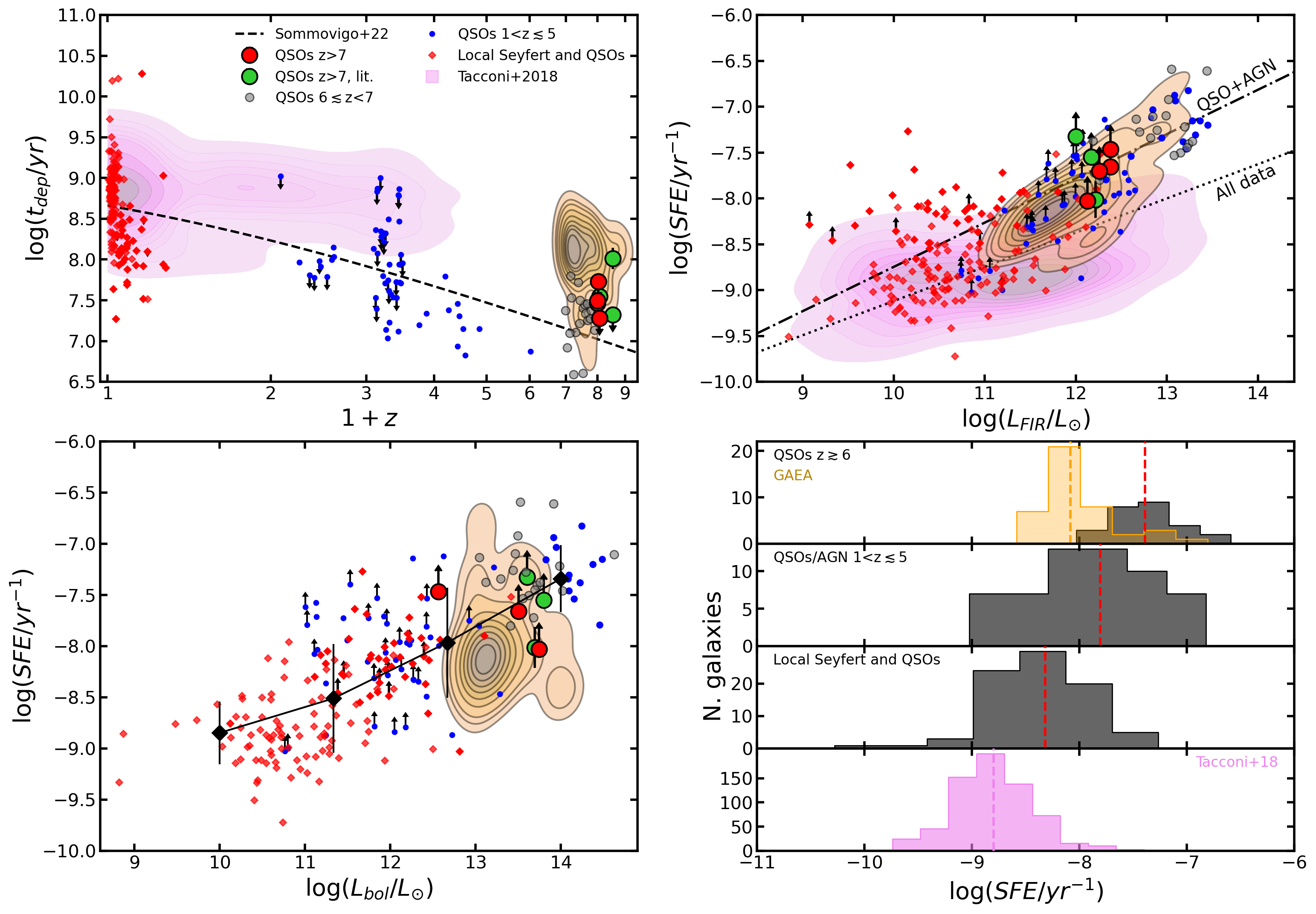}
\caption{Molecular gas consumption timescales and SFEs. 
\emph{Upper-left panel:} Depletion time as a function of ($1+z$), shown in logarithmic scale.
Sources are color-coded as follows: the $z>7$ quasars are red (this work) and green (\citealt{Venemans17a},  \citealt{Novak19},  and \citealt{Feruglio23}).
Literature samples include: $z<4$ star-forming galaxies \citep[pink-shaded region;][]{Tacconi18}; local Seyfert galaxies and quasars (red diamonds;  \citealt{Bischetti19a},  \citealt{Shangguan20},  \citealt{Salvestrini20},  \citealt{Salvestrini22}, and  \citealt{Salome23}); quasars at Cosmic Noon (blue dots;  \citealt{Bischetti21} and  \citealt{Bertola24}); and quasars at $6\lesssim z<7$ (gray circles; see Table~\ref{tab:comp_sample_z6}).
The parameter space occupied by the simulated quasars from {\sc GAEA} is shown as an orange-shaded region, with 10\% isodensity lines.
The dashed line represents the evolution of the depletion time with redshift \citep{Sommovigo22a}.
\emph{Upper-right panel:} SFE displayed as a function of $L_{IR}$.
The best-fit lines are shown for illustration purposes only and were obtained from a linear fit of the data shown in the legend.
Data points are color-coded as in the upper-left panel. 
\emph{Lower-left panel:} SFE vs. $L_{bol}$. The black diamonds represent the median SFE value in each of the equally spaced \lbol~bins. The vertical lines represent the distance between the 84th and 16th percentile of the SFE distribution in each bin.
Data points are color-coded as in the upper-left panel. 
\emph{Lower-right panel:} The histograms of the SFE distribution of each sample. The dashed line represents each distribution's median value.}
          \label{fig:SFE}%
\end{figure*}
Even if the role of this production channel is still debated (e.g., \citealt{Pipino11}), recent results from hydrodynamical simulations \citep{Sarangi19} suggest that this mechanism can be quite efficient (up to few \msun~yr$^{-1}$ for a $M_{BH}=10^8$ \msun) in luminous quasars such as our targets.
This means this channel can contribute to at least a few percent of the total dust budget, considering an AGN cycle of a million years.
Although this makes dust production in quasar winds less significant than the other mechanisms just described, it could still play a crucial role in dust enrichment in nuclear regions.
To conclude, building dust masses that exceed $10^8$~\msun~in the host galaxies of luminous quasars at redshift $z>7$ requires a combination of the physical processes described above.

As discussed above, the amount of dust traces the metal enrichment of the ISM.
For this reason, the gas-to-dust ratio (GDR=\mh/\mdust) is a crucial parameter to understand the rapid evolution and growth of galaxies at EoR.
Here, we adopted the \mh~measurement and upper limits listed in Table~\ref{tab:properties} to derive the GDR for 7 out of 8 quasars at $z>7$, among which J1007 is the only object with a GDR estimate.
We similarly derived GDRs for the quasars listed in Table~\ref{tab:comp_sample_z6} and quasars with similar luminosity at Cosmic Noon from \cite{Bischetti21}.
Since the high-z quasar population with dust mass estimates are biased toward luminous objects at FIR wavelength, we also considered a sample of local Seyfert galaxies ($L_{FIR}\sim10^{10-12}$ \lsun) to check for any potential selection bias.
In Fig.~\ref{fig:gdr}, the resulting GDRs are shown as a function of \mdust. 
Globally, the population of quasars at the EoR shows GDRs that are consistent (GDR$_{\rm mean}=80\pm40$) with the local value GDR=100 \citep{DraineLi07}, similar to that measured in local star-forming galaxies (e.g., \citealt{Casasola20}).
This local value is also commonly adopted at high redshifts to derive the molecular gas mass given the dust mass (e.g., \citealt{Neeleman21}).
Quasars at $z>7$ show upper limits on GDR that are consistent or just below the local value.
J0252 shows the tighter constraint of GDR among $z>7$ targets, but it is still consistent with GDR values observed in similarly luminous quasars.
We recall that assuming a higher (lower) \tdust~value would result in a higher (lower) GDR up to a factor of $\sim2$ with \tdust=65~K (45~K).
Looking at Fig.~\ref{fig:gdr}, GDR decreases at increasing \mdust, with a similar trend in quasars at different epochs and Seyfert galaxies.
Indeed, by fitting the GDR versus \mdust~with {\sc linmix}, we find a negative slope ($=-0.41\pm0.16$) when considering only quasars at the EoR.
Consistent results can be obtained by including either lower-redshift quasars \citep{Bischetti21} or local Seyfert galaxies (\citealt{Salvestrini22}; in the former case, there is a shift in the normalization).
The decreasing trend is likely driven by the level of metal enrichment of the ISM, which is higher at higher dust reservoirs.
This confirms that dust-rich systems are already enriched in metals, which are key ingredients of dust particles \citep{RemyRuyer14}.
However, we recall that our \mh~estimates are derived with a constant \aco~factor, while an assumption of a \aco~dependence on metallicity (e.g., \citealt{Amorin16}) would have balanced out the GDR anticorrelation with \mdust.
We conclude that quasars at the EoR exhibit GDR values (both detection and upper limits) consistent with those measured in quasars at lower redshifts and in local objects.
This confirms that assuming a local GDR value \citep{DraineLi07} is a reasonable choice for quasar host galaxies up to $z \sim 7.5$.

\subsection{Gas depletion and star formation efficiency}
\label{sec:sfe}
In the top-left panel of Fig.~\ref{fig:SFE}, we show the depletion time ($t_{dep}=M_{H2}/SFR$) as a function of redshift.
As a comparison, we include a collection of quasars spanning a wide range of redshift, namely: quasars at $6\lesssim z<7$ (see Table~\ref{tab:comp_sample_z6}), luminous objects at Cosmic Noon from \cite{Bischetti18},  \cite{Bischetti21},  \cite{Decarli22},  \cite{Bertola24}, and Salvestrini et al., in prep, and in the nearby Universe \citep[PDS456 and the PG survey, respectively]{Bischetti19a, Shangguan20}.
We also extend the comparison to AGN at intermediate luminosities ($L_{bol}\sim10^{44-45}$ erg/s), with samples of local AGN and Seyfert galaxies from \cite{Koss21} and  \cite{Salvestrini22} and highly accreting AGN hosts from \cite{Salome23}.
Eventually, we added star-forming and luminous infrared galaxies with CO detections from \cite{Tacconi18}, which cover a wide range of redshift, from the nearby Universe up to redshift $z \sim 4$.
As is clearly visible, the quasars at the EoR show relatively low $t_{dep}$ values ($\lesssim0.1$ Gyr), with few objects reaching depletion times of a few tens of megayears.
If we limit the analysis to the EoR, there is no evidence of the evolution of $t_{dep}$ in quasar host galaxies.

To further investigate this scenario, we contrasted our findings against theoretical predictions from the GAlaxy Evolution and Assembly ({\sc GAEA}) model.
In particular, the latest rendition of the model \citep{DeLucia24} combines the explicit partitioning of the cold gas into its molecular and neutral phases (the molecular gas ratio depends on the mid-plane pressure \citealt{BlitzRosolowsky06}, more details in \citealt{Xie17}), with improved modeling of AGN activity (both in terms of cold gas accretion onto SMBHs and AGN-driven feedback \citealt{Fontanot20}).
The {\sc GAEA} SAM has recently been coupled (Fontanot et al., in prep.) to the {\sc P-Millennium} simulation \citep{Baugh19}, which spans a volume of 800$^3$ Mpc$^3$, with a mass resolution of $m_p = 1.06 \times 10^8$ \msun, and assuming cosmological parameters consistent with the first year results from the \textit{Planck} satellite \citep{Planck_cosmpar}, to predict the evolution of galaxy properties from z$\sim$20 to the local Universe (Cantarella et al., in prep.). 
We extracted all predicted AGN galaxies from the simulated box with a $L_{\rm bol} > 3\times10^{46}$ erg s$^{-1}$ at $z>6$, resulting in a sample of 42 individual sources.
For each of these sources, we studied their mean molecular gas content and SFR by considering the evolution of these quantities over a timescale compatible with the Eddington time ($\sim$4 $\times$ 10$^7$ yrs), before the peak of the predicted AGN activity.
The synthetic objects show a distribution of $t_{dep}$ that covers a larger range of values, with the bulk of the population of 42 {\sc GAEA} quasars that have depletion times of a few hundred million years.
This can be due to the prescription used in the {\sc GAEA} SAM that instantaneously triggers the quasar phase when a sufficient amount of gas is concentrated in a certain radius in the host galaxy.
This favors the emergence of bright quasars, as well as the case of less massive objects, with respect to our current picture of bright quasars.
Indeed, {\sc GAEA} quasars show $M_{BH}$ that is roughly two orders of magnitude smaller ($\log(M_{BH}/M_{\odot}\sim7.40\pm0.23$)
than that observed in the quasar at redshift $z>6$ ($\log(M_{BH}/M_{\odot}\sim9.33\pm0.24$).
This is likely due to the flat prescription adopted for black hole seeding, which does not consider the hypothesis that massive ($M_{BH}>10^6$ \msun) can form via the direct collapse of giant gas clouds in the early Universe (for an alternative approach, see \citealt{Trinca22}).
Furthermore, the different distribution between synthetic and observed quasars can be partially due to an observational bias because we only detect the brightest sources at the EoR.
Indeed, considering the quasars at Cosmic Noon that span a larger regime \lbol~and \lfir~(see the upper-right and lower-left panels of Fig.~\ref{fig:SFE}), quasars from the SUPER (SINFONI Survey for Unveiling the Physics and Effect of Radiative feedback) and KASHz (KMOS AGN Survey at High redshift) samples from \cite{Bertola24} show depletion times that are consistent with the bulk of {\sc GAEA} objects.
Eventually, AGN and star-forming galaxies at lower redshifts show a wider range of $t_{dep}$ value, but the bulk of the distributions peak close to $t_{dep}\sim1$ Gyr.

An interpretation of this result requires that the feedback from the SMBH in high-redshift and luminous quasars can remove or heat the cold gas reservoir from their host \citep{Brusa18, Fiore17,Fluetsch19}.
However, several studies of SMBH-driven winds in quasars at different redshifts have shown that they are not powerful enough to impact the star formation ( \citealt{Bischetti19b},  \citealt{Tripodi23b}, and  \citealt{Novak19}), at least in the cold phase of the gas.
Simulations suggest that the quenching effect of the accreting SMBH is not instantaneously effective and, hence, that accretion onto the SMBH and massive star formation coexist ( \citealt{Costa20}, \citealt{Costa22}, and  \citealt{Valentini21}).

A different interpretation proposes that the relatively low depletion times are likely due to the highly efficient star formation of the quasar host that is favored by the concentration of cold gas in a relatively compact size.
To examine this scenario, we then derived the gas SFE, defined as $\rm SFE_{gas}=1/t_{dep}=SFR/M(H_2)$, which is represented as a function of $L_{IR}$ in the upper-right  panel of Fig.~\ref{fig:SFE}.
The trend between SFE and \lfir~shown in the upper-right  panel of Fig.~\ref{fig:SFE} suggests that the galaxies that are brighter at FIR wavelengths (hence have higher SFR), are more efficient at forming stars compared to galaxies at lower \lfir.
This is true, irrespective of the presence of an AGN at their center. 
Nonetheless, quasar hosts exceed up to one order of magnitude the SFE measured in star-forming and luminous galaxies (\citealt{Tacconi18}; dash-doted and dotted lines in the upper panel of Fig.~\ref{fig:SFE}) at high luminosities ($L_{FIR}>10^{12}$ \lsun).
In particular, quasars at the EoR show the highest efficiencies among the objects considered in our analysis, except for the hyper-luminous quasars at Cosmic Noon from the WISE/SDSS-selected hyper-luminous quasar (WISSH) sample (\citealt{Bischetti21} and Salvestrini et al. in prep.).
Regarding $z>7$ massive quasar hosts, we find lower SFE\ limits of $\rm >10^{-8}$ yr$^{-1}$, slightly higher than the SFE measured in J1007: $\rm SFE=9.7\pm2.5\times 10^{-9}$ yr$^{-1}$.

As discussed in Sect. \ref{sec:mh}, a highly star-forming host galaxy is a common characteristic among bright ($L_{bol}>10^{12.5}$ \lsun) quasars at high redshifts (e.g., \citealt{Venemans17c},  \citealt{Decarli22},  \citealt{Izumi21}, and \citealt{Tripodi24b}).
However, this effect could result from a combination of observational bias and the evolution of the MS (\citealt{RenziniPeng15}) at high redshifts.
Regarding the first point, we still lack a complete sampling, especially at (sub)millimeter wavelengths, of galaxy populations at intermediate luminosities ($L_{FIR}<10^{12}$ \lsun).
Concerning the second point, the SFR is expected to increase for MS galaxies with redshift as a result of the higher cosmological accretion rate at early times ($0<z<6$; \citealt{Tacconi20},  \citealt{Walter20}, and \citealt{Popesso23}).
This is confirmed by observational studies (e.g., \citealt{Tacconi20} and  \citealt{Sommovigo22a}) that found that the depletion time decreases with redshift, with values of $\sim0.01$ Gyr at $z\sim7$, consistent with those observed in quasars at the EoR (see the upper-left panel of Fig.\ref{fig:SFE}).
These two points should be taken into account when interpreting the lower-right panel of Fig.~\ref{fig:SFE}.
In each histogram, AGN and quasar sources are divided by redshift, while the luminous and star-forming galaxies from \cite{Tacconi18} are shown for comparison in the lower panel.
Quasars at the EoR represent only the population of the brightest sources at this epoch, and we are likely missing the rich population of intermediate luminosity AGN recently discovered with \textit{James Webb} Space Telescope (JWST; e.g., \cite{Harikane23},  \citealt{Maiolino24a}, and  \citealt{Maiolino24b}).
In this regard, the results from the {\sc GAEA} simulations predict a broader SFE distribution, with a peak at lower SFEs with respect to observations of quasars at similar redshifts.

Thanks to JWST, it is now possible to detect the stellar light from the quasar host \citep{Ding23}, and to model the corresponding emission to derive stellar masses (e.g., \citealt{Ding23} and  \citealt{Yue24}).
We took the estimate and upper limit of the stellar mass from \cite{Yue24} for four quasars at the EoR (J1120 and three other quasars), which are listed in Table~\ref{tab:comp_sample_z6}.
We then calculated the specific SFR (sSFR$=SFR/M_{\star}$) and molecular gas fraction ($f_{gas}=M_{H2}/M_{\star}$.
The lower limit and estimate of the sSFR ($>10^{-8}$ yr$^{-1}$) are way above the expected value for MS galaxies at the EoR. 
Their molecular gas fraction ($f_{gas}\sim0.05-0.5$) is lower than what observed in local quasars and AGN ( \citealt{Shangguan20} and \citealt{Salvestrini22}), but consistent with that derived from dynamical measurements in luminous quasars at Cosmic Noon \citep{Bischetti21}.
This confirms that these objects are experiencing an intense starburst phase that is coeval with the bright phase of the quasars, justifying the shift between the median value of the SFE distribution between AGN and non-AGN host galaxies in the histograms of Fig.~\ref{fig:SFE}.

In the lower-left panel of Fig.~\ref{fig:SFE}, we show the SFE as a function of the bolometric luminosity of the accreting SMBH.
The solid black line connects the median value of the SFE of quasars and AGN divided into four bins of \lbol.
The increasing trend between the SFE and \lbol\footnote{The Pearson correlation coefficient for the SFE-\lbol~ relation is 0.53 and 0.46 considering all quasars at $z>1$ or the all AGN and quasars shown in the plot, respectively.} shown in the lower-left panel of Fig.~\ref{fig:SFE} suggests that the processes that convey a significant amount of gas toward the nuclear region (e.g., disk instabilities) are more efficient than the effect of the feedback (e.g., outflows) produced by the accreting SMBH, which is expected to be proportional to the quasar luminosity \citep{Fiore17, Fluetsch19}.
We then conclude that among the known population of bright quasars at the EoR there is no clear evidence of an efficient quenching of the host-galaxy star formation due to quasar feedback.
On the contrary, at such early epochs, a bright quasar phase is likely coeval with an intense buildup phase in the host galaxy.
The coexistence between SMBH growth and intense star formation in the host is visible at all redshifts, when considering luminous quasars, suggesting that quasar activity benefits from the gas concentration in the nuclear region of the galaxy.
An alternative scenario can be that the quasar activity triggers the star formation in the host, by compressing the gas in the surrounding medium \citep{Cresci18}.

\section{Conclusions}
\label{sec:concl}
In this work we present new NOEMA observations targeting CO emission lines for five $z > 7$ quasars, completing the survey of molecular gas properties for all eight known quasars at this epoch. These observations represent the highest-redshift investigation of cold dust and gas to date in a sample of quasar host galaxies at the EoR.
By modeling the FIR emission with a MBB using {\sc EOS-Dustfit}, we derived dust properties and SFR estimates for our sample. Although no statistically significant CO emission lines were detected in the five targets, we derived upper limits on the molecular gas mass.
We compared the molecular gas and dust properties of all known $z > 7$ quasars with a compilation of quasar hosts and star-forming galaxies at different redshifts.
Our results are:
\begin{itemize} 
\item Among the eight known quasars at $z>7$, only one object (J1007; \citealt{Feruglio23}) has a statistically significant molecular gas mass estimate. We find no massive gas reservoirs ($M_{H2} <$ a few times $10^{10}$ \msun) at $z>7$. 
\item Combining the new observations presented in this work with measurements from the literature, we provide a new calibration to derive \mh~from \lcii~for $z\gtrsim6$ quasars. 
\item Quasar host galaxies at $z>7$ had already accumulated large dust reservoirs ($M_{d}\sim10^8$ \msun) in a relatively short time (a few hundred megayears) after the Big Bang. This suggests that the physical processes responsible for dust enrichment are very efficient and may include contributions from evolved stellar populations, SN, and reprocessing within the ISM.
The GDR estimate and upper limits for $z > 7$ quasars presented here align with the mean GDR for luminous quasars and AGN at later epochs, which is consistent with the local value (GDR=100; \citealt{DraineLi07}).
\item Quasars at the EoR are hosted in galaxies undergoing intense starburst phases, with SFEs up to an order of magnitude higher than those expected for non-AGN host galaxies. This suggests that the emergence of a luminous quasar phase is coeval with the rapid buildup of the host galaxy.
\item Semi-analytic models of quasar host galaxies from {\sc GAEA} at the EoR also support the idea that the quasar phase is triggered during periods of efficient star formation.
\end{itemize}

\begin{acknowledgements}
FS and CF thank J.M. Winters for his help in data reduction and calibration.
FS thanks V. D'Odorico for the useful discussion and suggestions.
The project leading to this publication has received support from ORP, that is funded by the European Union’s Horizon 2020 research and innovation programme under grant agreement No 101004719 [ORP].
This work is based on observations carried out under project number S23CX with the IRAM NOEMA Interferometer. IRAM is supported by INSU/CNRS (France), MPG (Germany) and IGN (Spain).
LZ, EP, AT, and FS acknowledge financial support from the Bando Ricerca Fondamentale INAF 2022 Large Grant ``Toward an holistic view of the Titans: multi-band observations of $z>6$ QSOs powered by greedy supermassive black holes''.
CF, and FS acknowledge financial support from PRIN MUR 2022 2022TKPB2P - BIG-z and Ricerca Fondamentale INAF 2023 Data Analysis grant ``ARCHIE ARchive Cosmic HI \& ISM  Evolution''.
M.B. acknowledges support from INAF under project 1.05.12.04.01 - 431 MINI-GRANTS di RSN1 "Mini-feedback"  and support from UniTs under project DF-microgrants23 "Hyper-Gal".
This paper makes use of the following ALMA data: ADS/JAO.ALMA\#2018.1.01188.S, \#2019.A.00017.S, \#2019.1.00074.S, \#2019.1.01025.S, \#2021.1.00934.S, \#2021.1.01262.S. ALMA is a partnership of ESO (representing its member states), NFS (USA) and NINS (Japan), together with NRC (Canada), MOST and ASIAA (Taiwan) and KASI (Republic of Korea), in cooperation with the Republic of Chile. The Joint ALMA Observatory is operated by ESO, AUI/NRAO and NAOJ.

\emph{Software:} {\sc Astropy} \citep{astropy1, astropy2, astropy3};  {\sc CASA} \citep{casa}; {\sc emcee} \citep{emcee1, emcee2}; {\sc GILDAS} \url{https://www.iram.fr/IRAMFR/GILDAS/}; {\sc linmix} \citep{linmix}; {\sc Matplotlib} \citep{matplotlib}; {\sc NumPy} \citep{numpy}; {\sc Seaborn} \citep{seaborn}; {\sc Scipy} \citep{scipy}.

\end{acknowledgements}

\bibliographystyle{aa}
\bibliography{biblio_s23cx} 

\begin{thebibliography}{143}
\expandafter\ifx\csname natexlab\endcsname\relax\def\natexlab#1{#1}\fi

\bibitem[{{Algera} {et~al.}(2024){Algera}, {Inami}, {Sommovigo}, {Fudamoto},
  {Schneider}, {Graziani}, {Dayal}, {Bouwens}, {Aravena}, {da Cunha},
  {Ferrara}, {Hygate}, {van Leeuwen}, {De Looze}, {Palla}, {Pallottini},
  {Smit}, {Stefanon}, {Topping}, \& {van der Werf}}]{Algera24}
{Algera}, H. S.~B., {Inami}, H., {Sommovigo}, L., {et~al.} 2024, \mnras, 527,
  6867

\bibitem[{{Amor{\'\i}n} {et~al.}(2016){Amor{\'\i}n}, {Mu{\~n}oz-Tu{\~n}{\'o}n},
  {Aguerri}, \& {Planesas}}]{Amorin16}
{Amor{\'\i}n}, R., {Mu{\~n}oz-Tu{\~n}{\'o}n}, C., {Aguerri}, J.~A.~L., \&
  {Planesas}, P. 2016, \aap, 588, A23

\bibitem[{{Astropy Collaboration} {et~al.}(2022){Astropy Collaboration},
  {Price-Whelan}, {Lim}, {Earl}, {Starkman}, {Bradley}, {Shupe}, {Patil},
  {Corrales}, {Brasseur}, {N{\"o}the}, {Donath}, {Tollerud}, {Morris},
  {Ginsburg}, {Vaher}, {Weaver}, {Tocknell}, {Jamieson}, {van Kerkwijk},
  {Robitaille}, {Merry}, {Bachetti}, {G{\"u}nther}, {Aldcroft},
  {Alvarado-Montes}, {Archibald}, {B{\'o}di}, {Bapat}, {Barentsen},
  {Baz{\'a}n}, {Biswas}, {Boquien}, {Burke}, {Cara}, {Cara}, {Conroy},
  {Conseil}, {Craig}, {Cross}, {Cruz}, {D'Eugenio}, {Dencheva}, {Devillepoix},
  {Dietrich}, {Eigenbrot}, {Erben}, {Ferreira}, {Foreman-Mackey}, {Fox},
  {Freij}, {Garg}, {Geda}, {Glattly}, {Gondhalekar}, {Gordon}, {Grant},
  {Greenfield}, {Groener}, {Guest}, {Gurovich}, {Handberg}, {Hart},
  {Hatfield-Dodds}, {Homeier}, {Hosseinzadeh}, {Jenness}, {Jones}, {Joseph},
  {Kalmbach}, {Karamehmetoglu}, {Ka{\l}uszy{\'n}ski}, {Kelley}, {Kern},
  {Kerzendorf}, {Koch}, {Kulumani}, {Lee}, {Ly}, {Ma}, {MacBride}, {Maljaars},
  {Muna}, {Murphy}, {Norman}, {O'Steen}, {Oman}, {Pacifici}, {Pascual},
  {Pascual-Granado}, {Patil}, {Perren}, {Pickering}, {Rastogi}, {Roulston},
  {Ryan}, {Rykoff}, {Sabater}, {Sakurikar}, {Salgado}, {Sanghi}, {Saunders},
  {Savchenko}, {Schwardt}, {Seifert-Eckert}, {Shih}, {Jain}, {Shukla}, {Sick},
  {Simpson}, {Singanamalla}, {Singer}, {Singhal}, {Sinha}, {Sip{\H{o}}cz},
  {Spitler}, {Stansby}, {Streicher}, {{\v{S}}umak}, {Swinbank}, {Taranu},
  {Tewary}, {Tremblay}, {de Val-Borro}, {Van Kooten}, {Vasovi{\'c}}, {Verma},
  {de Miranda Cardoso}, {Williams}, {Wilson}, {Winkel}, {Wood-Vasey}, {Xue},
  {Yoachim}, {Zhang}, {Zonca}, \& {Astropy Project Contributors}}]{astropy3}
{Astropy Collaboration}, {Price-Whelan}, A.~M., {Lim}, P.~L., {et~al.} 2022,
  \apj, 935, 167

\bibitem[{{Astropy Collaboration} {et~al.}(2018){Astropy Collaboration},
  {Price-Whelan}, {Sip{\H{o}}cz}, {G{\"u}nther}, {Lim}, {Crawford}, {Conseil},
  {Shupe}, {Craig}, {Dencheva}, {Ginsburg}, {VanderPlas}, {Bradley},
  {P{\'e}rez-Su{\'a}rez}, {de Val-Borro}, {Aldcroft}, {Cruz}, {Robitaille},
  {Tollerud}, {Ardelean}, {Babej}, {Bach}, {Bachetti}, {Bakanov}, {Bamford},
  {Barentsen}, {Barmby}, {Baumbach}, {Berry}, {Biscani}, {Boquien}, {Bostroem},
  {Bouma}, {Brammer}, {Bray}, {Breytenbach}, {Buddelmeijer}, {Burke},
  {Calderone}, {Cano Rodr{\'\i}guez}, {Cara}, {Cardoso}, {Cheedella}, {Copin},
  {Corrales}, {Crichton}, {D'Avella}, {Deil}, {Depagne}, {Dietrich}, {Donath},
  {Droettboom}, {Earl}, {Erben}, {Fabbro}, {Ferreira}, {Finethy}, {Fox},
  {Garrison}, {Gibbons}, {Goldstein}, {Gommers}, {Greco}, {Greenfield},
  {Groener}, {Grollier}, {Hagen}, {Hirst}, {Homeier}, {Horton}, {Hosseinzadeh},
  {Hu}, {Hunkeler}, {Ivezi{\'c}}, {Jain}, {Jenness}, {Kanarek}, {Kendrew},
  {Kern}, {Kerzendorf}, {Khvalko}, {King}, {Kirkby}, {Kulkarni}, {Kumar},
  {Lee}, {Lenz}, {Littlefair}, {Ma}, {Macleod}, {Mastropietro}, {McCully},
  {Montagnac}, {Morris}, {Mueller}, {Mumford}, {Muna}, {Murphy}, {Nelson},
  {Nguyen}, {Ninan}, {N{\"o}the}, {Ogaz}, {Oh}, {Parejko}, {Parley}, {Pascual},
  {Patil}, {Patil}, {Plunkett}, {Prochaska}, {Rastogi}, {Reddy Janga},
  {Sabater}, {Sakurikar}, {Seifert}, {Sherbert}, {Sherwood-Taylor}, {Shih},
  {Sick}, {Silbiger}, {Singanamalla}, {Singer}, {Sladen}, {Sooley},
  {Sornarajah}, {Streicher}, {Teuben}, {Thomas}, {Tremblay}, {Turner},
  {Terr{\'o}n}, {van Kerkwijk}, {de la Vega}, {Watkins}, {Weaver}, {Whitmore},
  {Woillez}, {Zabalza}, \& {Astropy Contributors}}]{astropy2}
{Astropy Collaboration}, {Price-Whelan}, A.~M., {Sip{\H{o}}cz}, B.~M., {et~al.}
  2018, \aj, 156, 123

\bibitem[{{Astropy Collaboration} {et~al.}(2013){Astropy Collaboration},
  {Robitaille}, {Tollerud}, {Greenfield}, {Droettboom}, {Bray}, {Aldcroft},
  {Davis}, {Ginsburg}, {Price-Whelan}, {Kerzendorf}, {Conley}, {Crighton},
  {Barbary}, {Muna}, {Ferguson}, {Grollier}, {Parikh}, {Nair}, {Unther},
  {Deil}, {Woillez}, {Conseil}, {Kramer}, {Turner}, {Singer}, {Fox}, {Weaver},
  {Zabalza}, {Edwards}, {Azalee Bostroem}, {Burke}, {Casey}, {Crawford},
  {Dencheva}, {Ely}, {Jenness}, {Labrie}, {Lim}, {Pierfederici}, {Pontzen},
  {Ptak}, {Refsdal}, {Servillat}, \& {Streicher}}]{astropy1}
{Astropy Collaboration}, {Robitaille}, T.~P., {Tollerud}, E.~J., {et~al.} 2013,
  \aap, 558, A33

\bibitem[{{Ba{\~n}ados} {et~al.}(2016){Ba{\~n}ados}, {Venemans}, {Decarli},
  {Farina}, {Mazzucchelli}, {Walter}, {Fan}, {Stern}, {Schlafly}, {Chambers},
  {Rix}, {Jiang}, {McGreer}, {Simcoe}, {Wang}, {Yang}, {Morganson}, {De Rosa},
  {Greiner}, {Balokovi{\'c}}, {Burgett}, {Cooper}, {Draper}, {Flewelling},
  {Hodapp}, {Jun}, {Kaiser}, {Kudritzki}, {Magnier}, {Metcalfe}, {Miller},
  {Schindler}, {Tonry}, {Wainscoat}, {Waters}, \& {Yang}}]{Banados16}
{Ba{\~n}ados}, E., {Venemans}, B.~P., {Decarli}, R., {et~al.} 2016, \apjs, 227,
  11

\bibitem[{{Baugh} {et~al.}(2019){Baugh}, {Gonzalez-Perez}, {Lagos}, {Lacey},
  {Helly}, {Jenkins}, {Frenk}, {Benson}, {Bower}, \& {Cole}}]{Baugh19}
{Baugh}, C.~M., {Gonzalez-Perez}, V., {Lagos}, C. D.~P., {et~al.} 2019, \mnras,
  483, 4922

\bibitem[{{Beelen} {et~al.}(2006){Beelen}, {Cox}, {Benford}, {Dowell},
  {Kov{\'a}cs}, {Bertoldi}, {Omont}, \& {Carilli}}]{Beelen06}
{Beelen}, A., {Cox}, P., {Benford}, D.~J., {et~al.} 2006, \apj, 642, 694

\bibitem[{{Bertola} {et~al.}(2024){Bertola}, {Circosta}, {Ginolfi}, {Mainieri},
  {Vignali}, {Calistro Rivera}, {Ward}, {Lopez}, {Pensabene}, {Alexander},
  {Bischetti}, {Brusa}, {Cappi}, {Comastri}, {Contursi}, {Cicone}, {Cresci},
  {Dadina}, {D'Amato}, {Feltre}, {Harrison}, {Kakkad}, {Lamperti}, {Lanzuisi},
  {Mannucci}, {Marconi}, {Perna}, {Piconcelli}, {Puglisi}, {Ricci}, {Scholtz},
  {Tozzi}, {Vietri}, {Zamorani}, \& {Zappacosta}}]{Bertola24}
{Bertola}, E., {Circosta}, C., {Ginolfi}, M., {et~al.} 2024, \aap, 691, A178

\bibitem[{{Bischetti} {et~al.}(2024){Bischetti}, {Choi}, {Fiore}, {Feruglio},
  {Carniani}, {D'Odorico}, {Ba{\~n}ados}, {Chen}, {Decarli}, {Gallerani},
  {Hlavacek-Larrondo}, {Lai}, {Leighly}, {Mazzucchelli}, {Perreault-Levasseur},
  {Tripodi}, {Walter}, {Wang}, {Yang}, {Zanchettin}, \& {Zhu}}]{Bischetti24}
{Bischetti}, M., {Choi}, H., {Fiore}, F., {et~al.} 2024, \apj, 970, 9

\bibitem[{{Bischetti} {et~al.}(2021){Bischetti}, {Feruglio}, {Piconcelli},
  {Duras}, {P{\'e}rez-Torres}, {Herrero}, {Venturi}, {Carniani}, {Bruni},
  {Gavignaud}, {Testa}, {Bongiorno}, {Brusa}, {Circosta}, {Cresci},
  {D'Odorico}, {Maiolino}, {Marconi}, {Mingozzi}, {Pappalardo}, {Perna},
  {Traianou}, {Travascio}, {Vietri}, {Zappacosta}, \& {Fiore}}]{Bischetti21}
{Bischetti}, M., {Feruglio}, C., {Piconcelli}, E., {et~al.} 2021, \aap, 645,
  A33

\bibitem[{{Bischetti} {et~al.}(2019{\natexlab{a}}){Bischetti}, {Maiolino},
  {Carniani}, {Fiore}, {Piconcelli}, \& {Fluetsch}}]{Bischetti19b}
{Bischetti}, M., {Maiolino}, R., {Carniani}, S., {et~al.} 2019{\natexlab{a}},
  \aap, 630, A59

\bibitem[{{Bischetti} {et~al.}(2018){Bischetti}, {Piconcelli}, {Feruglio},
  {Duras}, {Bongiorno}, {Carniani}, {Marconi}, {Pappalardo}, {Schneider},
  {Travascio}, {Valiante}, {Vietri}, {Zappacosta}, \& {Fiore}}]{Bischetti18}
{Bischetti}, M., {Piconcelli}, E., {Feruglio}, C., {et~al.} 2018, \aap, 617,
  A82

\bibitem[{{Bischetti} {et~al.}(2019{\natexlab{b}}){Bischetti}, {Piconcelli},
  {Feruglio}, {Fiore}, {Carniani}, {Brusa}, {Cicone}, {Vignali}, {Bongiorno},
  {Cresci}, {Mainieri}, {Maiolino}, {Marconi}, {Nardini}, \&
  {Zappacosta}}]{Bischetti19a}
{Bischetti}, M., {Piconcelli}, E., {Feruglio}, C., {et~al.} 2019{\natexlab{b}},
  \aap, 628, A118

\bibitem[{{Blitz} \& {Rosolowsky}(2006)}]{BlitzRosolowsky06}
{Blitz}, L. \& {Rosolowsky}, E. 2006, \apj, 650, 933

\bibitem[{{Bolatto} {et~al.}(2013){Bolatto}, {Wolfire}, \& {Leroy}}]{Bolatto13}
{Bolatto}, A.~D., {Wolfire}, M., \& {Leroy}, A.~K. 2013, \araa, 51, 207

\bibitem[{{Brusa} {et~al.}(2018){Brusa}, {Cresci}, {Daddi}, {Paladino},
  {Perna}, {Bongiorno}, {Lusso}, {Sargent}, {Casasola}, {Feruglio},
  {Fraternali}, {Georgiev}, {Mainieri}, {Carniani}, {Comastri}, {Duras},
  {Fiore}, {Mannucci}, {Marconi}, {Piconcelli}, {Zamorani}, {Gilli}, {La
  Franca}, {Lanzuisi}, {Lutz}, {Santini}, {Scoville}, {Vignali}, {Vito},
  {Rabien}, {Busoni}, \& {Bonaglia}}]{Brusa18}
{Brusa}, M., {Cresci}, G., {Daddi}, E., {et~al.} 2018, \aap, 612, A29

\bibitem[{{Carilli} \& {Walter}(2013)}]{CarilliWalter13}
{Carilli}, C.~L. \& {Walter}, F. 2013, \araa, 51, 105

\bibitem[{{Carniani} {et~al.}(2019){Carniani}, {Gallerani}, {Vallini},
  {Pallottini}, {Tazzari}, {Ferrara}, {Maiolino}, {Cicone}, {Feruglio}, {Neri},
  {D'Odorico}, {Wang}, \& {Li}}]{Carniani19}
{Carniani}, S., {Gallerani}, S., {Vallini}, L., {et~al.} 2019, \mnras, 489,
  3939

\bibitem[{{CASA Team} {et~al.}(2022){CASA Team}, {Bean}, {Bhatnagar}, {Castro},
  {Donovan Meyer}, {Emonts}, {Garcia}, {Garwood}, {Golap}, {Gonzalez Villalba},
  {Harris}, {Hayashi}, {Hoskins}, {Hsieh}, {Jagannathan}, {Kawasaki},
  {Keimpema}, {Kettenis}, {Lopez}, {Marvil}, {Masters}, {McNichols},
  {Mehringer}, {Miel}, {Moellenbrock}, {Montesino}, {Nakazato}, {Ott}, {Petry},
  {Pokorny}, {Raba}, {Rau}, {Schiebel}, {Schweighart}, {Sekhar}, {Shimada},
  {Small}, {Steeb}, {Sugimoto}, {Suoranta}, {Tsutsumi}, {van Bemmel},
  {Verkouter}, {Wells}, {Xiong}, {Szomoru}, {Griffith}, {Glendenning}, \&
  {Kern}}]{casa}
{CASA Team}, {Bean}, B., {Bhatnagar}, S., {et~al.} 2022, \pasp, 134, 114501

\bibitem[{{Casasola} {et~al.}(2020){Casasola}, {Bianchi}, {De Vis}, {Magrini},
  {Corbelli}, {Clark}, {Fritz}, {Nersesian}, {Viaene}, {Baes}, {Cassar{\`a}},
  {Davies}, {De Looze}, {Dobbels}, {Galametz}, {Galliano}, {Jones}, {Madden},
  {Mosenkov}, {Tr{\v{c}}ka}, \& {Xilouris}}]{Casasola20}
{Casasola}, V., {Bianchi}, S., {De Vis}, P., {et~al.} 2020, \aap, 633, A100

\bibitem[{{Casavecchia} {et~al.}(2025){Casavecchia}, {Maio}, {P{\'e}roux}, \&
  {Ciardi}}]{Casavecchia24b}
{Casavecchia}, B., {Maio}, U., {P{\'e}roux}, C., \& {Ciardi}, B. 2025, \aap,
  693, A119

\bibitem[{{Chabrier}(2003)}]{Chabrier03}
{Chabrier}, G. 2003, \pasp, 115, 763

\bibitem[{{Costa} {et~al.}(2022){Costa}, {Arrigoni Battaia}, {Farina},
  {Keating}, {Rosdahl}, \& {Kimm}}]{Costa22}
{Costa}, T., {Arrigoni Battaia}, F., {Farina}, E.~P., {et~al.} 2022, \mnras,
  517, 1767

\bibitem[{{Costa} {et~al.}(2020){Costa}, {Pakmor}, \& {Springel}}]{Costa20}
{Costa}, T., {Pakmor}, R., \& {Springel}, V. 2020, \mnras, 497, 5229

\bibitem[{{Cresci} \& {Maiolino}(2018)}]{Cresci18}
{Cresci}, G. \& {Maiolino}, R. 2018, Nature Astronomy, 2, 179

\bibitem[{{De Looze} {et~al.}(2014){De Looze}, {Cormier}, {Lebouteiller},
  {Madden}, {Baes}, {Bendo}, {Boquien}, {Boselli}, {Clements}, {Cortese},
  {Cooray}, {Galametz}, {Galliano}, {Graci{\'a}-Carpio}, {Isaak}, {Karczewski},
  {Parkin}, {Pellegrini}, {R{\'e}my-Ruyer}, {Spinoglio}, {Smith}, \&
  {Sturm}}]{DeLooze14}
{De Looze}, I., {Cormier}, D., {Lebouteiller}, V., {et~al.} 2014, \aap, 568,
  A62

\bibitem[{{De Lucia} {et~al.}(2024){De Lucia}, {Fontanot}, {Xie}, \&
  {Hirschmann}}]{DeLucia24}
{De Lucia}, G., {Fontanot}, F., {Xie}, L., \& {Hirschmann}, M. 2024, \aap, 687,
  A68

\bibitem[{{Decarli} {et~al.}(2024){Decarli}, {Loiacono}, {Farina}, {Dotti},
  {Lupi}, {Meyer}, {Mignoli}, {Pensabene}, {Strauss}, {Venemans}, {Yang},
  {Walter}, {Wolf}, {Ba{\~n}ados}, {Blecha}, {Bosman}, {Carilli}, {Comastri},
  {Connor}, {Costa}, {Eilers}, {Fan}, {Gilli}, {Jun}, {Liu}, {Marshall},
  {Mazzucchelli}, {Neeleman}, {Onoue}, {Overzier}, {Pudoka}, {Riechers}, {Rix},
  {Schindler}, {Trakhtenbrot}, {Trebitsch}, {Vestergaard}, {Volonteri}, {Wang},
  {Zhang}, \& {Zou}}]{Decarli24}
{Decarli}, R., {Loiacono}, F., {Farina}, E.~P., {et~al.} 2024, \aap, 689, A219

\bibitem[{{Decarli} {et~al.}(2023){Decarli}, {Pensabene}, {Diaz-Santos},
  {Ferkinhoff}, {Strauss}, {Venemans}, {Walter}, {Ba{\~n}ados}, {Bertoldi},
  {Fan}, {Farina}, {Riechers}, {Rix}, \& {Wang}}]{Decarli23}
{Decarli}, R., {Pensabene}, A., {Diaz-Santos}, T., {et~al.} 2023, \aap, 673,
  A157

\bibitem[{{Decarli} {et~al.}(2022){Decarli}, {Pensabene}, {Venemans}, {Walter},
  {Ba{\~n}ados}, {Bertoldi}, {Carilli}, {Cox}, {Fan}, {Farina}, {Ferkinhoff},
  {Groves}, {Li}, {Mazzucchelli}, {Neri}, {Riechers}, {Uzgil}, {Wang}, {Wang},
  {Weiss}, {Winters}, \& {Yang}}]{Decarli22}
{Decarli}, R., {Pensabene}, A., {Venemans}, B., {et~al.} 2022, \aap, 662, A60

\bibitem[{{Decarli} {et~al.}(2017){Decarli}, {Walter}, {Venemans},
  {Ba{\~n}ados}, {Bertoldi}, {Carilli}, {Fan}, {Farina}, {Mazzucchelli},
  {Riechers}, {Rix}, {Strauss}, {Wang}, \& {Yang}}]{Decarli17}
{Decarli}, R., {Walter}, F., {Venemans}, B.~P., {et~al.} 2017, \nat, 545, 457

\bibitem[{{Decarli} {et~al.}(2018){Decarli}, {Walter}, {Venemans},
  {Ba{\~n}ados}, {Bertoldi}, {Carilli}, {Fan}, {Farina}, {Mazzucchelli},
  {Riechers}, {Rix}, {Strauss}, {Wang}, \& {Yang}}]{Decarli18}
{Decarli}, R., {Walter}, F., {Venemans}, B.~P., {et~al.} 2018, \apj, 854, 97

\bibitem[{{Dell'Agli} {et~al.}(2019){Dell'Agli}, {Valiante}, {Kamath},
  {Ventura}, \& {Garc{\'\i}a-Hern{\'a}ndez}}]{DellAgli19b}
{Dell'Agli}, F., {Valiante}, R., {Kamath}, D., {Ventura}, P., \&
  {Garc{\'\i}a-Hern{\'a}ndez}, D.~A. 2019, \mnras, 486, 4738

\bibitem[{{Di Mascia} {et~al.}(2021){Di Mascia}, {Gallerani}, {Behrens},
  {Pallottini}, {Carniani}, {Ferrara}, {Barai}, {Vito}, \& {Zana}}]{DiMascia21}
{Di Mascia}, F., {Gallerani}, S., {Behrens}, C., {et~al.} 2021, \mnras, 503,
  2349

\bibitem[{{Dietrich} \& {Hamann}(2004)}]{DietrichHamann04}
{Dietrich}, M. \& {Hamann}, F. 2004, \apj, 611, 761

\bibitem[{{Ding} {et~al.}(2023){Ding}, {Onoue}, {Silverman}, {Matsuoka},
  {Izumi}, {Strauss}, {Jahnke}, {Phillips}, {Li}, {Volonteri}, {Haiman},
  {Andika}, {Aoki}, {Baba}, {Bieri}, {Bosman}, {Bottrell}, {Eilers},
  {Fujimoto}, {Habouzit}, {Imanishi}, {Inayoshi}, {Iwasawa}, {Kashikawa},
  {Kawaguchi}, {Kohno}, {Lee}, {Lupi}, {Lyu}, {Nagao}, {Overzier}, {Schindler},
  {Schramm}, {Shimasaku}, {Toba}, {Trakhtenbrot}, {Trebitsch}, {Treu},
  {Umehata}, {Venemans}, {Vestergaard}, {Walter}, {Wang}, \& {Yang}}]{Ding23}
{Ding}, X., {Onoue}, M., {Silverman}, J.~D., {et~al.} 2023, \nat, 621, 51

\bibitem[{{Draine}(2003)}]{Draine03}
{Draine}, B.~T. 2003, \araa, 41, 241

\bibitem[{{Draine}(2009)}]{Draine09}
{Draine}, B.~T. 2009, in Astronomical Society of the Pacific Conference Series,
  Vol. 414, Cosmic Dust - Near and Far, ed. T.~{Henning}, E.~{Gr{\"u}n}, \&
  J.~{Steinacker}, 453

\bibitem[{{Draine} \& {Li}(2007)}]{DraineLi07}
{Draine}, B.~T. \& {Li}, A. 2007, \apj, 657, 810

\bibitem[{{Duras} {et~al.}(2017){Duras}, {Bongiorno}, {Piconcelli}, {Bianchi},
  {Pappalardo}, {Valiante}, {Bischetti}, {Feruglio}, {Martocchia}, {Schneider},
  {Vietri}, {Vignali}, {Zappacosta}, {La Franca}, \& {Fiore}}]{Duras17}
{Duras}, F., {Bongiorno}, A., {Piconcelli}, E., {et~al.} 2017, \aap, 604, A67

\bibitem[{{Elvis} {et~al.}(2002){Elvis}, {Marengo}, \& {Karovska}}]{Elvis02}
{Elvis}, M., {Marengo}, M., \& {Karovska}, M. 2002, \apjl, 567, L107

\bibitem[{{Fan} {et~al.}(2023){Fan}, {Ba{\~n}ados}, \& {Simcoe}}]{Fan23}
{Fan}, X., {Ba{\~n}ados}, E., \& {Simcoe}, R.~A. 2023, \araa, 61, 373

\bibitem[{{Feruglio} {et~al.}(2018){Feruglio}, {Fiore}, {Carniani}, {Maiolino},
  {D'Odorico}, {Luminari}, {Barai}, {Bischetti}, {Bongiorno}, {Cristiani},
  {Ferrara}, {Gallerani}, {Marconi}, {Pallottini}, {Piconcelli}, \&
  {Zappacosta}}]{Feruglio18}
{Feruglio}, C., {Fiore}, F., {Carniani}, S., {et~al.} 2018, \aap, 619, A39

\bibitem[{{Feruglio} {et~al.}(2023){Feruglio}, {Maio}, {Tripodi}, {Winters},
  {Zappacosta}, {Bischetti}, {Civano}, {Carniani}, {D'Odorico}, {Fiore},
  {Gallerani}, {Ginolfi}, {Maiolino}, {Piconcelli}, {Valiante}, \&
  {Zanchettin}}]{Feruglio23}
{Feruglio}, C., {Maio}, U., {Tripodi}, R., {et~al.} 2023, \apjl, 954, L10

\bibitem[{{Fiore} {et~al.}(2017){Fiore}, {Feruglio}, {Shankar}, {Bischetti},
  {Bongiorno}, {Brusa}, {Carniani}, {Cicone}, {Duras}, {Lamastra}, {Mainieri},
  {Marconi}, {Menci}, {Maiolino}, {Piconcelli}, {Vietri}, \&
  {Zappacosta}}]{Fiore17}
{Fiore}, F., {Feruglio}, C., {Shankar}, F., {et~al.} 2017, \aap, 601, A143

\bibitem[{{Fluetsch} {et~al.}(2019){Fluetsch}, {Maiolino}, {Carniani},
  {Marconi}, {Cicone}, {Bourne}, {Costa}, {Fabian}, {Ishibashi}, \&
  {Venturi}}]{Fluetsch19}
{Fluetsch}, A., {Maiolino}, R., {Carniani}, S., {et~al.} 2019, \mnras, 483,
  4586

\bibitem[{{Fontanot} {et~al.}(2020){Fontanot}, {De Lucia}, {Hirschmann}, {Xie},
  {Monaco}, {Menci}, {Fiore}, {Feruglio}, {Cristiani}, \&
  {Shankar}}]{Fontanot20}
{Fontanot}, F., {De Lucia}, G., {Hirschmann}, M., {et~al.} 2020, \mnras, 496,
  3943

\bibitem[{{Foreman-Mackey} {et~al.}(2019){Foreman-Mackey}, {Farr}, {Sinha},
  {Archibald}, {Hogg}, {Sanders}, {Zuntz}, {Williams}, {Nelson}, {de
  Val-Borro}, {Erhardt}, {Pashchenko}, \& {Pla}}]{emcee2}
{Foreman-Mackey}, D., {Farr}, W., {Sinha}, M., {et~al.} 2019, The Journal of
  Open Source Software, 4, 1864

\bibitem[{{Foreman-Mackey} {et~al.}(2013){Foreman-Mackey}, {Hogg}, {Lang}, \&
  {Goodman}}]{emcee1}
{Foreman-Mackey}, D., {Hogg}, D.~W., {Lang}, D., \& {Goodman}, J. 2013, \pasp,
  125, 306

\bibitem[{{Gallerani} {et~al.}(2010){Gallerani}, {Maiolino}, {Juarez}, {Nagao},
  {Marconi}, {Bianchi}, {Schneider}, {Mannucci}, {Oliva}, {Willott}, {Jiang},
  \& {Fan}}]{Gallerani10}
{Gallerani}, S., {Maiolino}, R., {Juarez}, Y., {et~al.} 2010, \aap, 523, A85

\bibitem[{{Galliano} {et~al.}(2021){Galliano}, {Nersesian}, {Bianchi}, {De
  Looze}, {Roychowdhury}, {Baes}, {Casasola}, {Cassar{\'a}}, {Dobbels},
  {Fritz}, {Galametz}, {Jones}, {Madden}, {Mosenkov}, {Xilouris}, \&
  {Ysard}}]{Galliano21}
{Galliano}, F., {Nersesian}, A., {Bianchi}, S., {et~al.} 2021, \aap, 649, A18

\bibitem[{{Gunn} \& {Peterson}(1965)}]{GunnPeterson65}
{Gunn}, J.~E. \& {Peterson}, B.~A. 1965, \apj, 142, 1633

\bibitem[{{Harikane} {et~al.}(2023){Harikane}, {Zhang}, {Nakajima}, {Ouchi},
  {Isobe}, {Ono}, {Hatano}, {Xu}, \& {Umeda}}]{Harikane23}
{Harikane}, Y., {Zhang}, Y., {Nakajima}, K., {et~al.} 2023, \apj, 959, 39

\bibitem[{{Harris} {et~al.}(2020){Harris}, {Millman}, {van der Walt},
  {Gommers}, {Virtanen}, {Cournapeau}, {Wieser}, {Taylor}, {Berg}, {Smith},
  {Kern}, {Picus}, {Hoyer}, {van Kerkwijk}, {Brett}, {Haldane}, {del R{\'\i}o},
  {Wiebe}, {Peterson}, {G{\'e}rard-Marchant}, {Sheppard}, {Reddy}, {Weckesser},
  {Abbasi}, {Gohlke}, \& {Oliphant}}]{numpy}
{Harris}, C.~R., {Millman}, K.~J., {van der Walt}, S.~J., {et~al.} 2020, \nat,
  585, 357

\bibitem[{{Herrera-Camus} {et~al.}(2018){Herrera-Camus}, {Sturm},
  {Graci{\'a}-Carpio}, {Lutz}, {Contursi}, {Veilleux}, {Fischer},
  {Gonz{\'a}lez-Alfonso}, {Poglitsch}, {Tacconi}, {Genzel}, {Maiolino},
  {Sternberg}, {Davies}, \& {Verma}}]{HerreraCamus18}
{Herrera-Camus}, R., {Sturm}, E., {Graci{\'a}-Carpio}, J., {et~al.} 2018, \apj,
  861, 95

\bibitem[{Hunter(2007)}]{matplotlib}
Hunter, J.~D. 2007, Computing in Science \& Engineering, 9, 90

\bibitem[{{Hygate} {et~al.}(2023){Hygate}, {Hodge}, {da Cunha}, {Rybak},
  {Schouws}, {Inami}, {Stefanon}, {Graziani}, {Schneider}, {Dayal}, {Bouwens},
  {Smit}, {Bowler}, {Endsley}, {Gonzalez}, {Oesch}, {Stark}, {Algera},
  {Aravena}, {Barrufet}, {Ferrara}, {Fudamoto}, {Hilhorst}, {De Looze},
  {Nanayakkara}, {Pallottini}, {Riechers}, {Sommovigo}, {Topping}, \& {van der
  Werf}}]{Hygate23}
{Hygate}, A.~P.~S., {Hodge}, J.~A., {da Cunha}, E., {et~al.} 2023, \mnras, 524,
  1775

\bibitem[{{Izumi} {et~al.}(2021){Izumi}, {Matsuoka}, {Fujimoto}, {Onoue},
  {Strauss}, {Umehata}, {Imanishi}, {Kohno}, {Kawaguchi}, {Kawamuro}, {Baba},
  {Nagao}, {Toba}, {Inayoshi}, {Silverman}, {Inoue}, {Ikarashi}, {Iwasawa},
  {Kashikawa}, {Hashimoto}, {Nakanishi}, {Ueda}, {Schramm}, {Lee}, \&
  {Suh}}]{Izumi21}
{Izumi}, T., {Matsuoka}, Y., {Fujimoto}, S., {et~al.} 2021, \apj, 914, 36

\bibitem[{{Izumi} {et~al.}(2024){Izumi}, {Matsuoka}, {Onoue}, {Strauss},
  {Umehata}, {Silverman}, {Nagao}, {Imanishi}, {Kohno}, {Toba}, {Iwasawa},
  {Nakanishi}, {Sawamura}, {Fujimoto}, {Kikuta}, {Kawaguchi}, {Aoki}, \&
  {Goto}}]{Izumi24}
{Izumi}, T., {Matsuoka}, Y., {Onoue}, M., {et~al.} 2024, \apj, 972, 116

\bibitem[{{Kaasinen} {et~al.}(2024){Kaasinen}, {Venemans}, {Harrington},
  {Boogaard}, {Meyer}, {Ba{\~n}ados}, {Decarli}, {Walter}, {Neeleman},
  {Rivera}, \& {da Cunha}}]{Kaasinen24}
{Kaasinen}, M., {Venemans}, B., {Harrington}, K.~C., {et~al.} 2024, \aap, 684,
  A33

\bibitem[{{Kashikawa} {et~al.}(2015){Kashikawa}, {Ishizaki}, {Willott},
  {Onoue}, {Im}, {Furusawa}, {Toshikawa}, {Ishikawa}, {Niino}, {Shimasaku},
  {Ouchi}, \& {Hibon}}]{Kashikawa15}
{Kashikawa}, N., {Ishizaki}, Y., {Willott}, C.~J., {et~al.} 2015, \apj, 798, 28

\bibitem[{{Kelly}(2007)}]{linmix}
{Kelly}, B.~C. 2007, \apj, 665, 1489

\bibitem[{{Kennicutt}(1998)}]{Kennicutt98}
{Kennicutt}, Robert~C., J. 1998, \araa, 36, 189

\bibitem[{{Kirchschlager} {et~al.}(2019){Kirchschlager}, {Schmidt}, {Barlow},
  {Fogerty}, {Bevan}, \& {Priestley}}]{Kirchschlager19}
{Kirchschlager}, F., {Schmidt}, F.~D., {Barlow}, M.~J., {et~al.} 2019, \mnras,
  489, 4465

\bibitem[{{Koss} {et~al.}(2021){Koss}, {Strittmatter}, {Lamperti}, {Shimizu},
  {Trakhtenbrot}, {Saintonge}, {Treister}, {Cicone}, {Mushotzky}, {Oh},
  {Ricci}, {Stern}, {Ananna}, {Bauer}, {Privon}, {B{\"a}r}, {De Breuck},
  {Harrison}, {Ichikawa}, {Powell}, {Rosario}, {Sanders}, {Schawinski}, {Shao},
  {Megan Urry}, \& {Veilleux}}]{Koss21}
{Koss}, M.~J., {Strittmatter}, B., {Lamperti}, I., {et~al.} 2021, \apjs, 252,
  29

\bibitem[{{Kroupa}(2001)}]{Kroupa01}
{Kroupa}, P. 2001, \mnras, 322, 231

\bibitem[{{Le{\'s}niewska} \& {Micha{\l}owski}(2019)}]{Lesniewska19}
{Le{\'s}niewska}, A. \& {Micha{\l}owski}, M.~J. 2019, \aap, 624, L13

\bibitem[{{Madden} {et~al.}(2020){Madden}, {Cormier}, {Hony}, {Lebouteiller},
  {Abel}, {Galametz}, {De Looze}, {Chevance}, {Polles}, {Lee}, {Galliano},
  {Lambert-Huyghe}, {Hu}, \& {Ramambason}}]{Madden20}
{Madden}, S.~C., {Cormier}, D., {Hony}, S., {et~al.} 2020, \aap, 643, A141

\bibitem[{{Maio} {et~al.}(2022){Maio}, {P{\'e}roux}, \& {Ciardi}}]{Maio22}
{Maio}, U., {P{\'e}roux}, C., \& {Ciardi}, B. 2022, \aap, 657, A47

\bibitem[{{Maio} \& {Viel}(2023)}]{Maio23}
{Maio}, U. \& {Viel}, M. 2023, \aap, 672, A71

\bibitem[{{Maiolino} {et~al.}(2005){Maiolino}, {Cox}, {Caselli}, {Beelen},
  {Bertoldi}, {Carilli}, {Kaufman}, {Menten}, {Nagao}, {Omont}, {Wei{\ss}},
  {Walmsley}, \& {Walter}}]{Maiolino05}
{Maiolino}, R., {Cox}, P., {Caselli}, P., {et~al.} 2005, \aap, 440, L51

\bibitem[{{Maiolino} {et~al.}(2024{\natexlab{a}}){Maiolino}, {Risaliti},
  {Signorini}, {Trefoloni}, {Juodzbalis}, {Scholtz}, {Uebler}, {D'Eugenio},
  {Carniani}, {Fabian}, {Ji}, {Mazzolari}, {Bertola}, {Brusa}, {Bunker},
  {Charlot}, {Comastri}, {Cresci}, {DeCoursey}, {Egami}, {Fiore}, {Gilli},
  {Perna}, {Tacchella}, \& {Venturi}}]{Maiolino24b}
{Maiolino}, R., {Risaliti}, G., {Signorini}, M., {et~al.} 2024{\natexlab{a}},
  MNRAS, arXiv:2405.00504

\bibitem[{{Maiolino} {et~al.}(2024{\natexlab{b}}){Maiolino}, {Scholtz},
  {Curtis-Lake}, {Carniani}, {Baker}, {de Graaff}, {Tacchella}, {{\"U}bler},
  {D'Eugenio}, {Witstok}, {Curti}, {Arribas}, {Bunker}, {Charlot},
  {Chevallard}, {Eisenstein}, {Egami}, {Ji}, {Jones}, {Lyu}, {Rawle},
  {Robertson}, {Rujopakarn}, {Perna}, {Sun}, {Venturi}, {Williams}, \&
  {Willott}}]{Maiolino24a}
{Maiolino}, R., {Scholtz}, J., {Curtis-Lake}, E., {et~al.} 2024{\natexlab{b}},
  \aap, 691, A145

\bibitem[{{Mancini} {et~al.}(2015){Mancini}, {Schneider}, {Graziani},
  {Valiante}, {Dayal}, {Maio}, {Ciardi}, \& {Hunt}}]{Mancini15}
{Mancini}, M., {Schneider}, R., {Graziani}, L., {et~al.} 2015, \mnras, 451, L70

\bibitem[{{Mazzucchelli} {et~al.}(2017){Mazzucchelli}, {Ba{\~n}ados},
  {Venemans}, {Decarli}, {Farina}, {Walter}, {Eilers}, {Rix}, {Simcoe},
  {Stern}, {Fan}, {Schlafly}, {De Rosa}, {Hennawi}, {Chambers}, {Greiner},
  {Burgett}, {Draper}, {Kaiser}, {Kudritzki}, {Magnier}, {Metcalfe}, {Waters},
  \& {Wainscoat}}]{Mazzucchelli17}
{Mazzucchelli}, C., {Ba{\~n}ados}, E., {Venemans}, B.~P., {et~al.} 2017, \apj,
  849, 91

\bibitem[{{McMullin} {et~al.}(2007){McMullin}, {Waters}, {Schiebel}, {Young},
  \& {Golap}}]{McMullin07}
{McMullin}, J.~P., {Waters}, B., {Schiebel}, D., {Young}, W., \& {Golap}, K.
  2007, in Astronomical Society of the Pacific Conference Series, Vol. 376,
  Astronomical Data Analysis Software and Systems XVI, ed. R.~A. {Shaw},
  F.~{Hill}, \& D.~J. {Bell}, 127

\bibitem[{{Micelotta} {et~al.}(2018){Micelotta}, {Matsuura}, \&
  {Sarangi}}]{Micelotta18}
{Micelotta}, E.~R., {Matsuura}, M., \& {Sarangi}, A. 2018, \ssr, 214, 53

\bibitem[{Neeleman {et~al.}(2019)Neeleman, Bañados, Walter, Decarli, Venemans,
  Carilli, Fan, Farina, Mazzucchelli, Novak, Riechers, Rix, \&
  Wang}]{Neeleman19}
Neeleman, M., Bañados, E., Walter, F., {et~al.} 2019, The Astrophysical
  Journal, 882, 10

\bibitem[{{Neeleman} {et~al.}(2021){Neeleman}, {Novak}, {Venemans}, {Walter},
  {Decarli}, {Kaasinen}, {Schindler}, {Ba{\~n}ados}, {Carilli}, {Drake}, {Fan},
  \& {Rix}}]{Neeleman21}
{Neeleman}, M., {Novak}, M., {Venemans}, B.~P., {et~al.} 2021, \apj, 911, 141

\bibitem[{{Novak} {et~al.}(2019){Novak}, {Ba{\~n}ados}, {Decarli}, {Walter},
  {Venemans}, {Neeleman}, {Farina}, {Mazzucchelli}, {Carilli}, {Fan}, {Rix}, \&
  {Wang}}]{Novak19}
{Novak}, M., {Ba{\~n}ados}, E., {Decarli}, R., {et~al.} 2019, \apj, 881, 63

\bibitem[{{Pensabene} {et~al.}(2022){Pensabene}, {van der Werf}, {Decarli},
  {Ba{\~n}ados}, {Meyer}, {Riechers}, {Venemans}, {Walter}, {Wei{\ss}},
  {Brusa}, {Fan}, {Wang}, \& {Yang}}]{Pensabene22}
{Pensabene}, A., {van der Werf}, P., {Decarli}, R., {et~al.} 2022, \aap, 667,
  A9

\bibitem[{{Pineda} {et~al.}(2013){Pineda}, {Langer}, {Velusamy}, \&
  {Goldsmith}}]{Pineda13}
{Pineda}, J.~L., {Langer}, W.~D., {Velusamy}, T., \& {Goldsmith}, P.~F. 2013,
  \aap, 554, A103

\bibitem[{{Pipino} {et~al.}(2011){Pipino}, {Fan}, {Matteucci}, {Calura},
  {Silva}, {Granato}, \& {Maiolino}}]{Pipino11}
{Pipino}, A., {Fan}, X.~L., {Matteucci}, F., {et~al.} 2011, \aap, 525, A61

\bibitem[{{Planck Collaboration} {et~al.}(2014){Planck Collaboration}, {Ade},
  {Aghanim}, {Armitage-Caplan}, {Arnaud}, {Ashdown}, {Atrio-Barandela},
  {Aumont}, {Baccigalupi}, {Banday}, {Barreiro}, {Bartlett}, {Battaner},
  {Benabed}, {Beno{\^\i}t}, {Benoit-L{\'e}vy}, {Bernard}, {Bersanelli},
  {Bielewicz}, {Bobin}, {Bock}, {Bonaldi}, {Bond}, {Borrill}, {Bouchet},
  {Bridges}, {Bucher}, {Burigana}, {Butler}, {Calabrese}, {Cappellini},
  {Cardoso}, {Catalano}, {Challinor}, {Chamballu}, {Chary}, {Chen}, {Chiang},
  {Chiang}, {Christensen}, {Church}, {Clements}, {Colombi}, {Colombo},
  {Couchot}, {Coulais}, {Crill}, {Curto}, {Cuttaia}, {Danese}, {Davies},
  {Davis}, {de Bernardis}, {de Rosa}, {de Zotti}, {Delabrouille}, {Delouis},
  {D{\'e}sert}, {Dickinson}, {Diego}, {Dolag}, {Dole}, {Donzelli}, {Dor{\'e}},
  {Douspis}, {Dunkley}, {Dupac}, {Efstathiou}, {Elsner}, {En{\ss}lin},
  {Eriksen}, {Finelli}, {Forni}, {Frailis}, {Fraisse}, {Franceschi}, {Gaier},
  {Galeotta}, {Galli}, {Ganga}, {Giard}, {Giardino}, {Giraud-H{\'e}raud},
  {Gjerl{\o}w}, {Gonz{\'a}lez-Nuevo}, {G{\'o}rski}, {Gratton}, {Gregorio},
  {Gruppuso}, {Gudmundsson}, {Haissinski}, {Hamann}, {Hansen}, {Hanson},
  {Harrison}, {Henrot-Versill{\'e}}, {Hern{\'a}ndez-Monteagudo}, {Herranz},
  {Hildebrandt}, {Hivon}, {Hobson}, {Holmes}, {Hornstrup}, {Hou}, {Hovest},
  {Huffenberger}, {Jaffe}, {Jaffe}, {Jewell}, {Jones}, {Juvela},
  {Keih{\"a}nen}, {Keskitalo}, {Kisner}, {Kneissl}, {Knoche}, {Knox}, {Kunz},
  {Kurki-Suonio}, {Lagache}, {L{\"a}hteenm{\"a}ki}, {Lamarre}, {Lasenby},
  {Lattanzi}, {Laureijs}, {Lawrence}, {Leach}, {Leahy}, {Leonardi},
  {Le{\'o}n-Tavares}, {Lesgourgues}, {Lewis}, {Liguori}, {Lilje},
  {Linden-V{\o}rnle}, {L{\'o}pez-Caniego}, {Lubin}, {Mac{\'\i}as-P{\'e}rez},
  {Maffei}, {Maino}, {Mandolesi}, {Maris}, {Marshall}, {Martin},
  {Mart{\'\i}nez-Gonz{\'a}lez}, {Masi}, {Massardi}, {Matarrese}, {Matthai},
  {Mazzotta}, {Meinhold}, {Melchiorri}, {Melin}, {Mendes}, {Menegoni},
  {Mennella}, {Migliaccio}, {Millea}, {Mitra}, {Miville-Desch{\^e}nes},
  {Moneti}, {Montier}, {Morgante}, {Mortlock}, {Moss}, {Munshi}, {Murphy},
  {Naselsky}, {Nati}, {Natoli}, {Netterfield}, {N{\o}rgaard-Nielsen},
  {Noviello}, {Novikov}, {Novikov}, {O'Dwyer}, {Osborne}, {Oxborrow}, {Paci},
  {Pagano}, {Pajot}, {Paladini}, {Paoletti}, {Partridge}, {Pasian},
  {Patanchon}, {Pearson}, {Pearson}, {Peiris}, {Perdereau}, {Perotto},
  {Perrotta}, {Pettorino}, {Piacentini}, {Piat}, {Pierpaoli}, {Pietrobon},
  {Plaszczynski}, {Platania}, {Pointecouteau}, {Polenta}, {Ponthieu}, {Popa},
  {Poutanen}, {Pratt}, {Pr{\'e}zeau}, {Prunet}, {Puget}, {Rachen}, {Reach},
  {Rebolo}, {Reinecke}, {Remazeilles}, {Renault}, {Ricciardi}, {Riller},
  {Ristorcelli}, {Rocha}, {Rosset}, {Roudier}, {Rowan-Robinson},
  {Rubi{\~n}o-Mart{\'\i}n}, {Rusholme}, {Sandri}, {Santos}, {Savelainen},
  {Savini}, {Scott}, {Seiffert}, {Shellard}, {Spencer}, {Starck}, {Stolyarov},
  {Stompor}, {Sudiwala}, {Sunyaev}, {Sureau}, {Sutton}, {Suur-Uski}, {Sygnet},
  {Tauber}, {Tavagnacco}, {Terenzi}, {Toffolatti}, {Tomasi}, {Tristram},
  {Tucci}, {Tuovinen}, {T{\"u}rler}, {Umana}, {Valenziano}, {Valiviita}, {Van
  Tent}, {Vielva}, {Villa}, {Vittorio}, {Wade}, {Wandelt}, {Wehus}, {White},
  {White}, {Wilkinson}, {Yvon}, {Zacchei}, \& {Zonca}}]{Planck_cosmpar}
{Planck Collaboration}, {Ade}, P.~A.~R., {Aghanim}, N., {et~al.} 2014, \aap,
  571, A16

\bibitem[{{Popesso} {et~al.}(2023){Popesso}, {Concas}, {Cresci}, {Belli},
  {Rodighiero}, {Inami}, {Dickinson}, {Ilbert}, {Pannella}, \&
  {Elbaz}}]{Popesso23}
{Popesso}, P., {Concas}, A., {Cresci}, G., {et~al.} 2023, \mnras, 519, 1526

\bibitem[{{Popping} {et~al.}(2017){Popping}, {Somerville}, \&
  {Galametz}}]{Popping17}
{Popping}, G., {Somerville}, R.~S., \& {Galametz}, M. 2017, \mnras, 471, 3152

\bibitem[{{Pozzi} {et~al.}(2021){Pozzi}, {Calura}, {Fudamoto},
  {Dessauges-Zavadsky}, {Gruppioni}, {Talia}, {Zamorani}, {Bethermin},
  {Cimatti}, {Enia}, {Khusanova}, {Decarli}, {Le F{\`e}vre}, {Capak},
  {Cassata}, {Faisst}, {Yan}, {Schaerer}, {Silverman}, {Bardelli}, {Boquien},
  {Enia}, {Narayanan}, {Ginolfi}, {Hathi}, {Jones}, {Koekemoer}, {Lemaux},
  {Loiacono}, {Maiolino}, {Riechers}, {Rodighiero}, {Romano}, {Vallini},
  {Vergani}, \& {Zucca}}]{Pozzi21}
{Pozzi}, F., {Calura}, F., {Fudamoto}, Y., {et~al.} 2021, \aap, 653, A84

\bibitem[{{R{\'e}my-Ruyer} {et~al.}(2014){R{\'e}my-Ruyer}, {Madden},
  {Galliano}, {Galametz}, {Takeuchi}, {Asano}, {Zhukovska}, {Lebouteiller},
  {Cormier}, {Jones}, {Bocchio}, {Baes}, {Bendo}, {Boquien}, {Boselli},
  {DeLooze}, {Doublier-Pritchard}, {Hughes}, {Karczewski}, \&
  {Spinoglio}}]{RemyRuyer14}
{R{\'e}my-Ruyer}, A., {Madden}, S.~C., {Galliano}, F., {et~al.} 2014, \aap,
  563, A31

\bibitem[{{Renzini} \& {Peng}(2015)}]{RenziniPeng15}
{Renzini}, A. \& {Peng}, Y.-j. 2015, \apjl, 801, L29

\bibitem[{{Riechers} {et~al.}(2009){Riechers}, {Walter}, {Bertoldi}, {Carilli},
  {Aravena}, {Neri}, {Cox}, {Wei{\ss}}, \& {Menten}}]{Riechers09}
{Riechers}, D.~A., {Walter}, F., {Bertoldi}, F., {et~al.} 2009, \apj, 703, 1338

\bibitem[{{Riechers} {et~al.}(2007){Riechers}, {Walter}, {Carilli}, \&
  {Bertoldi}}]{Riechers07}
{Riechers}, D.~A., {Walter}, F., {Carilli}, C.~L., \& {Bertoldi}, F. 2007,
  \apjl, 671, L13

\bibitem[{{Riechers} {et~al.}(2006){Riechers}, {Walter}, {Carilli}, {Knudsen},
  {Lo}, {Benford}, {Staguhn}, {Hunter}, {Bertoldi}, {Henkel}, {Menten},
  {Weiss}, {Yun}, \& {Scoville}}]{Riechers06}
{Riechers}, D.~A., {Walter}, F., {Carilli}, C.~L., {et~al.} 2006, \apj, 650,
  604

\bibitem[{{Rizzo} {et~al.}(2021){Rizzo}, {Vegetti}, {Fraternali}, {Stacey}, \&
  {Powell}}]{Rizzo21}
{Rizzo}, F., {Vegetti}, S., {Fraternali}, F., {Stacey}, H.~R., \& {Powell}, D.
  2021, \mnras, 507, 3952

\bibitem[{{Rizzo} {et~al.}(2020){Rizzo}, {Vegetti}, {Powell}, {Fraternali},
  {McKean}, {Stacey}, \& {White}}]{Rizzo20}
{Rizzo}, F., {Vegetti}, S., {Powell}, D., {et~al.} 2020, \nat, 584, 201

\bibitem[{{Romano} {et~al.}(2022){Romano}, {Morselli}, {Cassata}, {Ginolfi},
  {Schaerer}, {B{\'e}thermin}, {Capak}, {Faisst}, {Le F{\`e}vre}, {Silverman},
  {Yan}, {Bardelli}, {Boquien}, {Dessauges-Zavadsky}, {Fujimoto}, {Hathi},
  {Jones}, {Koekemoer}, {Lemaux}, {M{\'e}ndez-Hern{\'a}ndez}, {Narayanan},
  {Talia}, {Vergani}, {Zamorani}, \& {Zucca}}]{Romano22}
{Romano}, M., {Morselli}, L., {Cassata}, P., {et~al.} 2022, \aap, 660, A14

\bibitem[{{Ronconi} {et~al.}(2024){Ronconi}, {Lapi}, {Torsello}, {Bressan},
  {Donevski}, {Pantoni}, {Behiri}, {Boco}, {Cimatti}, {D'Amato}, {Danese},
  {Giulietti}, {Perrotta}, {Silva}, {Talia}, \& {Massardi}}]{Ronconi24}
{Ronconi}, T., {Lapi}, A., {Torsello}, M., {et~al.} 2024, \aap, 685, A161

\bibitem[{{Saintonge} {et~al.}(2018){Saintonge}, {Wilson}, {Xiao}, {Lin},
  {Hwang}, {Tosaki}, {Bureau}, {Cigan}, {Clark}, {Clements}, {De Looze},
  {Dharmawardena}, {Gao}, {Gear}, {Greenslade}, {Lamperti}, {Lee}, {Li},
  {Micha{\l}owski}, {Mok}, {Pan}, {Sansom}, {Sargent}, {Smith}, {Williams},
  {Yang}, {Zhu}, {Accurso}, {Barmby}, {Brinks}, {Bourne}, {Brown}, {Chung},
  {Chung}, {Cibinel}, {Coppin}, {Davies}, {Davis}, {Eales}, {Fanciullo},
  {Fang}, {Gao}, {Glass}, {Gomez}, {Greve}, {He}, {Ho}, {Huang}, {Jeong},
  {Jiang}, {Jiao}, {Kemper}, {Kim}, {Kim}, {Kim}, {Ko}, {Kong}, {Lacaille},
  {Lacey}, {Lee}, {Lee}, {Lee}, {Masters}, {Oh}, {Papadopoulos}, {Park},
  {Park}, {Parsons}, {Rowlands}, {Scicluna}, {Scudder}, {Sethuram}, {Serjeant},
  {Shao}, {Sheen}, {Shi}, {Shim}, {Smith}, {Spekkens}, {Tsai}, {Verma},
  {Urquhart}, {Violino}, {Viti}, {Wake}, {Wang}, {Wouterloot}, {Yang}, {Yim},
  {Yuan}, \& {Zheng}}]{Saintonge18}
{Saintonge}, A., {Wilson}, C.~D., {Xiao}, T., {et~al.} 2018, \mnras, 481, 3497

\bibitem[{{Salom{\'e}} {et~al.}(2023){Salom{\'e}}, {Krongold}, {Longinotti},
  {Bischetti}, {Garc{\'\i}a-Burillo}, {Vega}, {S{\'a}nchez-Portal}, {Feruglio},
  {Jim{\'e}nez-Donaire}, \& {Zanchettin}}]{Salome23}
{Salom{\'e}}, Q., {Krongold}, Y., {Longinotti}, A.~L., {et~al.} 2023, \mnras,
  524, 3130

\bibitem[{{Salvestrini} {et~al.}(2022){Salvestrini}, {Gruppioni},
  {Hatziminaoglou}, {Pozzi}, {Vignali}, {Casasola}, {Paladino}, {Aalto},
  {Andreani}, {Marchesi}, \& {Stanke}}]{Salvestrini22}
{Salvestrini}, F., {Gruppioni}, C., {Hatziminaoglou}, E., {et~al.} 2022, \aap,
  663, A28

\bibitem[{{Salvestrini} {et~al.}(2020){Salvestrini}, {Gruppioni}, {Pozzi},
  {Vignali}, {Giannetti}, {Paladino}, \& {Hatziminaoglou}}]{Salvestrini20}
{Salvestrini}, F., {Gruppioni}, C., {Pozzi}, F., {et~al.} 2020, \aap, 641, A151

\bibitem[{{Sarangi} {et~al.}(2019){Sarangi}, {Dwek}, \& {Kazanas}}]{Sarangi19}
{Sarangi}, A., {Dwek}, E., \& {Kazanas}, D. 2019, \apj, 885, 126

\bibitem[{{Schneider} \& {Maiolino}(2024)}]{SchneiderMaiolino24}
{Schneider}, R. \& {Maiolino}, R. 2024, \aapr, 32, 2

\bibitem[{{Shangguan} {et~al.}(2020){Shangguan}, {Ho}, {Bauer}, {Wang}, \&
  {Treister}}]{Shangguan20}
{Shangguan}, J., {Ho}, L.~C., {Bauer}, F.~E., {Wang}, R., \& {Treister}, E.
  2020, \apj, 899, 112

\bibitem[{{Shao} {et~al.}(2022){Shao}, {Wang}, {Weiss}, {Wagg}, {Carilli},
  {Strauss}, {Walter}, {Cox}, {Fan}, {Menten}, {Narayanan}, {Riechers},
  {Bertoldi}, {Omont}, \& {Jiang}}]{Shao22}
{Shao}, Y., {Wang}, R., {Weiss}, A., {et~al.} 2022, \aap, 668, A121

\bibitem[{{Shen} {et~al.}(2011){Shen}, {Richards}, {Strauss}, {Hall},
  {Schneider}, {Snedden}, {Bizyaev}, {Brewington}, {Malanushenko},
  {Malanushenko}, {Oravetz}, {Pan}, \& {Simmons}}]{Shen11}
{Shen}, Y., {Richards}, G.~T., {Strauss}, M.~A., {et~al.} 2011, \apjs, 194, 45

\bibitem[{{Shen} {et~al.}(2019){Shen}, {Wu}, {Jiang}, {Ba{\~n}ados}, {Fan},
  {Ho}, {Riechers}, {Strauss}, {Venemans}, {Vestergaard}, {Walter}, {Wang},
  {Willott}, {Wu}, \& {Yang}}]{Shen19}
{Shen}, Y., {Wu}, J., {Jiang}, L., {et~al.} 2019, \apj, 873, 35

\bibitem[{{Sommovigo} {et~al.}(2022){Sommovigo}, {Ferrara}, {Pallottini},
  {Dayal}, {Bouwens}, {Smit}, {da Cunha}, {De Looze}, {Bowler}, {Hodge},
  {Inami}, {Oesch}, {Endsley}, {Gonzalez}, {Schouws}, {Stark}, {Stefanon},
  {Aravena}, {Graziani}, {Riechers}, {Schneider}, {van der Werf}, {Algera},
  {Barrufet}, {Fudamoto}, {Hygate}, {Labb{\'e}}, {Li}, {Nanayakkara}, \&
  {Topping}}]{Sommovigo22a}
{Sommovigo}, L., {Ferrara}, A., {Pallottini}, A., {et~al.} 2022, \mnras, 513,
  3122

\bibitem[{{Stefan} {et~al.}(2015){Stefan}, {Carilli}, {Wagg}, {Walter},
  {Riechers}, {Bertoldi}, {Green}, {Fan}, {Menten}, \& {Wang}}]{Stefan15}
{Stefan}, I.~I., {Carilli}, C.~L., {Wagg}, J., {et~al.} 2015, \mnras, 451, 1713

\bibitem[{{Tacconi} {et~al.}(2018){Tacconi}, {Genzel}, {Saintonge}, {Combes},
  {Garc{\'\i}a-Burillo}, {Neri}, {Bolatto}, {Contini}, {F{\"o}rster Schreiber},
  {Lilly}, {Lutz}, {Wuyts}, {Accurso}, {Boissier}, {Boone}, {Bouch{\'e}},
  {Bournaud}, {Burkert}, {Carollo}, {Cooper}, {Cox}, {Feruglio}, {Freundlich},
  {Herrera-Camus}, {Juneau}, {Lippa}, {Naab}, {Renzini}, {Salome}, {Sternberg},
  {Tadaki}, {{\"U}bler}, {Walter}, {Weiner}, \& {Weiss}}]{Tacconi18}
{Tacconi}, L.~J., {Genzel}, R., {Saintonge}, A., {et~al.} 2018, \apj, 853, 179

\bibitem[{{Tacconi} {et~al.}(2020){Tacconi}, {Genzel}, \&
  {Sternberg}}]{Tacconi20}
{Tacconi}, L.~J., {Genzel}, R., \& {Sternberg}, A. 2020, \araa, 58, 157

\bibitem[{{Trinca} {et~al.}(2022){Trinca}, {Schneider}, {Valiante}, {Graziani},
  {Zappacosta}, \& {Shankar}}]{Trinca22}
{Trinca}, A., {Schneider}, R., {Valiante}, R., {et~al.} 2022, \mnras, 511, 616

\bibitem[{{Tripodi} {et~al.}(2024{\natexlab{a}}){Tripodi}, {Feruglio}, {Fiore},
  {Zappacosta}, {Piconcelli}, {Bischetti}, {Bongiorno}, {Carniani}, {Civano},
  {Chen}, {Cristiani}, {Cupani}, {Di Mascia}, {D'Odorico}, {Fan}, {Ferrara},
  {Gallerani}, {Ginolfi}, {Maiolino}, {Mainieri}, {Marconi}, {Saccheo},
  {Salvestrini}, {Tortosa}, \& {Valiante}}]{Tripodi24b}
{Tripodi}, R., {Feruglio}, C., {Fiore}, F., {et~al.} 2024{\natexlab{a}}, \aap,
  689, A220

\bibitem[{{Tripodi} {et~al.}(2023){Tripodi}, {Feruglio}, {Kemper}, {Civano},
  {Costa}, {Elvis}, {Bischetti}, {Carniani}, {Di Mascia}, {D'Odorico}, {Fiore},
  {Gallerani}, {Ginolfi}, {Maiolino}, {Piconcelli}, {Valiante}, \&
  {Zappacosta}}]{Tripodi23b}
{Tripodi}, R., {Feruglio}, C., {Kemper}, F., {et~al.} 2023, \apjl, 946, L45

\bibitem[{{Tripodi} {et~al.}(2024{\natexlab{b}}){Tripodi}, {Scholtz},
  {Maiolino}, {Fujimoto}, {Carniani}, {Silverman}, {Feruglio}, {Ginolfi},
  {Zappacosta}, {Costa}, {Jones}, {Piconcelli}, {Bischetti}, \&
  {Fiore}}]{Tripodi24a}
{Tripodi}, R., {Scholtz}, J., {Maiolino}, R., {et~al.} 2024{\natexlab{b}},
  \aap, 682, A54

\bibitem[{{Tsukui} {et~al.}(2023){Tsukui}, {Wisnioski}, {Krumholz}, \&
  {Battisti}}]{Tsukui23}
{Tsukui}, T., {Wisnioski}, E., {Krumholz}, M.~R., \& {Battisti}, A. 2023,
  \mnras, 523, 4654

\bibitem[{{Valentini} {et~al.}(2021){Valentini}, {Gallerani}, \&
  {Ferrara}}]{Valentini21}
{Valentini}, M., {Gallerani}, S., \& {Ferrara}, A. 2021, \mnras, 507, 1

\bibitem[{{Valiante} {et~al.}(2011){Valiante}, {Schneider}, {Salvadori}, \&
  {Bianchi}}]{Valiante11}
{Valiante}, R., {Schneider}, R., {Salvadori}, S., \& {Bianchi}, S. 2011,
  \mnras, 416, 1916

\bibitem[{{Valiante} {et~al.}(2014){Valiante}, {Schneider}, {Salvadori}, \&
  {Gallerani}}]{Valiante14}
{Valiante}, R., {Schneider}, R., {Salvadori}, S., \& {Gallerani}, S. 2014,
  \mnras, 444, 2442

\bibitem[{{Venemans} {et~al.}(2019){Venemans}, {Neeleman}, {Walter}, {Novak},
  {Decarli}, {Hennawi}, \& {Rix}}]{Venemans19}
{Venemans}, B.~P., {Neeleman}, M., {Walter}, F., {et~al.} 2019, \apjl, 874, L30

\bibitem[{{Venemans} {et~al.}(2017{\natexlab{a}}){Venemans}, {Walter},
  {Decarli}, {Ba{\~n}ados}, {Carilli}, {Winters}, {Schuster}, {da Cunha},
  {Fan}, {Farina}, {Mazzucchelli}, {Rix}, \& {Weiss}}]{Venemans17c}
{Venemans}, B.~P., {Walter}, F., {Decarli}, R., {et~al.} 2017{\natexlab{a}},
  \apjl, 851, L8

\bibitem[{{Venemans} {et~al.}(2017{\natexlab{b}}){Venemans}, {Walter},
  {Decarli}, {Ba{\~n}ados}, {Hodge}, {Hewett}, {McMahon}, {Mortlock}, \&
  {Simpson}}]{Venemans17a}
{Venemans}, B.~P., {Walter}, F., {Decarli}, R., {et~al.} 2017{\natexlab{b}},
  \apj, 837, 146

\bibitem[{{Venemans} {et~al.}(2017{\natexlab{c}}){Venemans}, {Walter},
  {Decarli}, {Ferkinhoff}, {Wei{\ss}}, {Findlay}, {McMahon}, {Sutherland}, \&
  {Meijerink}}]{Venemans17b}
{Venemans}, B.~P., {Walter}, F., {Decarli}, R., {et~al.} 2017{\natexlab{c}},
  \apj, 845, 154

\bibitem[{{Venemans} {et~al.}(2020){Venemans}, {Walter}, {Neeleman}, {Novak},
  {Otter}, {Decarli}, {Ba{\~n}ados}, {Drake}, {Farina}, {Kaasinen},
  {Mazzucchelli}, {Carilli}, {Fan}, {Rix}, \& {Wang}}]{Venemans20}
{Venemans}, B.~P., {Walter}, F., {Neeleman}, M., {et~al.} 2020, \apj, 904, 130

\bibitem[{{Ventura} {et~al.}(2018){Ventura}, {Karakas}, {Dell'Agli},
  {Garc{\'\i}a-Hern{\'a}ndez}, \& {Guzman-Ramirez}}]{Ventura18}
{Ventura}, P., {Karakas}, A., {Dell'Agli}, F., {Garc{\'\i}a-Hern{\'a}ndez},
  D.~A., \& {Guzman-Ramirez}, L. 2018, \mnras, 475, 2282

\bibitem[{{Virtanen} {et~al.}(2020){Virtanen}, {Gommers}, {Oliphant},
  {Haberland}, {Reddy}, {Cournapeau}, {Burovski}, {Peterson}, {Weckesser},
  {Bright}, {van der Walt}, {Brett}, {Wilson}, {Millman}, {Mayorov}, {Nelson},
  {Jones}, {Kern}, {Larson}, {Carey}, {Polat}, {Feng}, {Moore}, {VanderPlas},
  {Laxalde}, {Perktold}, {Cimrman}, {Henriksen}, {Quintero}, {Harris},
  {Archibald}, {Ribeiro}, {Pedregosa}, {van Mulbregt}, \& {SciPy 1. 0
  Contributors}}]{scipy}
{Virtanen}, P., {Gommers}, R., {Oliphant}, T.~E., {et~al.} 2020, Nature
  Methods, 17, 261

\bibitem[{{Vizgan} {et~al.}(2022){Vizgan}, {Greve}, {Olsen}, {Zanella},
  {Narayanan}, {Dav{\`e}}, {Magdis}, {Popping}, {Valentino}, \&
  {Heintz}}]{Vizgan22}
{Vizgan}, D., {Greve}, T.~R., {Olsen}, K.~P., {et~al.} 2022, \apj, 929, 92

\bibitem[{{Walter} {et~al.}(2020){Walter}, {Carilli}, {Neeleman}, {Decarli},
  {Popping}, {Somerville}, {Aravena}, {Bertoldi}, {Boogaard}, {Cox}, {da
  Cunha}, {Magnelli}, {Obreschkow}, {Riechers}, {Rix}, {Smail}, {Weiss},
  {Assef}, {Bauer}, {Bouwens}, {Contini}, {Cortes}, {Daddi}, {Diaz-Santos},
  {Gonz{\'a}lez-L{\'o}pez}, {Hennawi}, {Hodge}, {Inami}, {Ivison}, {Oesch},
  {Sargent}, {van der Werf}, {Wagg}, \& {Yung}}]{Walter20}
{Walter}, F., {Carilli}, C., {Neeleman}, M., {et~al.} 2020, \apj, 902, 111

\bibitem[{{Walter} {et~al.}(2022){Walter}, {Neeleman}, {Decarli}, {Venemans},
  {Meyer}, {Weiss}, {Ba{\~n}ados}, {Bosman}, {Carilli}, {Fan}, {Riechers},
  {Rix}, \& {Thompson}}]{Walter22}
{Walter}, F., {Neeleman}, M., {Decarli}, R., {et~al.} 2022, \apj, 927, 21

\bibitem[{{Wang} {et~al.}(2021){Wang}, {Yang}, {Fan}, {Hennawi}, {Barth},
  {Banados}, {Bian}, {Boutsia}, {Connor}, {Davies}, {Decarli}, {Eilers},
  {Farina}, {Green}, {Jiang}, {Li}, {Mazzucchelli}, {Nanni}, {Schindler},
  {Venemans}, {Walter}, {Wu}, \& {Yue}}]{WangF21}
{Wang}, F., {Yang}, J., {Fan}, X., {et~al.} 2021, \apjl, 907, L1

\bibitem[{{Wang} {et~al.}(2024){Wang}, {Yang}, {Fan}, {Venemans}, {Decarli},
  {Ba{\~n}ados}, {Walter}, {Barth}, {Bian}, {Davies}, {Eilers}, {Farina},
  {Hennawi}, {Li}, {Mazzucchelli}, {Wang}, {Wu}, \& {Yue}}]{WangF24}
{Wang}, F., {Yang}, J., {Fan}, X., {et~al.} 2024, \apj, 968, 9

\bibitem[{{Wang} {et~al.}(2007){Wang}, {Carilli}, {Beelen}, {Bertoldi}, {Fan},
  {Walter}, {Menten}, {Omont}, {Cox}, {Strauss}, \& {Jiang}}]{WangR07}
{Wang}, R., {Carilli}, C.~L., {Beelen}, A., {et~al.} 2007, \aj, 134, 617

\bibitem[{{Wang} {et~al.}(2010){Wang}, {Carilli}, {Neri}, {Riechers}, {Wagg},
  {Walter}, {Bertoldi}, {Menten}, {Omont}, {Cox}, \& {Fan}}]{WangR10}
{Wang}, R., {Carilli}, C.~L., {Neri}, R., {et~al.} 2010, \apj, 714, 699

\bibitem[{{Wang} {et~al.}(2011){Wang}, {Wagg}, {Carilli}, {Neri}, {Walter},
  {Omont}, {Riechers}, {Bertoldi}, {Menten}, {Cox}, {Strauss}, {Fan}, \&
  {Jiang}}]{WangR11b}
{Wang}, R., {Wagg}, J., {Carilli}, C.~L., {et~al.} 2011, \aj, 142, 101

\bibitem[{{Wang} {et~al.}(2013){Wang}, {Wagg}, {Carilli}, {Walter}, {Lentati},
  {Fan}, {Riechers}, {Bertoldi}, {Narayanan}, {Strauss}, {Cox}, {Omont},
  {Menten}, {Knudsen}, {Neri}, \& {Jiang}}]{WangR13}
{Wang}, R., {Wagg}, J., {Carilli}, C.~L., {et~al.} 2013, \apj, 773, 44

\bibitem[{Waskom(2021)}]{seaborn}
Waskom, M.~L. 2021, Journal of Open Source Software, 6, 3021

\bibitem[{{Witstok} {et~al.}(2023){Witstok}, {Jones}, {Maiolino}, {Smit}, \&
  {Schneider}}]{Witstok23}
{Witstok}, J., {Jones}, G.~C., {Maiolino}, R., {Smit}, R., \& {Schneider}, R.
  2023, \mnras, 523, 3119

\bibitem[{{Witstok} {et~al.}(2022){Witstok}, {Smit}, {Maiolino}, {Kumari},
  {Aravena}, {Boogaard}, {Bouwens}, {Carniani}, {Hodge}, {Jones}, {Stefanon},
  {van der Werf}, \& {Schouws}}]{Witstok22}
{Witstok}, J., {Smit}, R., {Maiolino}, R., {et~al.} 2022, \mnras, 515, 1751

\bibitem[{{Xie} {et~al.}(2017){Xie}, {De Lucia}, {Hirschmann}, {Fontanot}, \&
  {Zoldan}}]{Xie17}
{Xie}, L., {De Lucia}, G., {Hirschmann}, M., {Fontanot}, F., \& {Zoldan}, A.
  2017, \mnras, 469, 968

\bibitem[{{Yang} {et~al.}(2021){Yang}, {Wang}, {Fan}, {Barth}, {Hennawi},
  {Nanni}, {Bian}, {Davies}, {Farina}, {Schindler}, {Ba{\~n}ados}, {Decarli},
  {Eilers}, {Green}, {Guo}, {Jiang}, {Li}, {Venemans}, {Walter}, {Wu}, \&
  {Yue}}]{Yang21}
{Yang}, J., {Wang}, F., {Fan}, X., {et~al.} 2021, \apj, 923, 262

\bibitem[{{Yang} {et~al.}(2020){Yang}, {Wang}, {Fan}, {Hennawi}, {Davies},
  {Yue}, {Banados}, {Wu}, {Venemans}, {Barth}, {Bian}, {Boutsia}, {Decarli},
  {Farina}, {Green}, {Jiang}, {Li}, {Mazzucchelli}, \& {Walter}}]{Yang20}
{Yang}, J., {Wang}, F., {Fan}, X., {et~al.} 2020, \apjl, 897, L14

\bibitem[{{Yue} {et~al.}(2024){Yue}, {Eilers}, {Simcoe}, {Mackenzie},
  {Matthee}, {Kashino}, {Bordoloi}, {Lilly}, \& {Naidu}}]{Yue24}
{Yue}, M., {Eilers}, A.-C., {Simcoe}, R.~A., {et~al.} 2024, \apj, 966, 176

\bibitem[{{Zanella} {et~al.}(2018){Zanella}, {Daddi}, {Magdis}, {Diaz Santos},
  {Cormier}, {Liu}, {Cibinel}, {Gobat}, {Dickinson}, {Sargent}, {Popping},
  {Madden}, {Bethermin}, {Hughes}, {Valentino}, {Rujopakarn}, {Pannella},
  {Bournaud}, {Walter}, {Wang}, {Elbaz}, \& {Coogan}}]{Zanella18}
{Zanella}, A., {Daddi}, E., {Magdis}, G., {et~al.} 2018, \mnras, 481, 1976

\end{thebibliography}
\onecolumn
\begin{appendix}
\section{Far-infrared SED models}
\label{sec:app_corner}
\begin{figure*}[ht!]
\centering
\includegraphics[width=\textwidth]{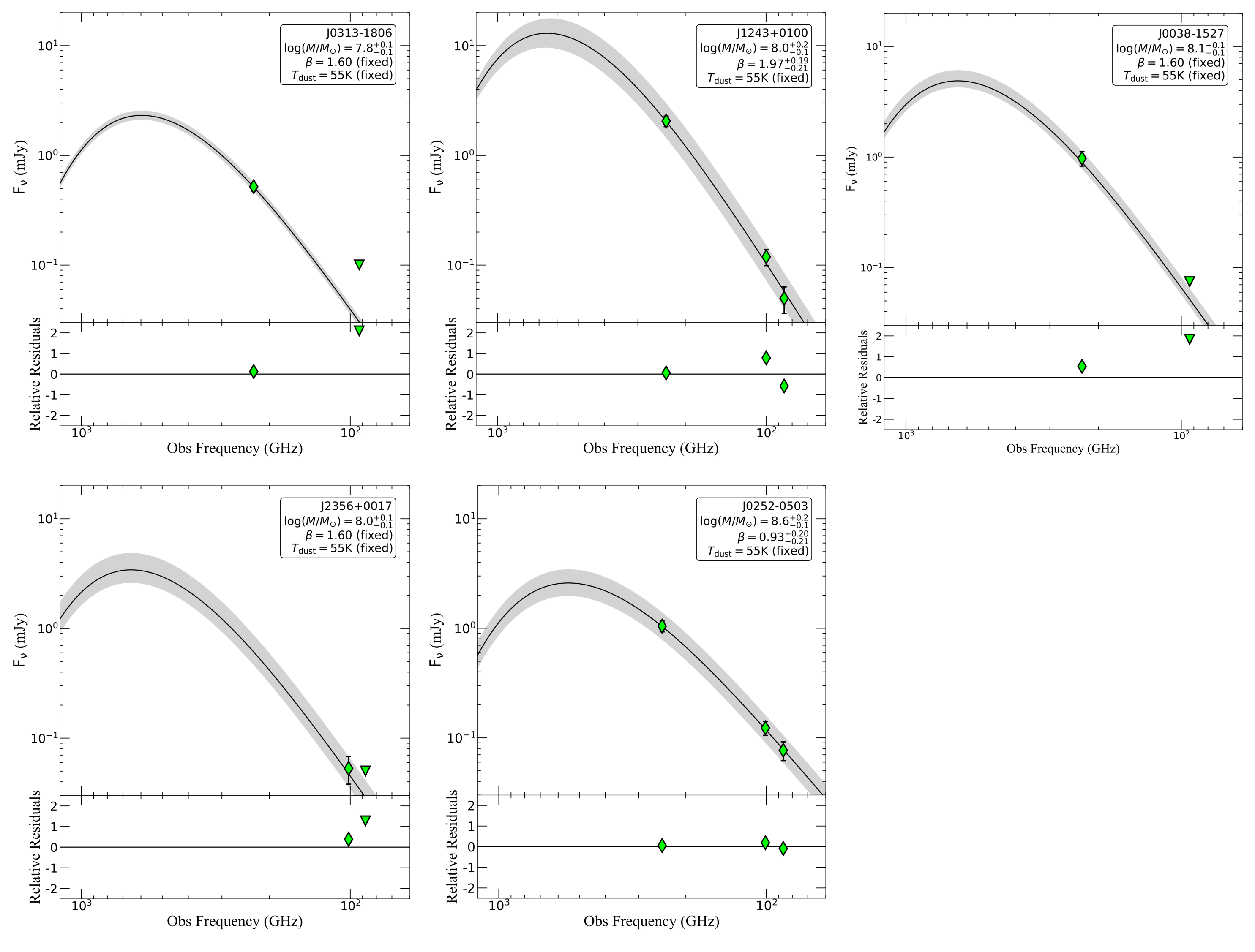}
\caption{FIR SED. Observed fluxes and upper limits are shown as green diamonds and triangles, respectively. The best-fit model is the black line, and the corresponding parameters (namely $\beta$, $T_d$, and $M_d$) are reported in each panel. The gray-shaded region represents the 68\% confidence interval of the best-fit parameters. Relative residuals are shown in the lower box of each panel.}
\label{fig:fir_sed}
\end{figure*}

\begin{figure*}[ht!]
    \centering
    \includegraphics[width=0.55\textwidth]{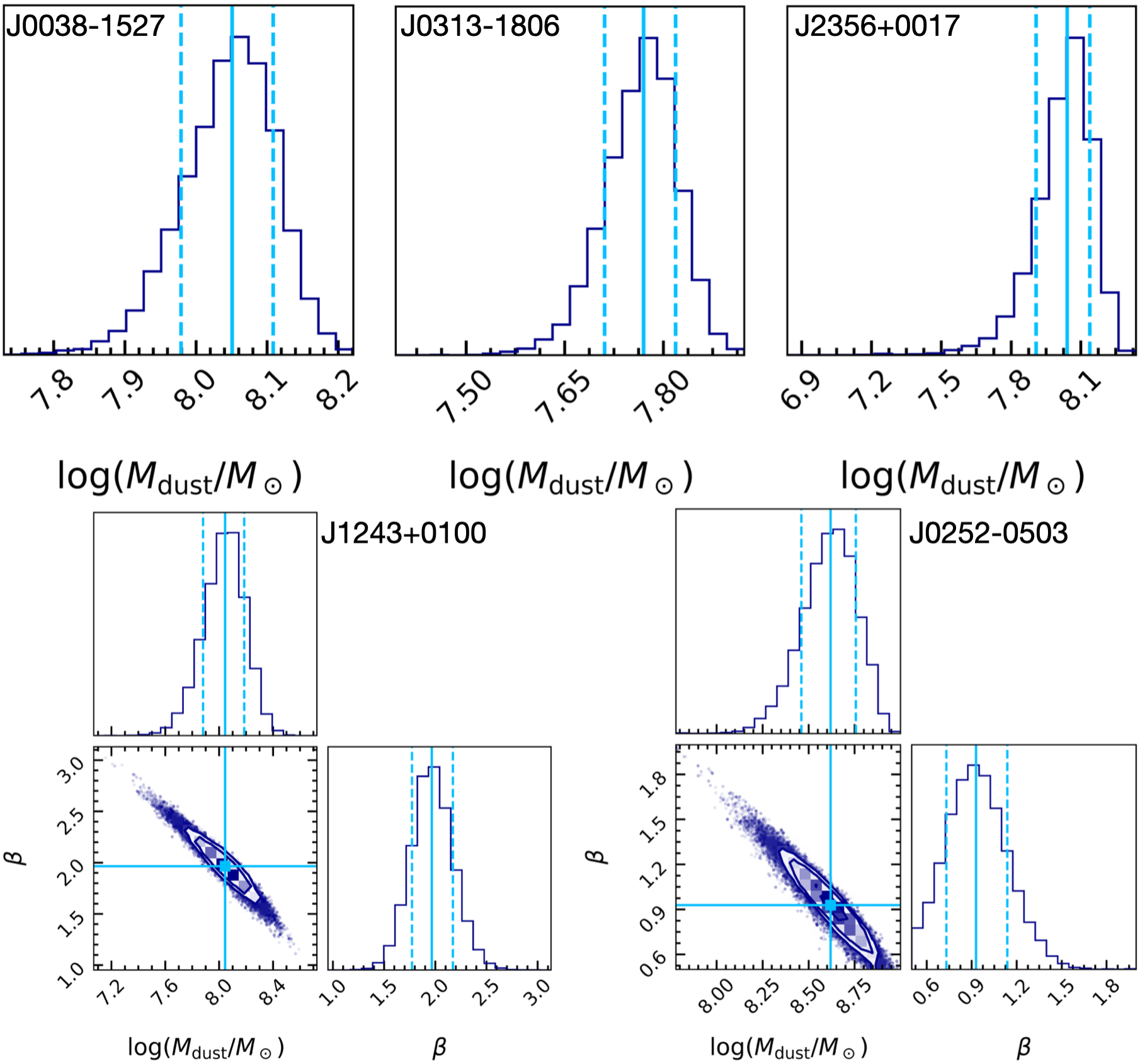}
    \caption{Corner plots showing the posterior probability distributions of the best fit parameters: $M_d$ for each source, and $\beta$  only for J0252 and J1243. Solid cyan lines indicate the best-fitting value for each parameter, while the dashed lines mark each parameter's 16th and 84th percentiles.}
\label{fig:corner_plots}
\end{figure*}

In Fig.~\ref{fig:fir_sed} we present the best-fit models of the FIR SED of the five quasars at $z>7$ in our sample. 
The corresponding corner plots of the free parameters obtained with the Bayesian analysis of the MBB with the {\sc EOS-Dustfit} code are shown in Fig.~\ref{fig:corner_plots}.
Details of the analysis are reported in Sect.~\ref{sec:dustsed}.

\begin{figure*}[ht!]
\centering
\includegraphics[width=0.85\textwidth]{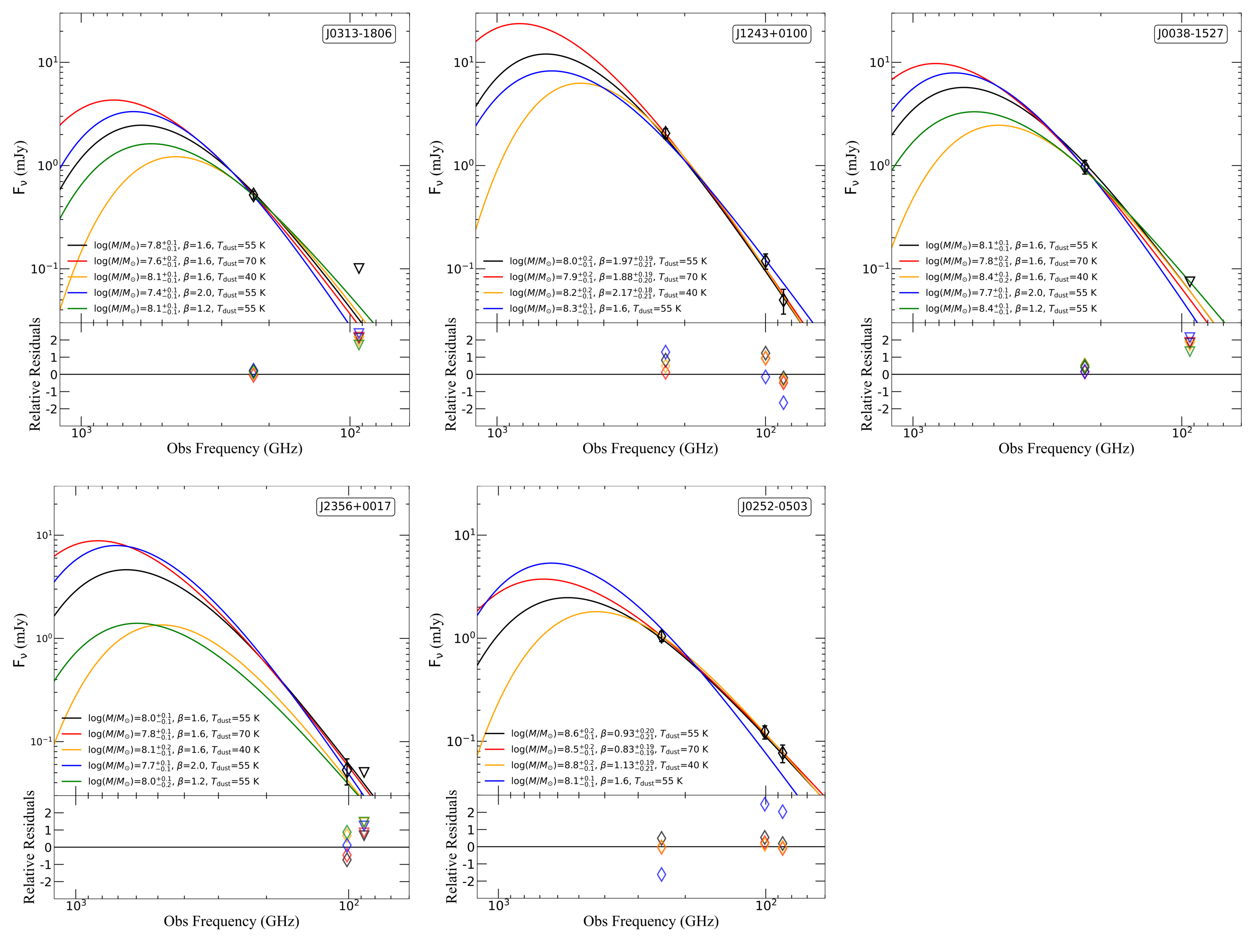}
\caption{FIR SED obtained with different parameter values. Observed fluxes and upper limits are shown as diamonds and triangles, respectively. The best-fit parameters and corresponding uncertainties are reported in the legend. Best-fit models and relative residuals in the lower box are color-coded accordingly.}
\label{fig:fir_sed_multi_par}
\end{figure*}

In Fig.~\ref{fig:fir_sed_multi_par} we show the results of the modeling of the continuum emission when assuming different values for $\beta$ and $T_d$ for each of the five quasars.
The baseline model (i.e., the one shown in Fig.~\ref{fig:fir_sed}) is represented with a solid black line in each panel, while observed fluxes and upper limits are black diamonds and triangles, respectively.
The impact of the assumption of a different dust temperature is shown as the best-fit model obtained with a $T_d=70$ (40)~K, which is displayed with a red (orange) curve.
In the case of J1243 and J0252, $\beta$ was left free to vary, while it is fixed to 1.6 for the remaining targets.

For J0313, J0038, and J2356, we also evaluate the impact of the assumption of $\beta=2$ and $\beta=1.2$ with a blue and solid green line, respectively.
For J1243 and J0252, the solid blue line represents the case of $\beta=1.6$ fixed.
When we varied $\beta$, we left $T_d=55$~K fixed.
Relative residuals are shown in the lower box of each panel, color-coded accordingly to the best-fit lines of the upper box.

\FloatBarrier %\usepackage{placeins}

\FloatBarrier %\usepackage{placeins}

\section{Continuum maps}
\label{sec:app_contmap}
\begin{figure*}[ht!]
    \centering
    \resizebox{0.8\textwidth}{!}{%
    \begin{tabular}{ccc}
        \includegraphics[width=0.3\textwidth]{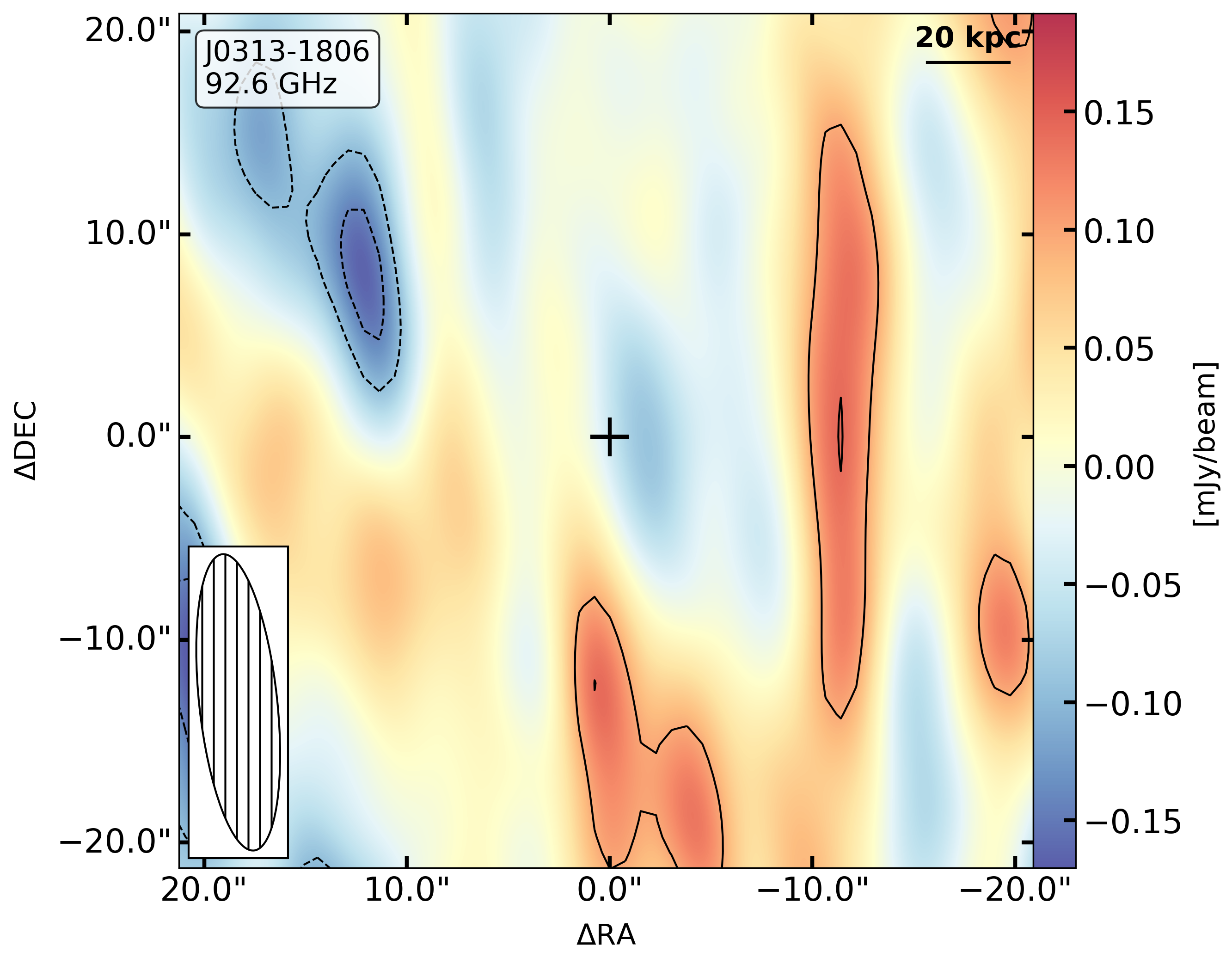} &
        \includegraphics[width=0.3\textwidth]{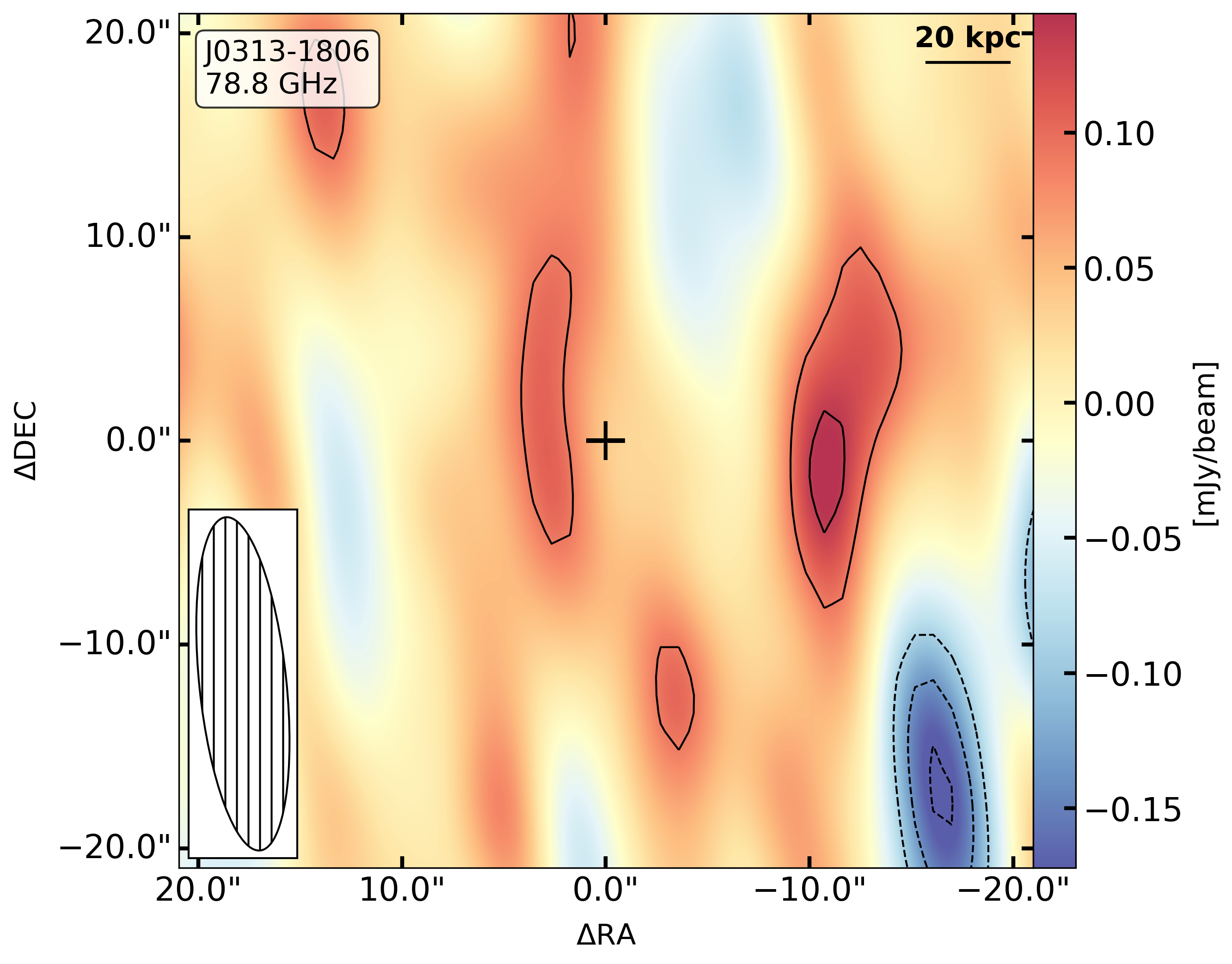} &
        \includegraphics[width=0.3\textwidth]{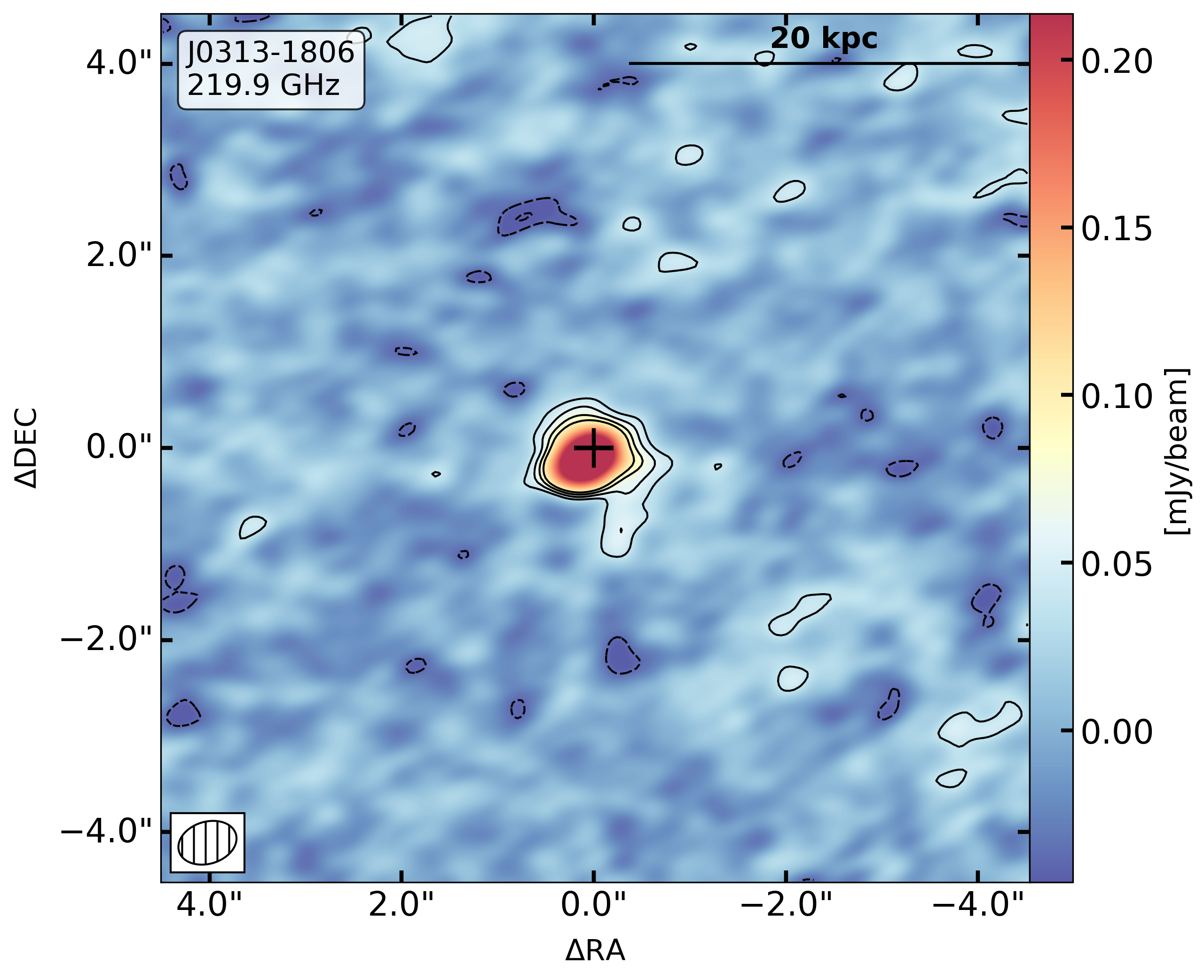} \\
        \includegraphics[width=0.3\textwidth]{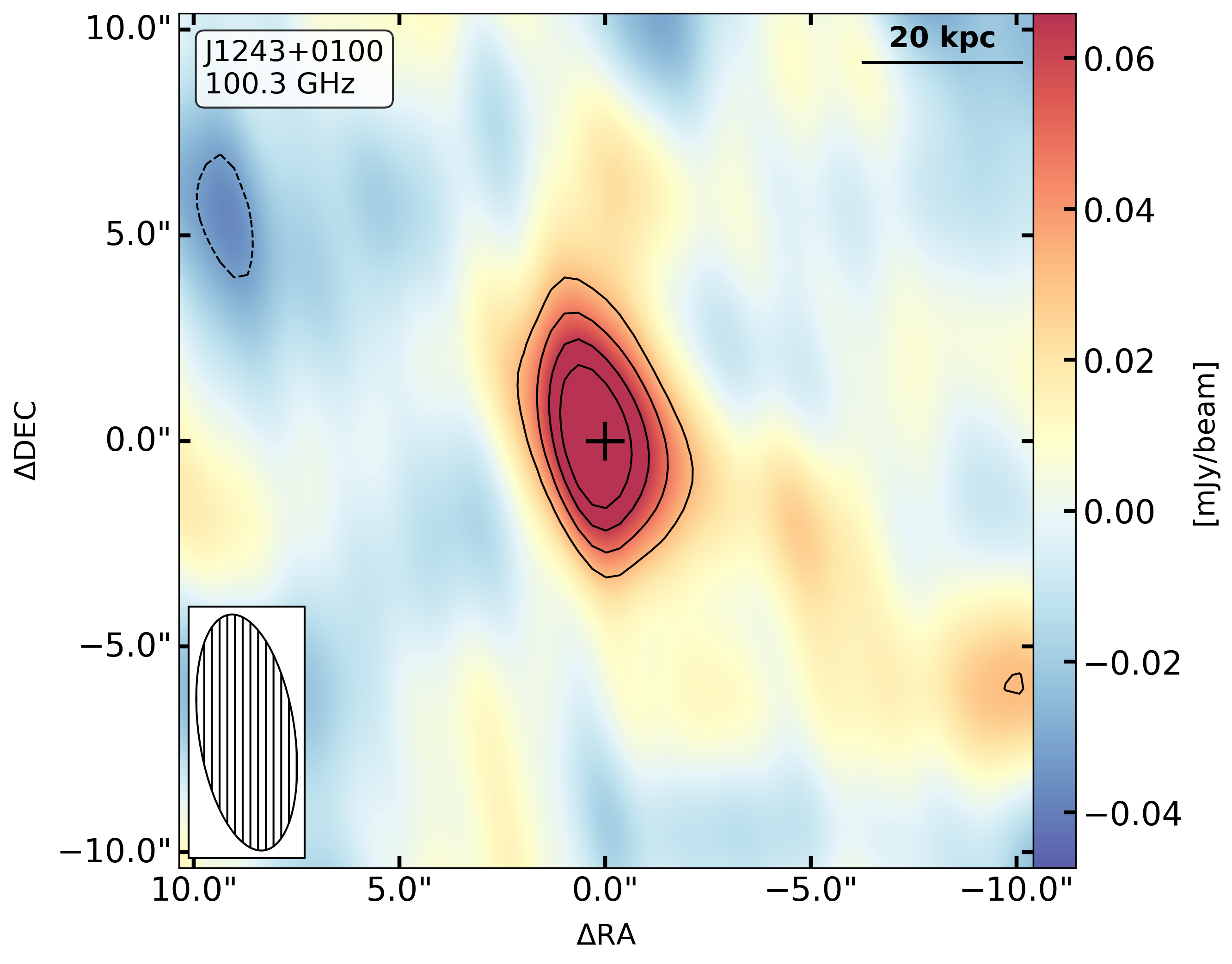} &
        \includegraphics[width=0.3\textwidth]{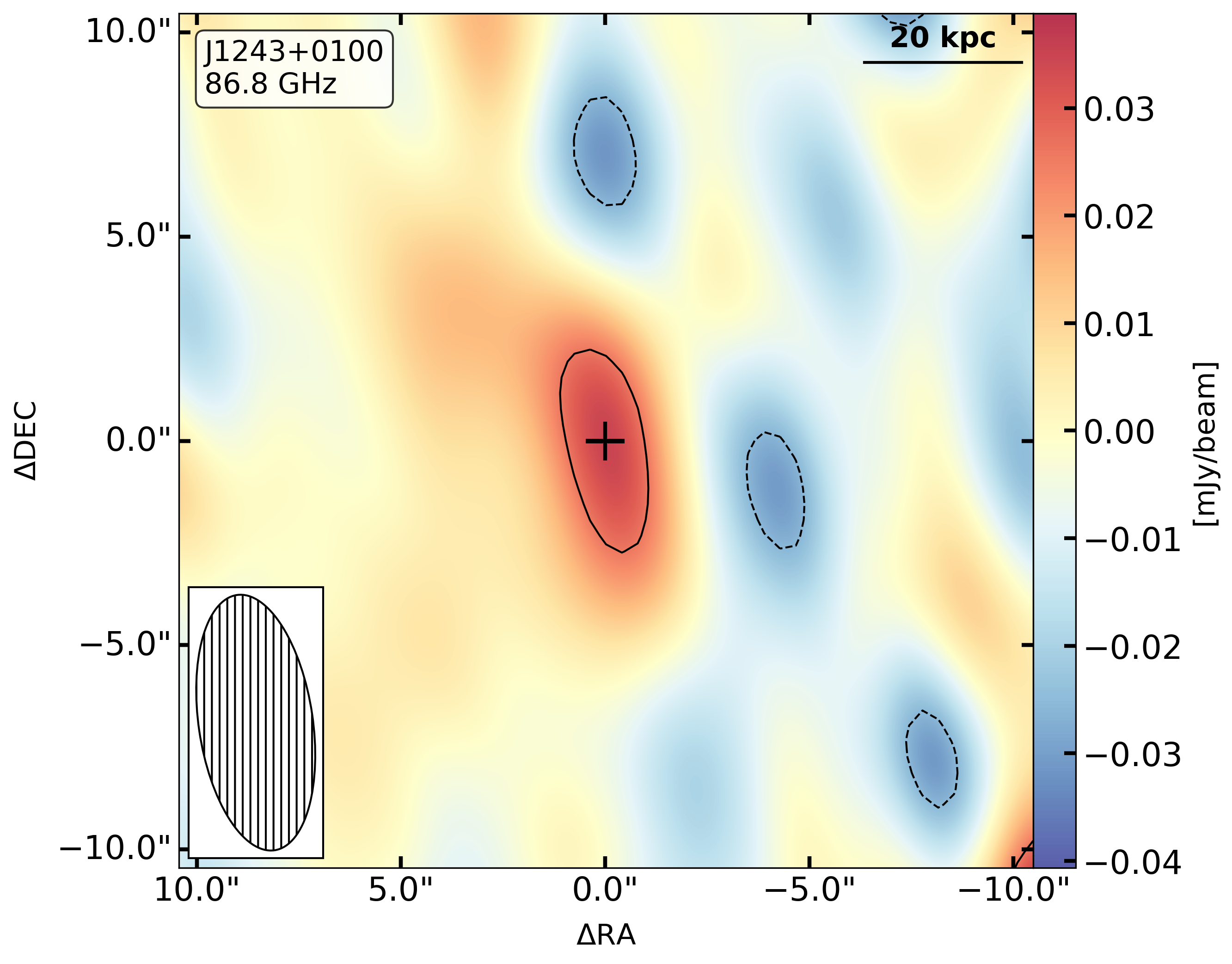} &
        \includegraphics[width=0.3\textwidth]{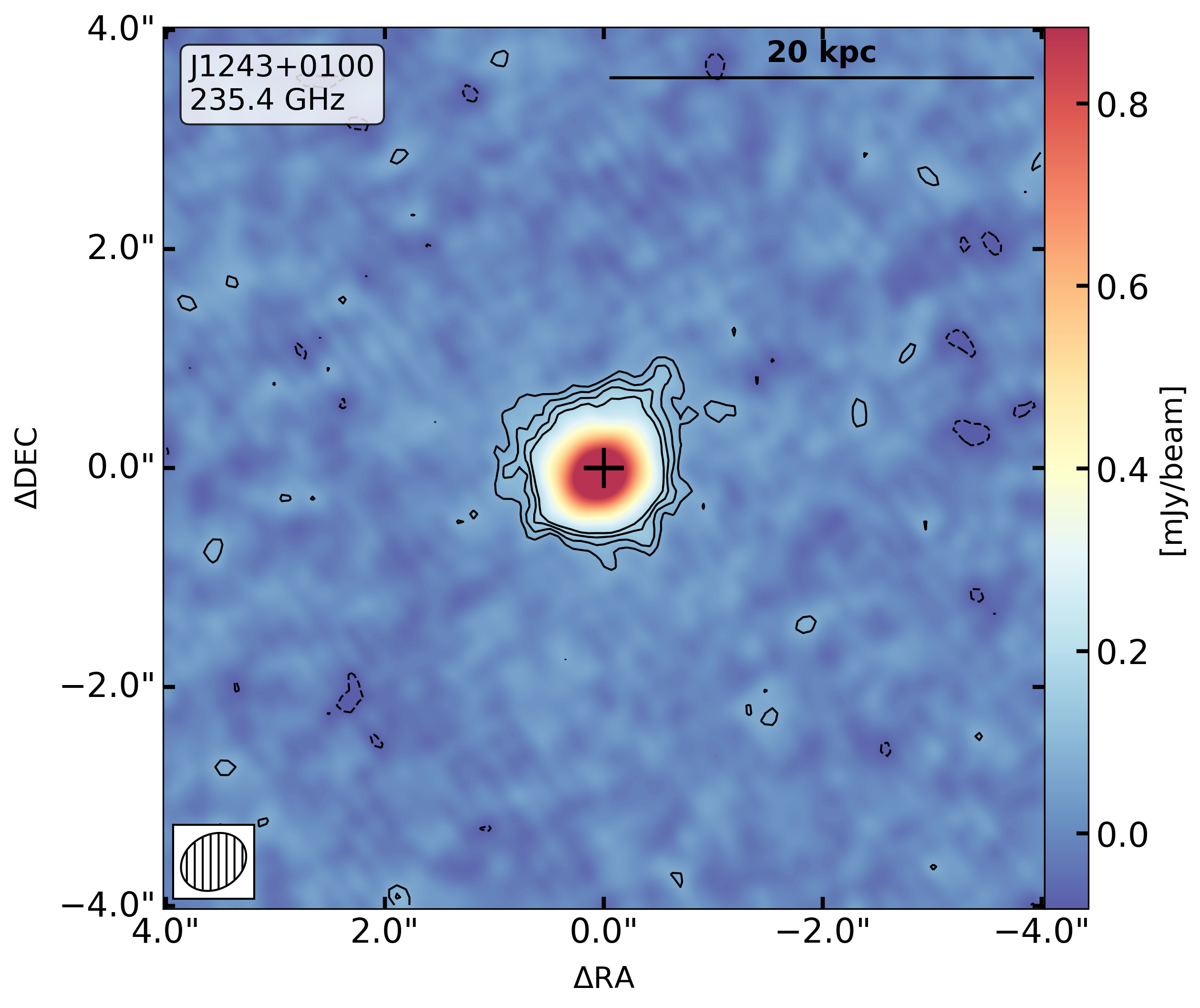} \\
        \includegraphics[width=0.3\textwidth]{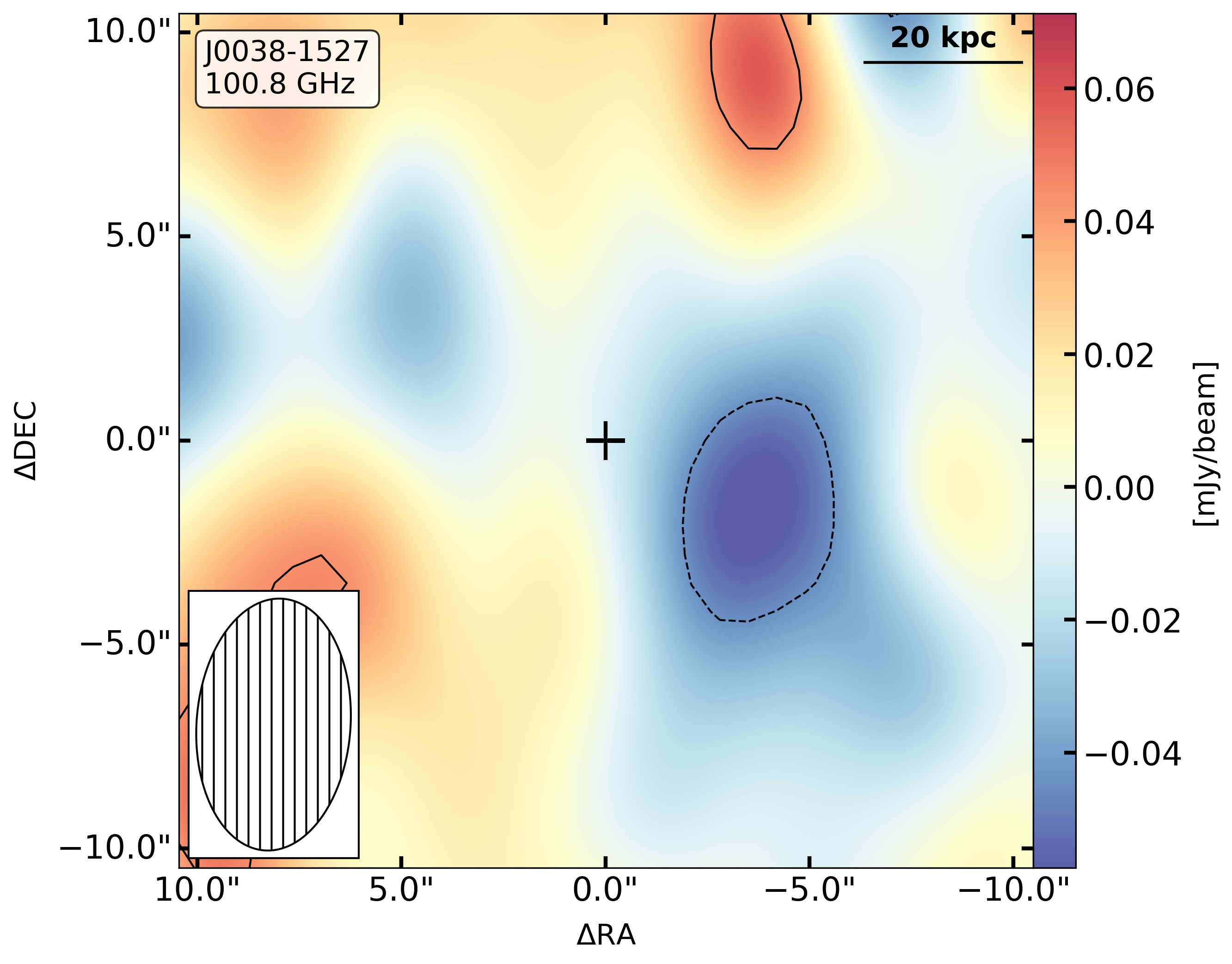} &
        \includegraphics[width=0.3\textwidth]{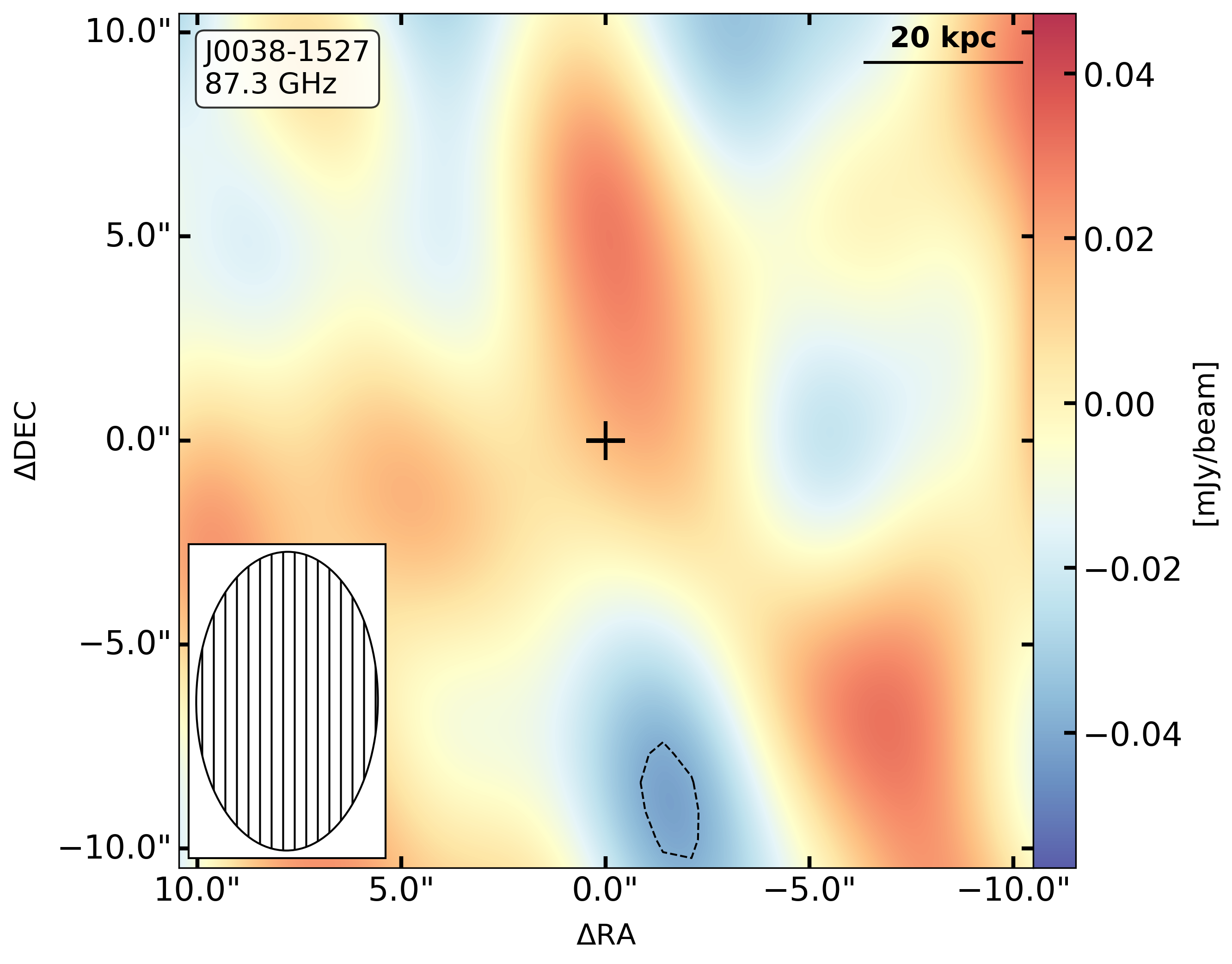} &
        \includegraphics[width=0.3\textwidth]{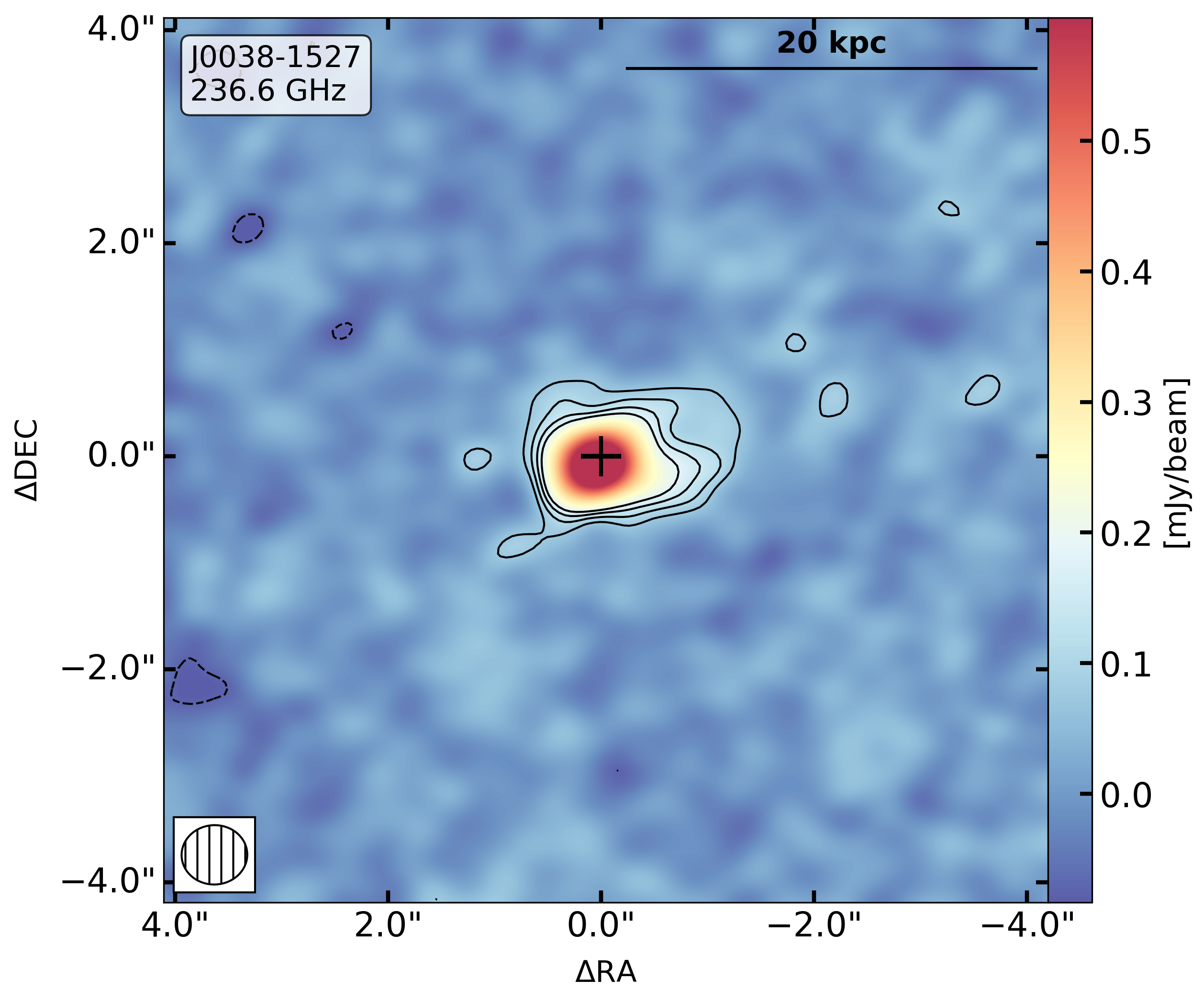} \\
        \includegraphics[width=0.3\textwidth]{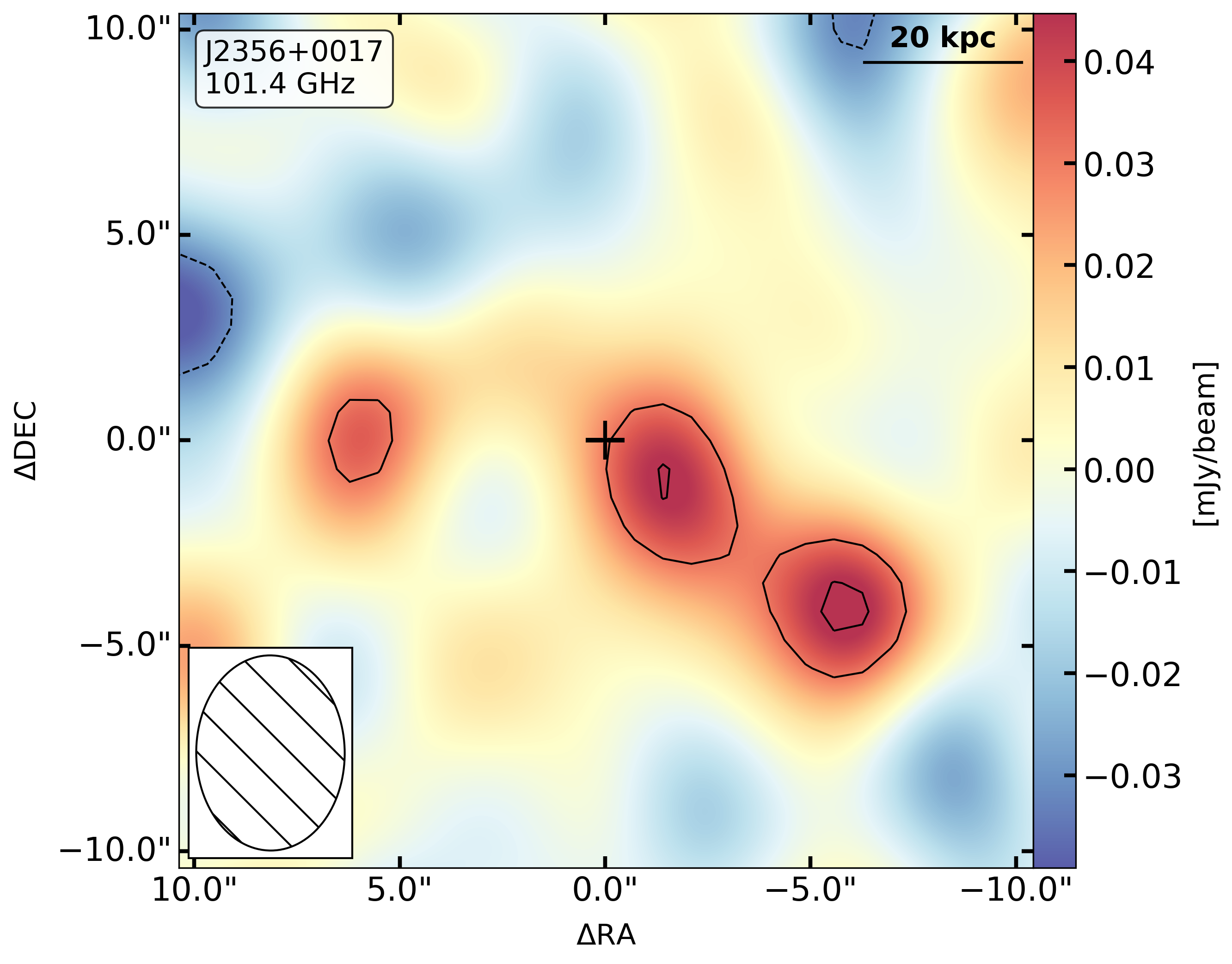} &
        \includegraphics[width=0.3\textwidth]{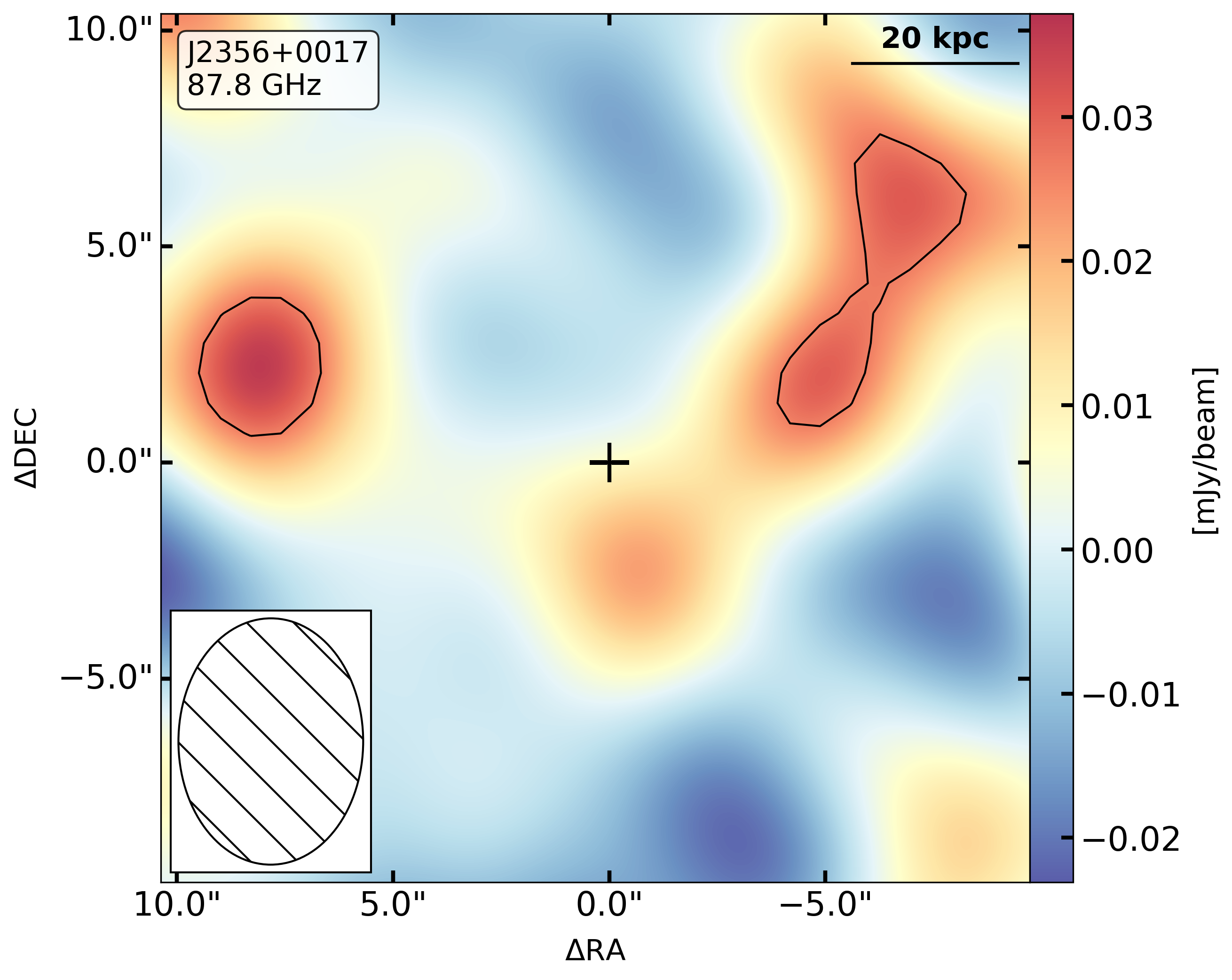} \\
        \includegraphics[width=0.3\textwidth]{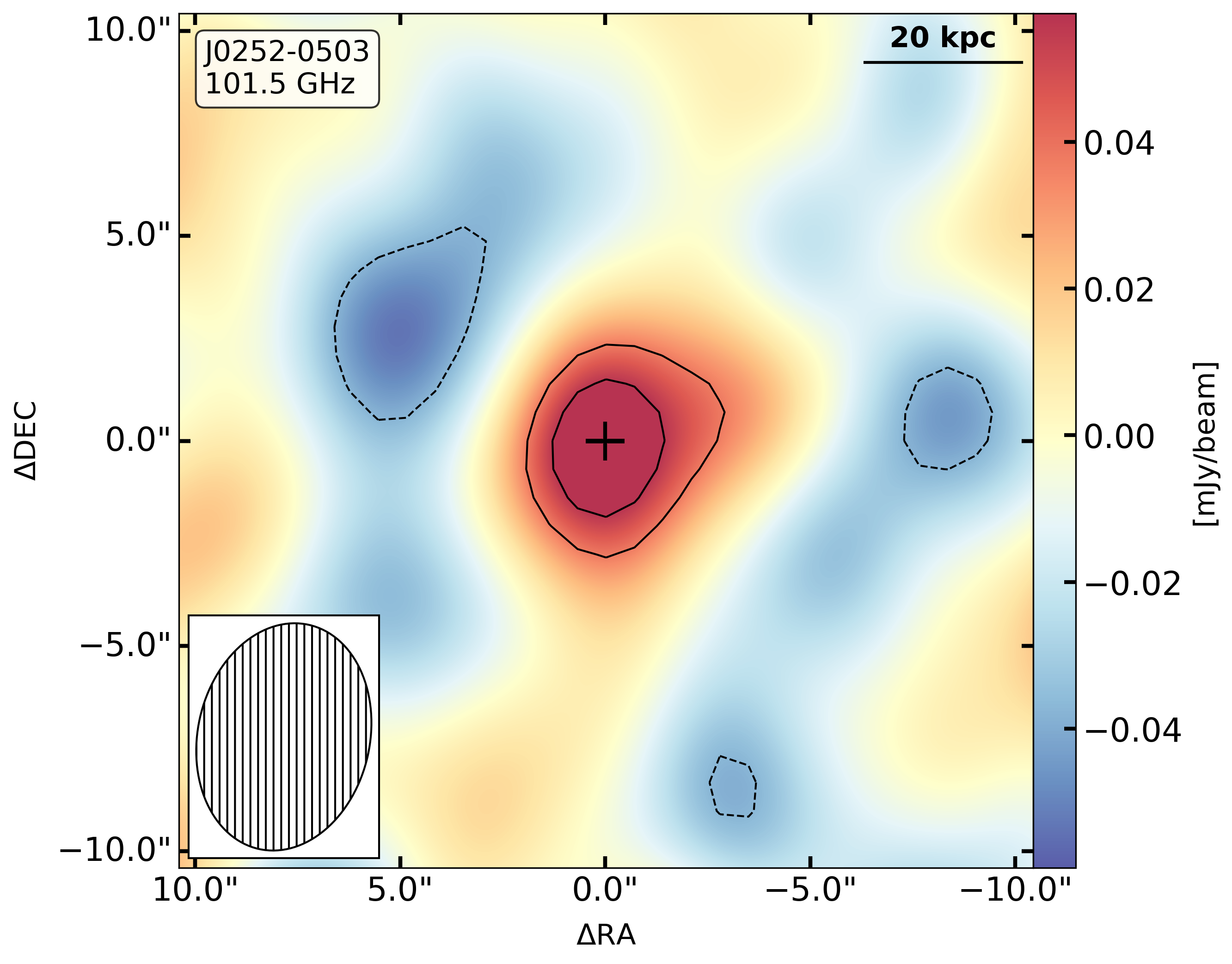} &
        \includegraphics[width=0.3\textwidth]{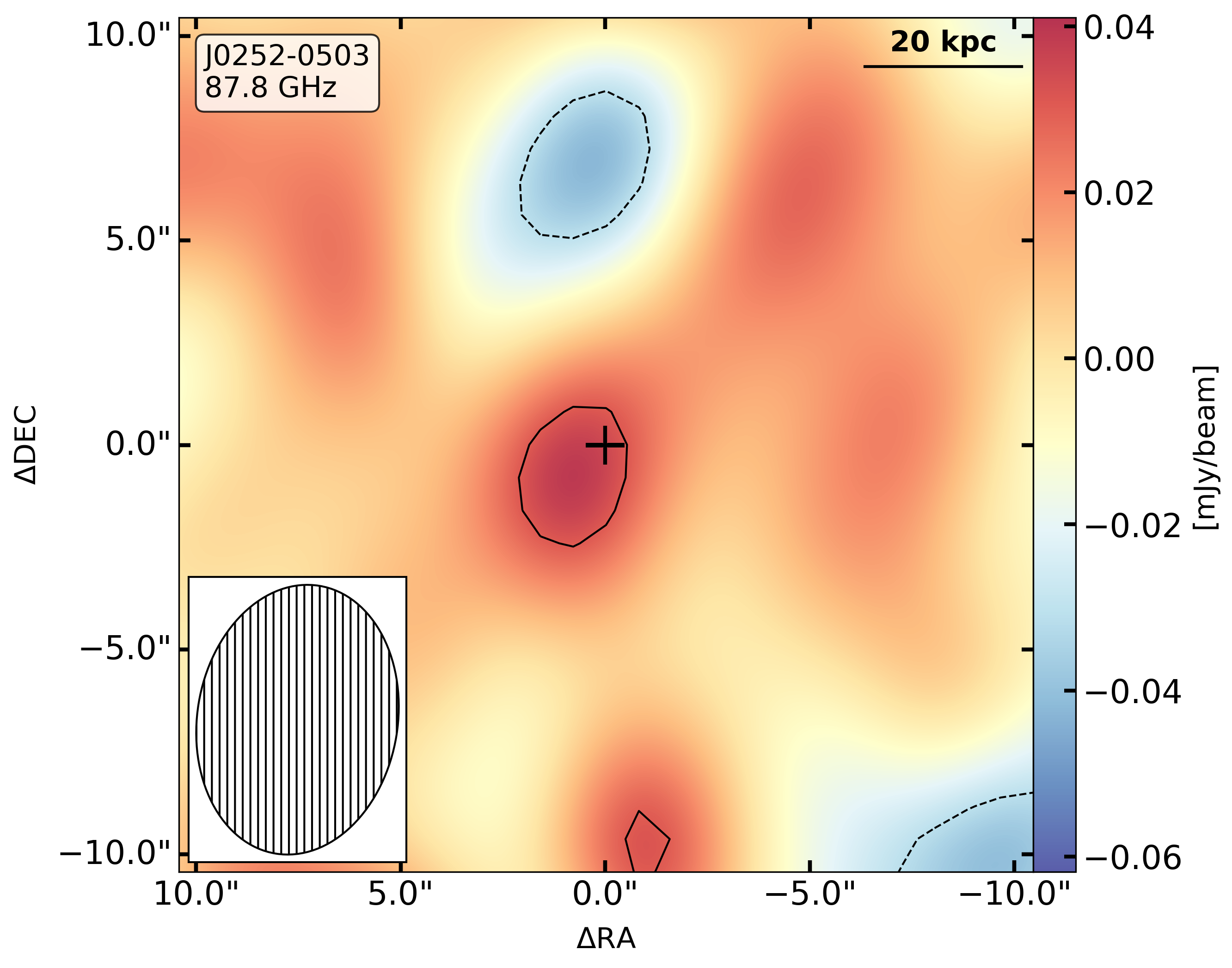} &
        \includegraphics[width=0.3\textwidth]{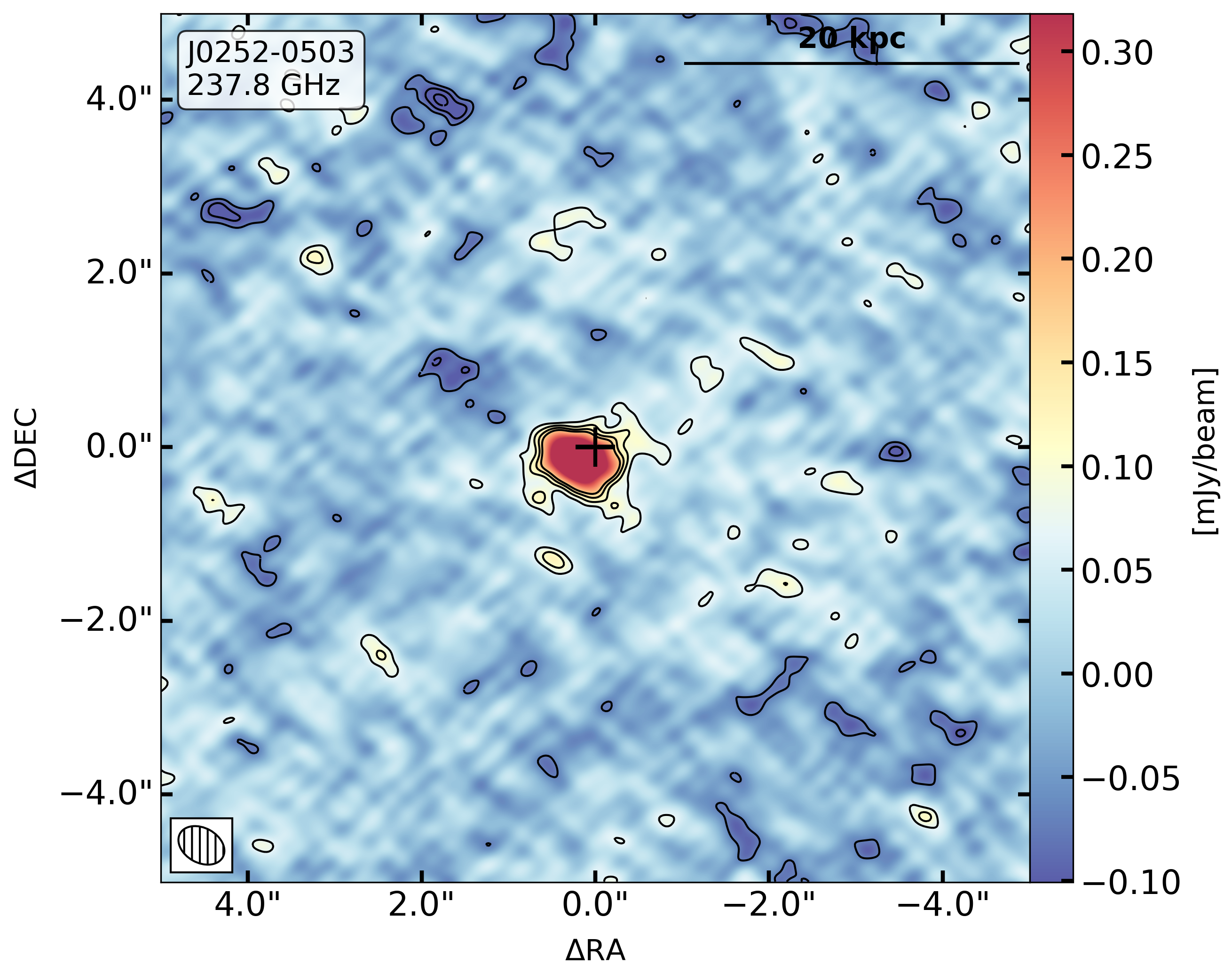} \\
    \end{tabular} 
    }
    \caption{Continuum maps. The source name and the observed frequency are in the upper-left corner. The optical position of the quasar is reported with a black cross.
Contour levels are -3, -2, 2, 3, 4, and 5 $\sigma$ significance; the RMS is listed in Table \ref{tab:observations}.
The clean beam for each observation is shown in the lower-left corner of the diagram, and the corresponding size is reported in Table \ref{tab:observations}.}
\label{fig:cont_maps}
\end{figure*}

In Fig.~\ref{fig:cont_maps} we present the continuum maps produced from the interferometric data collected and analyzed as described in Sect.~\ref{sec:obs}.
Flux densities were extracted from the continuum maps as described in Sect.~\ref{sec:obs}.

\FloatBarrier %\usepackage{placeins}

\newpage

\section{Comparison sample of high-z quasars}
\label{sec:app_pdr}
In Table~\ref{tab:comp_sample_z6} we report the information collected from the literature for 21 quasars at redshift $z\gtrsim6$ and a quasar at $z\sim5$. 
Quasars have been selected to have observations of CO emission lines, and properties derived from the FIR SED fitting.
In addition, 19 out of the 21 quasars at $z\gtrsim6$ have \lcii~measurement.
We report $L^{\prime}_{CO(1--0)}$ derived from $L^{\prime}_{CO(J+1--J)}$ available in the literature (see column $J_{up}$), assuming for all objects the same fixed CO SLED that we used in Sect.~\ref{sec:results}.
In particular, we assumed $r_{J+1,J}=L^{\prime}_{CO(J+1--J)}/L^{\prime}_{CO(1--0)}$ as follows: $r_{7,6}=0.65$, $r_{6,5}=0.9$, $r_{2,1}=1$.
The host-galaxy SFR is then derived from \lfir~by adopting a Chabrier IMF.
Moreover, we report only the measurement uncertainties for the sources for which these quantities were reported in the literature (see references).
During calculations and fitting procedures described in Sect.~\ref{sec:results}, for each measurement without uncertainty, we assigned an error that is equal to the median of the relative uncertainties available.

\begin{sidewaystable*}
\sisetup{separate-uncertainty}
\centering
\begin{tabular}{llllllllll} 
\hline\hline             
Source  &       RA, Dec  &       $z$     &       $\log(L_{bol})$ &       $\log(M_{BH})$  &       $(L_{[CII]})$   &       $\log(L_{TIR})$ &       $(L^{\prime}_{CO(1-0)})$        &       $\log(M_{dust})$    &   Ref     \\
\hline
        &       J2000   &               &       [erg/s]         &       [M$_{\odot}$]   &     $10^9$ L$_{\odot}$      &       [L$_{\odot}$]   &       $10^{10}$ K km/s pc$^2$ & [M$_{\odot}$]   &       \\
\hline
SDSSJ0100+2802  &       01:00:13.02 +28:02:25.80        &       6.3258  &       48.187  $\pm$   0.001   &       10.29   $\pm$   0.01    &       3.56    $\pm$   0.49    &       12.42   $\pm$   0.09    &       3.54    $\pm$   0.46    &       7.36    $\pm$   0.13    &       [1,8,13]        \\
J010953.13–304726.30  &       01:09:53.13     -30:47:26.32    &       6.7909  &       46.71   $\pm$   0.05    &       9.12    $\pm$   0.16    &       2.4     $\pm$   0.2     &       12.08   $\pm$   0.09    &       11.54   $\pm$   2       &       8.48    $\pm$   0.4     &       [2,9,15]        \\
VDESJ0224-4711  &       02:24:26.54     -47:11:29.40    &       6.522   &       47.53   $\pm$   0.03    &       9.11    $\pm$   0.06    &       5.43    $\pm$   0.15    &       13.52   $\pm$   0.2     &       1.48    $\pm$   0.22    &       7.97    $\pm$   0.11    &       [3,10]  \\
PSOJ036.5+03    &       02:26:01.87     +03:02:59.24    &       6.541   &       47.3    $\pm$   0.09    &       9.48    $\pm$   0.12    &       5.8     $\pm$   0.7     &       12.71   $\pm$   0.11    &       1.65    $\pm$   0.18    &       7.79    $\pm$   0.15    &       [2,8,16]        \\
J0305-3150      &       03:05:16.92     -31:50:55.90    &       6.6145  &       46.88   $\pm$   0.13    &       8.95    $\pm$   0.14    &       3.9     $\pm$   0.2     &       12.83   $\pm$   0.013   &       3.08    $\pm$   0.46    &       8.95    $\pm$   0.25    &       [2,9,16]        \\
SDSS J0338+0021 &       03:38:29.31     +00:21:56.30    &       5.0278  &       47.25   &                       9.4     $\pm$   0.12    &       5.69    $\pm$   0.83    &       13.08   &                       1.12    $\pm$   0.17    &       8.33    $\pm$   0.03    &       [4,11]  \\
J0840+5624      &       08:40:35.09     +56:24:19.90    &       5.8441  &       46.713  $\pm$   0.002   &       9.55    $\pm$   0.19    &               &               &       1.81    $\pm$   0.64$^a$        &                               &       [5,12]  \\
J0927+2001      &       09:27:21.82     +20:01:23.70    &       5.7716  &       46.99   $\pm$   0.003   &       9.69    $\pm$   0.12    &               &               &       4.75    $\pm$   1.15$^b$        &                               &       [5,12]  \\
PSO J159.2257–02.5438         &       10:36:54.19     -02:32:37.94    &       6.3822  &       47.26   $\pm$   0.008   &       9.68    $\pm$   0.05    &       1.19    $\pm$   0.07    &       12.2    &                       1.05    $\pm$   2       &       7.89    $\pm$   0.04    &       [1,11]  \\
VIK J1048–0109        &       10:48:19.09     -01:09:40.29    &       6.6766  &       47.023  $\pm$   0.012   &       9.53    $\pm$   0.11    &       2.77    $\pm$   0.08    &       12.89   &                       1.72    $\pm$   0.17    &       8.7     $\pm$   0.11    &       [1,11]  \\
DELS J1104+2134 &       11:04:21.58     +21:34:28.85    &       6.7672  &       47.18   $\pm$   0.03    &       9.23    $\pm$   0.04    &       1.82    $\pm$   0.28    &       12.7    &                       1.18    $\pm$   0.23    &       8.53    $\pm$   0.18    &       [6,11]  \\
SDSSJ1148+5251  &       11:48:16.10     +52:51:50.00    &       6.4189  &       47.538  $\pm$   0.002   &       9.94    $\pm$   0.02    &       4.2     $\pm$   0.35    &       13      $\pm$   0.06    &       3.2     $\pm$   0.3$^b$ &       8.51    $\pm$   0.1     &       [5,10,17]       \\
PSOJ183+05      &       12:12:26.98     +05:05:33.49    &       6.4386  &       47.351  $\pm$   0.006   &       9.67    $\pm$   0.06    &       25.3    $\pm$   0.4     &       13.08   $\pm$   0.15    &       3.51    $\pm$   0.39    &       8.65    $\pm$   0.08    &       [1,13,10]       \\
ULASJ1319+0950  &       13:19:11.29     +09:50:51.40    &       6.1336  &       47.247  $\pm$   0.001   &       9.46    $\pm$   0.03    &       4.4     $\pm$   0.9     &       12.99   $\pm$   0.008   &       1.8     $\pm$   0.23    &       8.8     $\pm$   0.16    &       [1,8,10]        \\
PSO231-20       &       15:26:37.84     -20:50:00.66    &       6.5864  &       47.28   $\pm$   0.09    &       9.48    $\pm$   0.19    &       2.87    $\pm$   0.15    &       13.28   $\pm$   0.09    &       2.15    $\pm$   0.2     &       8.72    $\pm$   0.04    &       [2,14,10]       \\
PJ308-21        &       20:32:10.00     -21:14:02.40    &       6.2354  &       47.34   $\pm$   0.004   &       9.41    $\pm$   0.07    &       1.73    $\pm$   0.14    &       12.42   $\pm$   0.18    &       <0.29                   &       7.7     $\pm$   0.2     &       [1,14,11]       \\
J2054-0005      &       20:54:06.49     -00:05:14.80    &       6.0397  &       47.092  $\pm$   0.013   &       9.34    $\pm$   0.05    &       3.3     $\pm$   0.5     &       12.73   $\pm$   0.007   &       0.86    $\pm$   0.31    &       8.03    $\pm$   0.09    &       [1,8,11]        \\
VIMOS2911001793 &       22:19:17.22     +01:02:48.90    &       6.1503  &       47      $\pm$   0.07    &       9.49    $\pm$   0.05    &       2.59    $\pm$   0.13    &       12.24   $\pm$   0.03    &       1.42    $\pm$   0.22    &       7.92    $\pm$   0.04    &       [7,8,11]        \\
PSO J338.2298+29.5089   &       22:32:55.15     +29:30:32.23    &       6.6668  &       47.61   $\pm$   0.16    &       9.43    $\pm$   0.15    &       2       $\pm$   0.1     &       12.31   $\pm$   0.11    &       0.94    $\pm$   0.23    &       7.96    $\pm$   0.05    &       [2,8,11]        \\
SDSSJ2310+1855  &       23:10:38.89     +18:55:19.70    &       6.0028  &       47.5    $\pm$   0.001   &       9.86    $\pm$   0.06    &       8.7     $\pm$   1.4     &       13.2    $\pm$   0.004   &       0       $\pm$   0       &       8.64    $\pm$   0.06    &       [5,8,18]        \\
J2348-3054      &       23:48:33.35     -30:54:10.28    &       6.9018  &       46.63   $\pm$   0.17    &       9.3     $\pm$   0.15    &       1.9     $\pm$   0.3     &       12.63   $\pm$   0.03    &       12.46   $\pm$   2.62    &       8.04    $\pm$   0.12    &       [2,9,16]        \\
PSOJ359-06      &       23:56:32.45     -06:22:59.26    &       6.1726  &       47.409  $\pm$   0.005   &       9.58    $\pm$   0.09    &       2.42    $\pm$   0.15    &       12.32   &                       0.89    $\pm$   0.25$^b$        &       7.97    $\pm$   0.04    &       [1,11]  \\
\hline
\end{tabular}
\flushleft 
\caption{Quasar properties. From left to right, columns include source name, coordinates, redshift, logarithm of bolometric luminosity, logarithm of the BH mass, [CII] luminosity, logarithm of the total infrared luminosity, CO(1-0) luminosity, upper J of the CO transition considered, logarithm of the dust mass.
CO(1-0) luminosities are derived by assuming a CO SLED from CO(7-6) emission line, apart from $^a$ and $^b$, where we used CO(2-1) and (6-5) transitions.
References for optical and FIR measurements are: [1] \cite{Shen11}; [2] \cite{Mazzucchelli17}; [3] \cite{WangF21}; [4] \cite{DietrichHamann04}; [5] \cite{Shen19}; [6] \cite{Yang21}; [7] \cite{Kashikawa15}; [8] \cite{Decarli18}; [9] \cite{Venemans17b}; [10] \cite{Tripodi24b}; [11] \cite{Decarli22}; [12] \cite{WangR11b}; [13] \cite{Decarli23}; [14] \cite{Pensabene22}; [15] \cite{Venemans17a}; [16] \cite{Kaasinen24}; [17] \cite{Stefan15}; [18] \cite{Feruglio18}.}
\label{tab:comp_sample_z6}
\end{sidewaystable*}

\end{appendix}
\end{document}